\documentclass[twocolumn]{aastex61}

\usepackage[switch]{lineno}

\usepackage{amsmath}
\usepackage{amssymb}

\newcommand\aastex{AAS\TeX}

\newcommand{\eqb}{\begin{eqnarray}}
\newcommand{\eqe}{\end{eqnarray}}

\newcommand{\sT}{\sigma_{\rm T}}
\newcommand{\p}{^\prime}

\newcommand{\g}{\gamma}
\newcommand{\gp}{\gamma^{\prime}}

\newcommand{\psim}{\lower.5ex\hbox{$\; \buildrel \propto \over\sim \;$}}
\newcommand{\lbar}{\lower.0ex\hbox{$\; \buildrel
{\lower0.0ex \hbox{-}} \over\lambda  \;$}}

\newcommand{\cm}{\mathrm{cm}}

\newcommand{\Gauss}{\mathrm{G}}

\newcommand{\fermi}{{\em Fermi}}


\shorttitle{\aastex\ Mrk\,421 2013 flare}
\shortauthors{Ahnen et al.}

\begin{document}


\title{Unravelling the complex behavior of Mrk\,421 with simultaneous X-ray and VHE observations during an extreme flaring activity in April 2013}

\correspondingauthor{David Paneque, Ana Babic, Justin Finke, Tarek Hassan, Maria Petropoulou}
\email{dpaneque@mppmu.mpg.de, ana.babic@fer.hr, justin.finke@nrl.navy.mil, thassan@ifae.es,   m.petropoulou@astro.princeton.edu}

%
\collaboration{MAGIC collaboration:}
\author{V.~A.~Acciari}
\affiliation{Inst. de Astrofísica de Canarias, E-38200 La Laguna, and Universidad de La Laguna, Dpto. Astrofísica, E-38206 La Laguna, Tenerife, Spain}
\author{S.~Ansoldi}
\affiliation{Università di Udine, and INFN Trieste, I-33100 Udine, Italy}
\affiliation{Japanese MAGIC Consortium: ICRR, The University of Tokyo, 277-8582 Chiba, Japan; Department of Physics, Kyoto University, 606-8502 Kyoto, Japan; Tokai University, 259-1292 Kanagawa, Japan; RIKEN, 351-0198 Saitama, Japan}
\author{L.~A.~Antonelli}
\affiliation{National Institute for Astrophysics (INAF), I-00136 Rome, Italy}
\author{A.~Arbet Engels}
\affiliation{ETH Zurich, CH-8093 Zurich, Switzerland}
\author{D.~Baack}
\affiliation{Technische Universität Dortmund, D-44221 Dortmund, Germany}
\author{A.~Babi\'c}
\affiliation{Croatian Consortium: University of Rijeka, Department of Physics, 51000 Rijeka; University of Split - FESB, 21000 Split; University of Zagreb - FER, 10000 Zagreb; University of Osijek, 31000 Osijek; Rudjer Boskovic Institute, 10000 Zagreb, Croatia}
\author{B.~Banerjee}
\affiliation{Saha Institute of Nuclear Physics, HBNI, 1/AF Bidhannagar, Salt Lake, Sector-1, Kolkata 700064, India}
\author{U.~Barres de Almeida}
\affiliation{Centro Brasileiro de Pesquisas Físicas (CBPF), 22290-180 URCA, Rio de Janeiro (RJ), Brasil}
\author{J.~A.~Barrio}
\affiliation{IPARCOS Institute and EMFTEL Department, Universidad Complutense de Madrid, E-28040 Madrid, Spain}
\author{J.~Becerra Gonz\'alez}
\affiliation{Inst. de Astrofísica de Canarias, E-38200 La Laguna, and Universidad de La Laguna, Dpto. Astrofísica, E-38206 La Laguna, Tenerife, Spain}
% ...
\collaboration{Other groups and collaborators:}
\author{M.\,Petropoulou}
\affiliation{Princeton University, Princeton NJ, USA}
\author{J.\,Finke}
\affiliation{US Naval Research Laboratory, Washington DC, USA}
\author{F.\,D'Ammando}
\affiliation{INAF—Istituto di Radioastronomia, Via P. Gobetti 101, I-40129 Bologna, Italy}
\author{M.\,Balokovi\'{c}}
\affiliation{Center for Astrophysics $\vert$ Harvard \& Smithsonian, 60 Garden Street, Cambridge, MA 02138, USA}
\affiliation{Black Hole Initiative at Harvard University, 20 Garden Street, Cambridge, MA 02138, USA}
\author{G.\,Madejski}
\affiliation{W.\,W. Hansen Experimental Physics Laboratory, Kavli Institute for Particle Astrophysics and Cosmology, Department of Physics and SLAC National Accelerator Laboratory, Stanford University, Stanford, CA 94305, USA}
\author{K.~Mori}
\affiliation{Columbia Astrophysics Laboratory, 550 W 120th St. New York, NY 10027, USA}
\author{Simonetta Puccetti}
\affiliation{Agenzia Spaziale Italiana (ASI)-Unita' di Ricerca Scientifica, Via del Politecnico, I-00133 Roma, Italy}
\author{C.\,Leto}
\affiliation{ASI Science Data Center, Via del Politecnico snc I-00133, Roma, Italy}
\affiliation{INAF - Osservatorio Astronomico di Roma, via di Frascati 33, I-00040 Monteporzio, Italy}
\author{M.\,Perri}
\affiliation{ASI Science Data Center, Via del Politecnico snc I-00133, Roma, Italy}
\affiliation{INAF - Osservatorio Astronomico di Roma, via di Frascati 33, I-00040 Monteporzio, Italy}
 \author{F.\,Verrecchia}
\affiliation{ASI Science Data Center, Via del Politecnico snc I-00133, Roma, Italy}
\affiliation{INAF - Osservatorio Astronomico di Roma, via di Frascati 33, I-00040 Monteporzio, Italy}
\author{M.\,Villata}
\affiliation{INAF - Osservatorio Astrofisico di Torino, 10025 Pino Torinese (TO), Italy}
\author{C.\,M.\,Raiteri}
\affiliation{INAF - Osservatorio Astrofisico di Torino, 10025 Pino Torinese (TO), Italy}

\author{I.\,Agudo}
\affiliation{Instituto de Astrof\'{i}sica de Andaluc\'{i}a (CSIC), Apartado 3004, E-18080 Granada, Spain}

\author{R.\,Bachev}
\affiliation{Institute of Astronomy and National Astronomical Observatory, Bulgarian Academy of Sciences, 72 Tsarigradsko shosse Blvd., 1784 Sofia, Bulgaria}

\author{A.\,Berdyugin}
\affiliation{Tuorla Observatory, Department of Physics and Astronomy, V\" ais\" al\" antie 20, FIN-21500 Piikki\"o, Finland}
\author{D.\,A.\,Blinov}
\affiliation{Astronomical Institute, St.\,Petersburg State University, Universitetskij Pr. 28, Petrodvorets, 198504 St.\,Petersburg, Russia}
\affiliation{Department of Physics and Institute for Plasma Physics, University of Crete, 71003, Heraklion, Greece}
\affiliation{Foundation for Research and Technology - Hellas, IESL, Voutes, 71110 Heraklion, Greece}

\author{R.\,Chanishvili}
\affiliation{Abastumani Observatory, Mt. Kanobili, 0301 Abastumani, Georgia}

\author{W.\,P.\,Chen}
\affiliation{Graduate Institute of Astronomy, National Central University, 300 Zhongda Road, Zhongli 32001, Taiwan}

\author{R.\,Chigladze}
\affiliation{Abastumani Observatory, Mt. Kanobili, 0301 Abastumani, Georgia}

\author{G.\,Damljanovic}
\affiliation{Astronomical Observatory, Volgina 7, 11060 Belgrade, Serbia}
\author{C.\,Eswaraiah}
\affiliation{Graduate Institute of Astronomy, National Central University, 300 Zhongda Road, Zhongli 32001, Taiwan}
\author{T.\,S.\,Grishina}
\affiliation{Astronomical Institute, St.\,Petersburg State University, Universitetskij Pr. 28, Petrodvorets, 198504 St.\,Petersburg, Russia}
\author{S.\,Ibryamov}
\affiliation{Institute of Astronomy and National Astronomical Observatory, Bulgarian Academy of Sciences, 72 Tsarigradsko shosse Blvd., 1784 Sofia, Bulgaria}
\author{B.\,Jordan}
\affiliation{School of Cosmic Physics, Dublin Institute For Advanced Studies, Ireland}
\author{S.\,G.\,Jorstad}
\affiliation{Astronomical Institute, St.\,Petersburg State University, Universitetskij Pr. 28, Petrodvorets, 198504 St.\,Petersburg, Russia}
\affiliation{Institute for Astrophysical Research, Boston University, 725 Commonwealth Avenue, Boston, MA 02215}
\author{M.\,Joshi}
\affiliation{Astronomical Institute, St.\,Petersburg State University, Universitetskij Pr. 28, Petrodvorets, 198504 St.\,Petersburg, Russia}
\author{E.\,N.\,Kopatskaya}
\affiliation{Astronomical Institute, St.\,Petersburg State University, Universitetskij Pr. 28, Petrodvorets, 198504 St.\,Petersburg, Russia}
\author{O.\,M.\,Kurtanidze}
\affiliation{Abastumani Observatory, Mt. Kanobili, 0301 Abastumani, Georgia}
\affiliation{Engelhardt Astronomical Observatory, Kazan Federal University, Tatarstan, Russia}
\affiliation{Center for Astrophysics, Guangzhou University, Guangzhou 510006, China}
\author{S.\,O.\,Kurtanidze}
\affiliation{Abastumani Observatory, Mt. Kanobili, 0301 Abastumani, Georgia}
\author{E.\,G.\,Larionova}
\affiliation{Astronomical Institute, St.\,Petersburg State University, Universitetskij Pr. 28, Petrodvorets, 198504 St.\,Petersburg, Russia}
\author{L.\,V.\,Larionova}
\affiliation{Astronomical Institute, St.\,Petersburg State University, Universitetskij Pr. 28, Petrodvorets, 198504 St.\,Petersburg, Russia}
\author{V.\,M.\,Larionov}
\affiliation{Astronomical Institute, St.\,Petersburg State University, Universitetskij Pr. 28, Petrodvorets, 198504 St.\,Petersburg, Russia}
\affiliation{Pulkovo Observatory, St.-Petersburg, Russia}
\author{G.\,Latev}
\affiliation{Institute of Astronomy and National Astronomical Observatory, Bulgarian Academy of Sciences, 72 Tsarigradsko shosse Blvd., 1784 Sofia, Bulgaria}
\author{H.\,C.\,Lin}
\affiliation{Graduate Institute of Astronomy, National Central University, 300 Zhongda Road, Zhongli 32001, Taiwan}
\author{A.\,P.\,Marscher}
\affiliation{Institute for Astrophysical Research, Boston University, 725 Commonwealth Avenue, Boston, MA 02215}
\author{A.\,A.\,Mokrushina}
\affiliation{Astronomical Institute, St.\,Petersburg State University, Universitetskij Pr. 28, Petrodvorets, 198504 St.\,Petersburg, Russia}
\affiliation{Pulkovo Observatory, St.-Petersburg, Russia}
\author{D.\,A.\,Morozova}
\affiliation{Astronomical Institute, St.\,Petersburg State University, Universitetskij Pr. 28, Petrodvorets, 198504 St.\,Petersburg, Russia}
\author{M.\,G.\,Nikolashvili}
\affiliation{Abastumani Observatory, Mt. Kanobili, 0301 Abastumani, Georgia}
\author{E.\,Semkov}
\affiliation{Institute of Astronomy and National Astronomical Observatory, Bulgarian Academy of Sciences, 72 Tsarigradsko shosse Blvd., 1784 Sofia, Bulgaria}
\author{P.S.\, Smith}
\affiliation{Steward Observatory, University of Arizona, 933 N. Cherry Ave., Tucson, AZ 85721, USA}
\author{A.\,Strigachev}
\affiliation{Institute of Astronomy and National Astronomical Observatory, Bulgarian Academy of Sciences, 72 Tsarigradsko shosse Blvd., 1784 Sofia, Bulgaria}
\author{Yu.\,V.\,Troitskaya}
\affiliation{Astronomical Institute, St.\,Petersburg State University, Universitetskij Pr. 28, Petrodvorets, 198504 St.\,Petersburg, Russia}
\author{I.\,S.\,Troitsky}
\affiliation{Astronomical Institute, St.\,Petersburg State University, Universitetskij Pr. 28, Petrodvorets, 198504 St.\,Petersburg, Russia}
\author{O.\,Vince}
\affiliation{Astronomical Observatory, Volgina 7, 11060 Belgrade, Serbia}
\author{J.\,Barnes}
\affiliation{Department of Physics, Salt Lake Community College, Salt Lake City, Utah 84070 USA}
\author{T.\,G\"uver}
\affiliation{Istanbul University, Science Faculty, Department  of Astronomy and Space Sciences, Beyaz\i t, 34119, Istanbul, Turkey}
\author{J.\,W.\,Moody}
\affiliation{Department of Physics and Astronomy, Brigham Young University, Provo Utah 84602 USA}
\author{A.\,C.\,Sadun}
\affiliation{Department of Physics, University of Colorado Denver, Denver, Colorado, CO 80217-3364, USA}
\author{T.\,Hovatta}
\affiliation{Finnish Centre for Astronomy with ESO (FINCA), University of Turku, FI-20014 Turku, Finland}
\affiliation{Aalto University, Mets\"ahovi Radio Observatory, Mets\"ahovintie 114, 02540 Kylm\"al\"a, Finland}
\author{J.\,L.\,Richards}
\affiliation{Department of Physics and Astronomy, Purdue University, West Lafayette, IN 47907, USA} 
\author{W.\,Max-Moerbeck}
\affiliation{Departamento de Astronomia, Universidad de Chile, Camino El Observatorio 1515, Las Condes, Santiago, Chile}
\author{A.\,C.\,R.\,Readhead}
\affiliation{Owens Valley Radio Observatory, California Institute of Technology, Pasadena, CA 91125, USA}
\author{A.\,L\"ahteenm\"aki}
\affiliation{Aalto University, Mets\"ahovi Radio Observatory, Mets\"ahovintie 114, 02540 Kylm\"al\"a, Finland}
\affiliation{Aalto University Department of Radio Science and Engineering, P.O. BOX 13000, FI-00076 Aalto, Finland}
\author{M.\,Tornikoski}
\affiliation{Aalto University, Mets\"ahovi Radio Observatory, Mets\"ahovintie 114, 02540 Kylm\"al\"a, Finland}
\author{J.\,Tammi}
\affiliation{Aalto University, Mets\"ahovi Radio Observatory, Mets\"ahovintie 114, 02540 Kylm\"al\"a, Finland}
\author{V.\,Ramakrishnan}
\affiliation{Aalto University, Mets\"ahovi Radio Observatory, Mets\"ahovintie 114, 02540 Kylm\"al\"a, Finland}
\author{R.\,Reinthal}
\affiliation{Tuorla Observatory, Department of Physics and Astronomy, V\" ais\" al\" antie 20, FIN-21500 Piikki\"o, Finland}
\AuthorCollaborationLimit=250



\begin{abstract}
We report on a multi-band variability and correlation study of the TeV blazar Mrk\,421 during an exceptional flaring activity observed from 2013 April 11 to 2013 April 19. The study uses, among others, data from GASP-WEBT, {\it Swift}, {\it NuSTAR}, {\it Fermi}-LAT, VERITAS, and MAGIC. The large blazar activity, and the 43~hours of  simultaneous {\it NuSTAR} and MAGIC/VERITAS observations, permitted variability studies on 15 minute time bins, and over three \mbox{X-ray} bands (3--7~keV, 7--30~keV and 30--80~keV) and three  very-high-energy ($>$0.1~TeV, hereafter VHE) \mbox{gamma-ray} bands \mbox{(0.2--0.4~TeV,} 0.4--0.8~TeV and $>$0.8~TeV). We detected substantial flux variations on multi-hour and sub-hour timescales in all the X-ray and VHE gamma-ray bands. The characteristics of the sub-hour flux variations are essentially energy-independent, while the multi-hour flux variations can have a strong dependence on the energy of the X-ray and the VHE gamma rays. The three VHE bands and the three X-ray bands are positively correlated with no time-lag, but the strength and the characteristics of the correlation changes substantially over time and across energy bands.  Our findings favour multi-zone scenarios for explaining the achromatic/chromatic variability of the fast/slow components of the  light  curves, as well as the changes in the flux-flux correlation on day-long timescales.  We interpret these results within a magnetic reconnection scenario, where the multi-hour flux variations are dominated by the combined emission from various plasmoids of different sizes and velocities, while the sub-hour flux variations are dominated by the emission from a single small plasmoid moving across the magnetic reconnection layer.

\end{abstract}




\section{Introduction}

Markarian\,421 (Mrk\,421), with a redshift of \mbox{$z=0.0308$}, is one of the closest BL Lac objects \citep[][]{1975ApJ...198..261U}, which happens to be also the first BL Lac object significantly detected at gamma-ray energies \citep[with EGRET,][]{1992ApJ...401L..61L}, and the first extragalactic object significantly detected at very-high-energy ($>$0.1~TeV, hereafter VHE) gamma rays \citep[with Whipple,][]{1992Natur.358..477P}. Mrk\,421 is also the brightest persistent \mbox{X-ray/TeV} blazar in the sky, and among the few sources whose spectral energy distribution (SED) can be accurately characterized by
current instruments from radio-to-VHE \citep{2011ApJ...736..131A}. Consequently, Mrk\,421 is among the few X-ray/TeV objects that can be studied with a great level of detail during both low and high activity \citep{fossati08,2015A&A...576A.126A,2016ApJ...819..156B}, and hence an object whose study maximizes our chances to understand the blazar phenomenon in general. 

Because of these reasons, every year since 2009, we organize 
extensive multiwavelength (MWL) observing campaigns where Mrk\,421 is monitored from radio-to-VHE gamma rays during the half year that it is visible with optical telescopes and Imaging Atmospheric Cherenkov Telescopes (IACTs). This multi-instrument and multi-year program provides a large {\it time} and {\it energy}
coverage that, owing to the brightness and proximity of Mrk\,421, yields the most detailed 
characterization of the broadband SED and its temporal evolution, compared to any other MWL campaign on any
other TeV target.

During the MWL campaign in the 2013 season, in the second week of 2013 April, we observed exceptionally high X-ray and VHE gamma-ray activity with the 
\textit{Neil Gehrels Swift Gamma-ray Burst Observatory} ({\it Swift}), the Nuclear Spectroscopic Telescope Array ({\it NuSTAR}), the Large Area Telescope onboard the {\it Fermi} Gamma-ray Space Telecope (\fermi-LAT), the Major Atmospheric Gamma Imaging Cherenkov telescope (MAGIC), and the Very Energetic Radiation Imaging Telescope Array System (VERITAS), as reported in various Astronomer's Telegrams \citep[e.g. see][]{2013ATel.4974...1B,2013ATel.4976...1C,2013ATel.4977...1P}. Among other things, the VHE gamma-ray flux was found to be two orders of magnitude larger than that measured during the first months of the MWL campaign in January and February 2013 \citep{2016ApJ...819..156B}. This enhanced activity triggered very deep observations with optical, X-ray and gamma-ray instruments, including a modified survey mode for \fermi\ from April 12 (23:00 UTC) until April 15th (18:00 UTC), which increased the LAT exposure on Mrk\,421 by about a factor of two. 

While Mrk\,421 has shown outstanding X-ray and VHE gamma-ray activity in the past \citep[e.g.][]{1996Natur.383..319G,fossati08,Mrk421Feb2010}, this is the most complete characterization of a flaring activity of Mrk\,421 to date. 
An extensive multi-instrument dataset was accumulated during nine consecutive days. It includes VHE observations with MAGIC, the use of public VHE data from VERITAS, and high-sensitive X-ray observations with {\it NuSTAR}. Notably, there are 43 hours of simultaneous VHE gamma-ray (MAGIC and VERITAS) and X-ray ({\it NuSTAR}) observations. A first evaluation of the X-ray activity measured with {\it Swift} and {\it NuSTAR} was reported in \citet{2015ApJ...811..143P}. This manuscript reports the full multi-band characterisation of this outstanding event, which includes, for the first time, a report of the VHE gamma-ray data, and it focuses on an unprecedented study of the X-ray-vs-VHE correlation in $3\times3$ energy bands. This study demonstrates that there is
a large degree of complexity in the variability in the X-ray and VHE gamma-ray domains, which relate to the most energetic and variable segments of Mrk\,421's SED, and indicates that the broadband emission of blazars require multi-zone theoretical models.

This paper is organized as follows: in Section~\ref{sec:obs} we briefly describe the observations that were performed, and in Section~\ref{sec:MWLCs} we report the measured multi-instrument light curves. Section~\ref{sec:variability} provides a detailed characterization of the multi-band variability, with special focus on the X-ray and VHE gamma-ray variations observed on April 15th. In Section~\ref{sec:correlations} we characterize the multi-band correlations observed when comparing the  X-ray emission in three energy bands, with that of VHE gamma rays in three energy bands. In Section~\ref{sec:discussion2} we discuss the implications of the observational results reported in this paper and finally, in Section~\ref{sec:conclusions}, we provide some concluding remarks.

\section{Observations and  Datasets}
\label{sec:obs}

\begin{deluxetable*}{   l c l c cc c}
\tablenum{1}
\tablecaption{MAGIC, VERITAS and {\it NuSTAR} Observations Overlap\label{tab:XVHEoverlap}}
\tablewidth{0pt}
\tablehead{
\colhead{Night} & \colhead{Date} & \colhead{Date} & \multicolumn2c{MAGIC + VERITAS} & \colhead{{\it NuSTAR}} & \colhead{VHE observations with }  \\
\nocolhead{  } & \colhead{ Apr 2013 } & \colhead{MJD} & \colhead{ bins\tablenotemark{a}}  & \colhead{simultaneous\tablenotemark{b}} & \colhead{bins\tablenotemark{a}} & \colhead{simultaneous X-ray coverage\tablenotemark{c} }  
}
\decimalcolnumbers
\startdata
1\dotfill & 10/11   & 56392/56393  &  21 + 15 &  0  &  43  &  30/36 (83\%)\\
2\dotfill & 11/12   & 56393/56394  &  24 + 25 &  2  &   65  &  33/47  (70\%)\\
3\dotfill & 12/13   & 56394/56395  &  23 + 14 &  4  &   17  &  14/33  (42\%)\\
4\dotfill & 13/14   & 56395/56396  &  25 + 19 &  3  &   30  &  29/41  (71\%)\\
5\dotfill & 14/15   & 56396/56397  &  25 + 24 &  2  &   30  &  31/47  (66\%)\\
6\dotfill & 15/16   & 56397/56398  &  16 + 11 &  0  &   30  &  20/27  (74\%)\\
7\dotfill & 16/17   & 56398/56399  &  10 +  0 &  0  &    32  &  4/10  (40\%)\\
8\dotfill & 17/18   & 56399/56400  &  13 +  0 &  0  &    20  &  6/13  (46\%)\\
9\dotfill & 18/19   & 56400/56401  &  10 +  0 &  0  &    19  &  6/10  (60\%)\\
all\dotfill & 10-19   & 56392 - 56401  &  167 + 108 &  11  &    286  &  173/264  (66\%)\\
\enddata
\tablenotetext{a}{Number of 15-minute time bins with observations  by the respective instrument.}
\tablenotetext{b}{Number of 15-minute time bins with measurements above 0.4 TeV in which MAGIC and VERITAS observed the source simultaneously.} 
\tablenotetext{c}{
The ratio of X-ray 15-bins simultaneous to the VHE 15-min bins (used as denominator), and percentage.} 
\end{deluxetable*}

The observations presented here are part of the multi-instrument
campaign  for Mrk\,421 that has occurred yearly since 2009
\citep{2011ApJ...736..131A}. The instruments that participate in this campaign can change somewhat from
year to year, but they are typically more than 20 covering energies from radio to VHE gamma rays. The 2013 campaign included observations from {\it NuSTAR} for the first
time, as a part of its primary mission \citep{2013ApJ...770..103H}. The instruments that participated in the 2013 campaign, as well as their performances and data analysis strategies, were reported in \citet{2016ApJ...819..156B}, which is our first publication with the 2013 multi-instrument data set, and focused on the low {\mbox X-ray/VHE} activity observed in January-to-March 2013.

During the first observations in April 2013, Mrk\,421 showed high X-ray and VHE gamma-ray activity, which triggered daily-few-hour-long multi-instrument observations that lasted from April 10 (MJD~56392) to April 19 (MJD~56401).
Among other instruments, this data set contains an exceptionally deep
temporal coverage at VHE gamma rays above 0.2~TeV, as
the source was observed with MAGIC during nine consecutive nights, and with VERITAS
during six nights \citep{2017AIPC.1792e0001B}. The geographical longitude of VERITAS is 93$^\circ$ (about six hours) west to that of MAGIC, and hence VERITAS
observations followed those from MAGIC, sometimes providing continuous VHE gamma-ray coverage during 10 hours in a single night.
 The total MAGIC observation time was nearly 42 hours, while for VERITAS it was 27 hours, yielding a total VHE
observation time of 69 hours in nine days (66 hours when counting once the MAGIC-VERITAS simultaneous observations). The time coverage of VHE data is slightly different for different VHE gamma-ray energies, being a few hours longer above 0.8~TeV in comparison to that below 0.4~TeV. This un-even coverage is due to the increased energy threshold associated with observations taken at large zenith angles. In the case of MAGIC, the energy threshold at zenith angles of about 60$^\circ$ is about 0.4 TeV, and hence the low-energy gamma-ray observations are not possible \citep[see ][for dependence of analysis energy threshold with the zenith angle of observations]{2016APh....72...76A}. MAGIC and
VERITAS observed Mrk\,421 simultaneously during 2hr 45min. The simultaneous observations between these two instruments occurred when MAGIC was observing at large zenith angle ($>$ 55$^\circ$), hence yielding simultaneous flux measurements only above 0.4~TeV. The extensive VHE coverage is particularly relevant since, as it will
be described in Section \ref{sec:MWLCs}, Mrk\,421 showed large variability and one of the brightest
VHE flaring activity recorded to date.  This unprecedented brightness allows us to match the sampling frequency of the simultaneous VHE (MAGIC and VERITAS) and X-ray ({\it NuSTAR}) observations to 15-min time intervals in three distinct energy bands: 0.2--0.4~TeV, 0.4--0.8~TeV and $>$0.8~TeV. 
The above-mentioned time-cadence and energy bands were chosen as a good compromise between having both, a good sampling of the multi-band VHE activity of Mrk\,421 during the April 2013 period, and reasonably accurate VHE flux measurements,  with the relative flux errors typically below 10\%.
See \citet{2017AIPC.1792e0001B} for the VERITAS photon fluxes.  In the case of the MAGIC light curves in the three energy bands, the small effect related to the event migration in energy was computed using the VHE gamma-ray spectrum from the full 9-day data set, which is well represented by the following log-parabola function 
\begin{equation}
\frac{d\Phi}{dE} = \left(\frac{E}{0.3\,\mathrm{TeV}}\right)^{
-(2.14)-(0.45)\cdot \log10 \left( \frac{E}{0.3\,\mathrm{TeV}}\right)}
\end{equation}

However, owing to the relatively narrow energy bands, the derived photon fluxes are not significantly affected by the specific choice of the used spectral shape: the photon fluxes derived with power-law spectra with indices $p=2$ and $p=3$ are in agreement, within the statistical uncertainties, with those derived with the 9-day log-parabolic spectral shape.

Using the simultaneous MAGIC and VERITAS observations, we noted a systematic offset of about 20\% in the VHE gamma-ray flux measurements derived with these two instruments. The VHE gamma-ray fluxes from VERITAS are systematically lower than those from MAGIC by a factor that is energy-dependent, being about 10\% in the 0.2--0.4~TeV band, and
about 30\% above 0.8~TeV. This offset, which is perfectly consistent with the known systematic uncertainties affecting each experiment \citep{Madhavan2013,2016APh....72...76A}, becomes evident due to the low statistical uncertainties associated with the flux measurements reported here. Appendix \ref{app:scalingfactors} reports a characterization of this offset, and describes the procedure that we followed for correcting it, scaling up the VERITAS fluxes to match those from MAGIC. The physics results reported in this manuscript do not depend on the absolute value of the VHE gamma-ray flux, and hence one could have scaled down the MAGIC fluxes to match those of VERITAS.  The correction applied is only relevant for the intra-night variability and correlation studies.

A key characteristic of this dataset is the extensive and
simultaneous coverage in the X-ray bands provided by {\it Swift} and,
especially, by {\it NuSTAR}. {\it Swift} observed Mrk\,421 for 18 hours, split in 63 observations spread over the nine days, and performed during the MAGIC and VERITAS observations. 
{\it NuSTAR} observed Mrk\,421 for 71 hours during the above-mentioned nine
days, out of which 43 hours were taken simultaneously to the
VHE observations from MAGIC and VERITAS.
The VHE and X-ray temporal coverage is summarized in Table~\ref{tab:XVHEoverlap}. 

The raw {\em NuSTAR} data were processed exactly as described in \citet{2016ApJ...819..156B}, except that in this study the {\em NuSTAR} analysis was performed separately for each 15-minute time bin with simultaneous VHE observations, as summarized in Table~\ref{tab:XVHEoverlap}. Using Xspec \citep{1996ASPC..101...17A}, we calculated fluxes in the 3--7~keV, 7--30~keV, and \mbox{30--80~keV} bands from a fit of a log-parabolic model to the data within each time bin. The cross-normalization between the two {\em NuSTAR} telescope modules was treated as a free parameter. The statistical uncertainties on fluxes were calculated as 68\,\% confidence intervals, and do not include the systematic uncertainty in absolute calibration, which is estimated to be 10--20\,\% \citep{2015ApJS..220....8M}.

The analysis procedures used to process the  {\it Swift}-XRT data are described in \citet{2016ApJ...819..156B}. In addition, in order to avoid additional flux uncertainties, we excluded 16 {\it Swift}-XRT observations in which Mrk\,421 was positioned near the CCD bad-columns \citep{2017AJ....153....2M}. Fig.~\ref{fig:mwl1} shows a comparison of the {\it Swift}-XRT and {\it NuSTAR} X-ray fluxes in the band \mbox{3--7~keV.} Overall, there is a good agreement between the two instruments, with flux differences typically smaller than 20\%. Such flux differences are within the systematic uncertainties in the absolute flux calibration of {\it NuSTAR}   \citep[][]{2015ApJS..220....8M} and {\it Swift}-XRT \citep[][]{2017AJ....153....2M}.

Differently to the \fermi-LAT analysis reported in \citet{2016ApJ...819..156B}, the LAT data results shown here were produced with events above 0.3~GeV (instead of 0.1~GeV) and with Pass8 (instead of Pass7). The analysis above 0.3~GeV is less affected by systematic uncertainties, and it is also less sensitive to possible contamination from non-accounted (transient) neighboring sources. The higher minimum energy reduces somewhat the detected number of photons from the source, but, owing to its hard gamma-ray spectrum (photon index$<$2.0), the effect is small. Specifically, we used the standard \fermi\ analysis software tools version v11r07p00, and the P8R3\_SOURCE\_V2 response function on events with energy above 0.3~GeV coming from a 10$^\circ$ region of interest (ROI) around Mrk\,421. We used a 100$^\circ$ zenith-angle cut to avoid contamination from the Earth's limb\footnote{A zenith-angle cut of 90$^\circ$ is needed if using events down to 0.1~GeV, but one can use a zenith-angle cut of 100$^\circ$ above 0.3~GeV without the need for using a dedicated Earth limb template.}, and modeled the diffuse Galactic and isotropic extragalactic background with the files gll\_iem\_v07.fits and iso\_P8R3\_SOURCE\_V2\_v1.txt respectively\footnote{\url{https://fermi.gsfc.nasa.gov/ssc/data/access/lat/BackgroundModels.html}}. All point sources in the fourth \fermi-LAT source catalog \citep[4FGL,][]{4FGLPaper} located in the 10$^\circ$ ROI and an additional surrounding 5$^\circ$-wide annulus were included in the model. In the unbinned likelihood fit, the spectral parameters were set to the values from the 4FGL, while the normalization of the diffuse components and the normalization parameters of the 16 sources (within the ROI) identified as variable were initially left free to vary. However, owing to the short timescales considered in this analysis, only two of these sources were significantly detected in 10 days: 4FGL~J1127.8+3618 and 4FGL~ J1139.0+4033, and hence we fixed the normalization of the other ones to the 4FGL catalog values.
 The \fermi-LAT spectrum from the 10-day time period considered here (from MJD~56392 to MJD~56402) is well described with a power-law function with a photon flux above 0.3~GeV of (23.3$\pm$1.6)$\times 10^{-8}$cm$^{-2}$s$^{-1}$ and photon index 1.79$\pm$0.05. A spectral analysis over 1-day and 12-hour time intervals shows that the photon index does not vary significantly throughout the 10-day period. The data was split into \mbox{12-hours-long} intervals centered at the VHE observations (e.g. simultaneous to the VHE) and their complementary time intervals (e.g. when there are no VHE observations), which are close to \mbox{12-hours-long} intervals. Owing to the limited event count in the \mbox{12-hour} time intervals, and the lack of spectral variability throughout the 10-day period, we fixed the shape to the power-law index to 1.79 (the value from the 10-day period) to derive the photon fluxes, keeping always the normalization factor for Mrk\,421 and the two above-mentioned 4FGL sources as a free parameters in the log-likelihood fit of each of the 12-hour time intervals.

The characterization of the activity of Mrk\,421 at optical frequencies was performed with many instruments from the GLAST-AGILE Support Program (GASP) of the Whole Earth Blazar Telescope (WEBT), here after GASP-WEBT
\citep[e.g.,][]{2008A&A...481L..79V, 2009A&A...504L...9V}, namely the observatory in Roque de los Muchachos (KVA telescope),  Lowell (Perkins telescope), Crimean, St. Petersburg, Abastumani, Rozhen (50/70 cm, 60 cm, and 200 cm telescopes), Vidojevica, and Lulin. Moreover, this study
also uses data from the iTelescopes, the Remote Observatory for Variable Object Research (ROVOR), and the TUBITAK National Observatory (TUG). The polarization measurements were performed with four observatories: Lowell (Perkins telescope), St. Petersburg, Crimean, and Steward (Bok telescope). The data reduction was done exactly as in \citet{2016ApJ...819..156B}.

Besides the 15~GHz and 37~GHz radio observations performed with the OVRO and Metsahovi telescopes, which were described in \citet{2016ApJ...819..156B}, here we also present a flux measurement performed with the IRAM~30m telescope at 86~GHz. This observation was performed under the Polarimetric Monitoring of AGN at Millimeter Wavelengths program \citep[POLAMI\footnote{http://polami.iaa.es},][]{2018MNRAS.474.1427A}, that regularly monitors Mrk\,421 in the short millimeter range. The POLAMI data was reduced and calibrated as described in \citet{2018MNRAS.473.1850A}.

\section{Multi-instrument light curves during the outstanding flaring activity in April 2013}
\label{sec:MWLCs}

\begin{figure*}[htp]
\begin{center}
\includegraphics[width=0.98\linewidth]{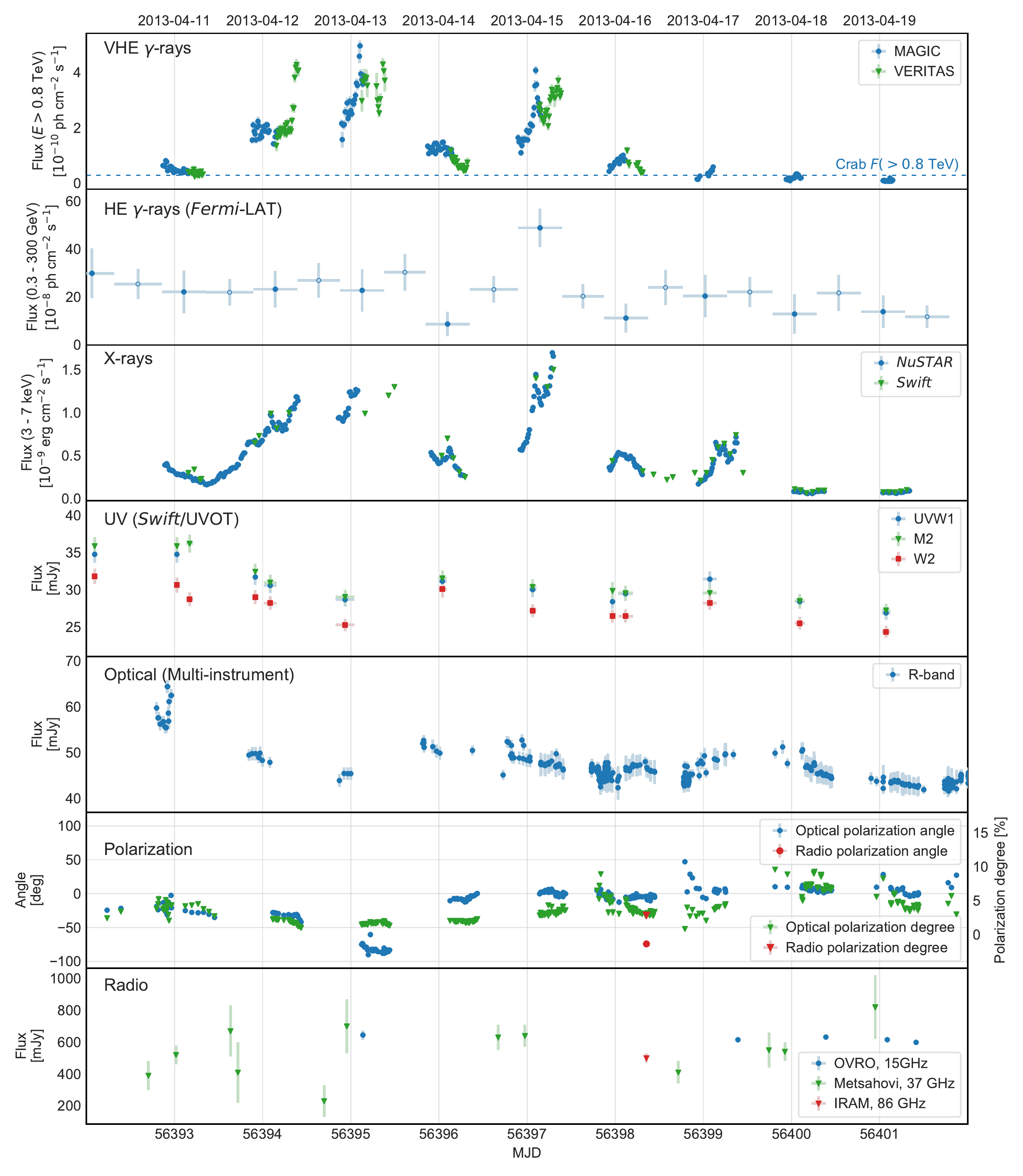}
\caption{Multiwavelength light curve for Mrk\,421 during the bright flaring activity in April 2013. The correspondence between the instruments and the measured fluxes is given in the legends. The horizontal dashed line in the VHE light curves represents the flux of the Crab Nebula, as reported in \citet{2016APh....72...76A}.
The VERITAS fluxes have been scaled using the coefficients described in Appendix~\ref{app:scalingfactors}. The filled markers in the \fermi-LAT panel depict the flux during the 12-hour time interval centered at the VHE observations, while the open markers denote the periods without corresponding observations in VHE bands. }
\label{fig:mwl1}
\end{center}
\end{figure*}

The multi-instrument light curves derived from all the observations spanning from radio to VHE gamma rays are shown in Fig.~\ref{fig:mwl1}. The top panel of Fig.~\ref{fig:mwl1} shows an excellent coverage of the 9-day flaring activity in the VHE regime, as a result of the combined
MAGIC and VERITAS observations. The peak flux at TeV energies, observed in April 13 (MJD~56395), reached up to 15 times the flux of the Crab Nebula, that is about 30 times the typical non-flaring activity of Mrk\,421, and about 150 times the activity shown a few months before, on January and February 2013, as reported in \citet{2016ApJ...819..156B}.
Moreover, this is the highest TeV flux ever measured with MAGIC for any blazar. This is also the third highest flux ever measured from a blazar with an IACT, after the extremely large outburst from Mrk\,421 detected with VERITAS in February 2010 \citep{Mrk421Feb2010} and the large flare from PKS~2155-304 detected by HESS in July 2006 \citep{2007ApJ...664L..71A}.

Figure~\ref{fig:mwl1} shows that the most extreme flux variations occur in the X-ray and the VHE gamma-ray bands.  At GeV energies, within the accuracy of the measurements, there is enhanced activity only on MJD~56397 (April 15th), when the flux is about a factor of two larger than the previous and following $\sim$12-hour time intervals. Interestingly, on April 15th we also find the highest X-ray flux, and the highest intra-night X-ray flux increase measured during this flaring activity in April 2013.

The R-band activity is comparable to the one measured in January-March 2013, when Mrk\,421 showed very low VHE and X-ray activity \citep{2016ApJ...819..156B}. The measured fluxes at optical wavelengths are large when compared to the flux levels typically seen during the period
of 2007-2015 \citep{2017MNRAS.472.3789C}. 
Generally, during the observations performed in January-April 2013,
Mrk\,421 was 4--5 times brighter in the optical than the photometric minima
that occurred in 2008-09 and at the end of 2011. Fig.~\ref{fig:mwl1} shows that
Mrk\,421 faded at R-band from
about 60~mJy on MJD~56393 to about 45~mJy two days later, and then varied
between 45-50~mJy during the following week, and appear decoupled from the VHE and X-ray activity.  The optical light curve
is in agreement with that of the less well sampled {\it Swift}/UVOT light curve.
Besides the optical brightness of Mrk\,421 in 2013, the object
showed a bluer optical continuum than average.  This was determined from
the differential spectrophotometry obtained by the Steward Observatory
monitoring program.  By comparing the instrumental spectrum of Mrk\,421 with
that of a nearby field comparison star, it is found that for the wavelengths 475~nm and 725~nm
$[F(475)/F(725)]_\mathrm{April-2013}/[F(475)/F(725)]_\mathrm{average}$ = 1.072~$\pm$~0.002,
where the average instrumental flux ratio is determined from all of the
available observations from 2008-2018.  The bluer color of Mrk\,421 is
consistent with a higher dominance of the non-thermal continuum over the host galaxy starlight included within the observing
aperture, which has a redder spectrum. This explanation for the observed
variations in the optical color of Mrk\,421 is further confirmed by the trend that the
continuum becomes slightly redder as the AGN generally fades during April 2013.  The same trend in color is also seen in the long-term near-IR data \citep{2017MNRAS.472.3789C}.

Optical linear polarization of Mrk\,421 was also monitored, and the
measurements are shown in Fig.~\ref{fig:mwl1}.  Again, the results are comparable to those measured during the first quarter in 2013, and reported in \citet{2016ApJ...819..156B}.
 Since 2008, the degree of polarization, $P$, has ranged
from 0\% to 15\%, although observations of $P>$~10\% are rare, about 10 out of around
1400 observations  \citep{2017MNRAS.472.3789C}.  During April 2013, the polarization ranged from
about 1--9\%, with a large majority of measurements showing $P<$~5\%.  The largest
changes in the degree of polarization on a daily time scale were an
increase from $P\sim$~3\% to $P\sim$~7\% on MJD~56399/400, followed by a decrease back
to about 4\% on the next day.  Changes of nearly as much as 5\% in polarization
are observed within a day, particularly on MJD~56398/99, but otherwise,
variations in $P$ are typically limited to $<$1\% over hour timescales.  
The electric vector position angle (EVPA) of the optical polarization was
at about -20$^\circ$ at the start of 2013 April.  Between MJD~56394
and MJD~56395 the EVPA rotated from about -30$^\circ$ to about -90$^\circ$ while generally
$P<$~2.5\%.  The largest daily rotation in EVPA occurs between MJD~56395 and MJD~56396, where the EVPA goes from about -90$^\circ$ to about -10$^\circ$.  Because of the
daily gap in the optical monitoring, it is unfortunately not clear if the
EVPA reversed its direction of rotation from MJD~56394 to MJD~56396 (i.e. 2 days), or
continued in the same direction requiring a rotation of $>$90$^\circ$ during one
of the two observing gaps on MJD~56394/6.  The variability of Mrk\,421
during the densely sampled portions of the optical monitoring does not
hint that such large changes in EVPA can take place on short time scales
until near the the end of MJD~56398 when a counterclockwise rotation of
about 50$^\circ$ is seen over a period of about 6 hours. Outside of this excursion,
the EVPA stays near 0$^\circ$ from MJD~56396 onward.  The single daily
deviation of EVPA to 90--100$^\circ$ on MJD~56395 coincides with brightest VHE
flare observed in April 2013.  However, no significant change in EVPA is
apparent during the sharp rise in VHE flux observed near the middle of
MJD~56394 or during the dramatic high-energy activity at the beginning of
MJD~56397.  For most of the monitoring period, the optical EVPA was near
the historical most likely angle for this object \citep[EVPA=0$^\circ$,][]{2017MNRAS.472.3789C}, although the one-day excursion on MJD~56395 brought the EVPA
nearly orthogonal to the most likely value.  For comparison, the 15 GHz
VLBI maps of Mrk\,421 show a jet detected out to about 5 mas at the position
angle of about -40$^\circ$ \citep{2019ApJ...874...43L}.
In the radio band the activity measured during the entire nine-day observing period is constant with a flux of about 0.6~Jy. The single 86~GHz measurement with IRAM~30m shows a polarization degree of about 3\%, which is similar to that of the optical frequencies; yet the polarization angle differs by about 70$^\circ$, which suggests that the optical and radio emission are being produced in different locations of the jet of Mrk\,421. Overall, the radio and optical fluxes, as well the optical polarization variations (polarization degree and EVPA) appear completely decoupled from the large X-ray and VHE gamma-ray activity seen in April 2013. In fact, the behavior observed at radio and optical during April 2013 is similar to the one observed during the previous months, when Mrk\,421 showed extremely low X-ray and VHE gamma-ray activity \citep[see][]{2016ApJ...819..156B}.

\begin{figure*}
\begin{center}
\includegraphics[width=\linewidth]{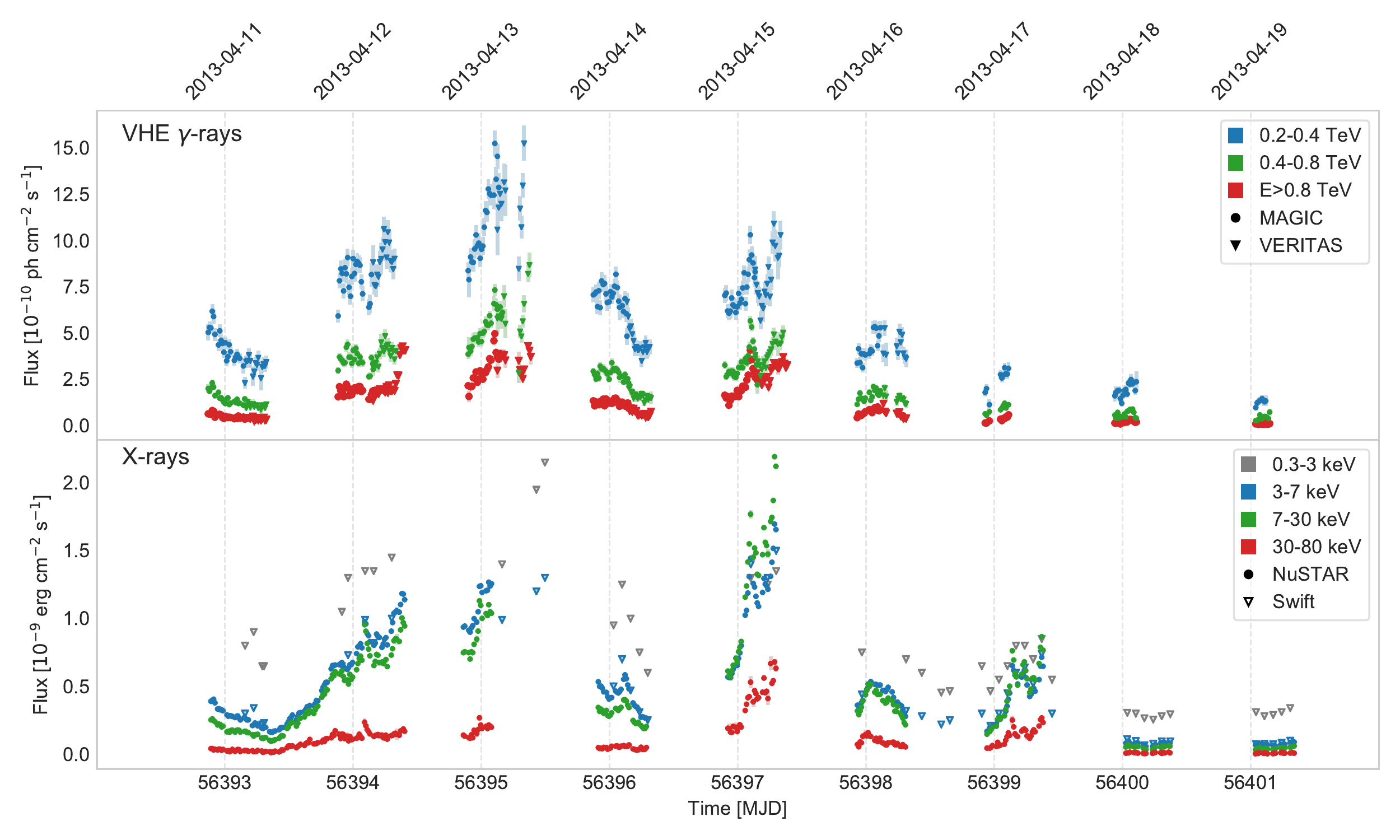}
\caption{Light curves in various VHE and X-ray energy bands obtained with data from MAGIC, VERITAS and {\it NuSTAR}  (split in 15 minute time bins) and {\it Swift}-XRT (from several observations with an average duration of about 17 minutes). For the sake of clarity, the 0.3-3~keV fluxes have been scaled with a factor 0.5. The statistical uncertainties are in most cases smaller than the size of the marker used to depict the VHE and X-ray fluxes.
\label{fig:mwl2}}
\end{center}
\end{figure*}

The hard X-ray and the VHE gamma-ray bands covered with {\it NuSTAR}, MAGIC, and VERITAS
are the most interesting ones because they exhibit the largest flux variations, and because of the
exquisite temporal coverage and the simultaneity in the dataset.  Fig.~\ref{fig:mwl2} reports the flux measurements in these bands,
 each split into three distinct bands, 3-7~keV, 7-30~keV and
30-80~keV\footnote{The upper edge of the NuSTAR energy range is actually 79~keV, but owing to the negligible impact on the flux values, in this paper we will use 80~keV for simplicity.} for {\it NuSTAR} and 0.2-0.4~TeV, 0.4-0.8~TeV and \mbox{$>$0.8~TeV} for MAGIC and VERITAS.  The temporal coverage for the band $>$0.8 TeV is a about 2 hours longer than for the band 0.2-0.4 TeV because of the increasing analysis energy threshold with the increasing zenith angle of the observations. The exquisite characterization of the multi-band flux variations in the X-ray and VHE gamma-ray bands reported in Fig.~\ref{fig:mwl2} will be used in the next sections for the broadband variability and correlation studies. 

\section{Multi-band \& multi-timescale variability}
\label{sec:variability}

\subsection{Fractional variability }
\label{subsec:fvar}

The flux variability reported in the multi-band light curves can be
quantified using the fractional variability
parameter $F_{\mathrm{var}}$, as prescribed in \citet{2003MNRAS.345.1271V}:

\begin{equation}
  F_{\mathrm{var}} = \sqrt{\frac{S^2 - <\sigma_{\mathrm{err}}^2 >}{<F_{\gamma}>^2}}
  \label{form_nva}
\end{equation}
$<F_{\gamma}>$ denotes the average photon flux, $S$ the
standard deviation of the $N$ flux measurements and
\mbox{$<\sigma_{\mathrm{err}}^2>$} the mean squared error, all
determined for a given instrument and energy band. 
The uncertainty on  $F_{\mathrm{var}}$ is calculated using the
prescription from \citet{Poutanen2008}, as described in
\citet{2015A&A...573A..50A}. 
This formalism allows one to quantify the variability amplitude, with
uncertainties dominated by the flux measurement errors, and the number
of measurements performed. 
The systematic uncertainties on the absolute flux
measurements\footnote{The systematic uncertainties in the flux
  measurements at the radio, optical, X-ray and GeV bands are of the
  order of 10--15\%, while for the VHE bands are $\sim$20--25\%.} do
not directly add to the uncertainty in $F_{\mathrm{var}}$.  The caveats in
the usage of $F_{\mathrm{var}}$ to quantify the variability in the flux
measurements performed with different instruments are described in
\citet{2014A&A...572A.121A,2015A&A...573A..50A,2015A&A...576A.126A}. The most important caveat is that the ability to quantify the variability depends on the temporal coverage (observing sampling) and the sensitivity of the instruments used, which is somewhat different across the electromagnetic spectrum. A big advantage of the study presented here is that the temporal coverage for three bands in X-rays (from {\it NuSTAR}) and three bands in VHE gamma rays (from MAGIC and VERITAS) is exactly the same, which allows us to make a more direct comparison of the variability in these energy bands.
                                                     
\begin{figure*}
\centering
\includegraphics[width=0.6\linewidth]{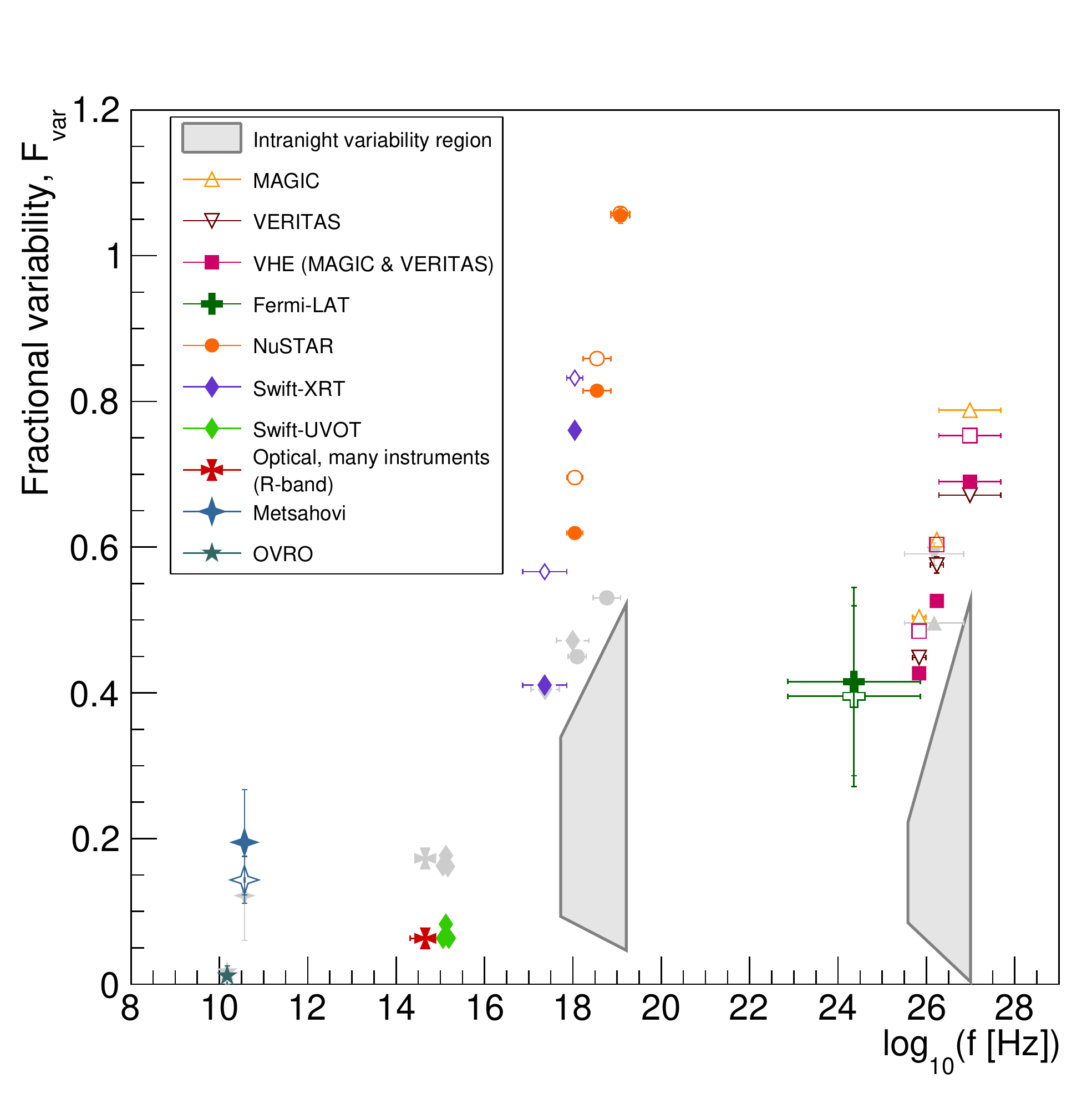}\\
\includegraphics[width=0.49\linewidth]{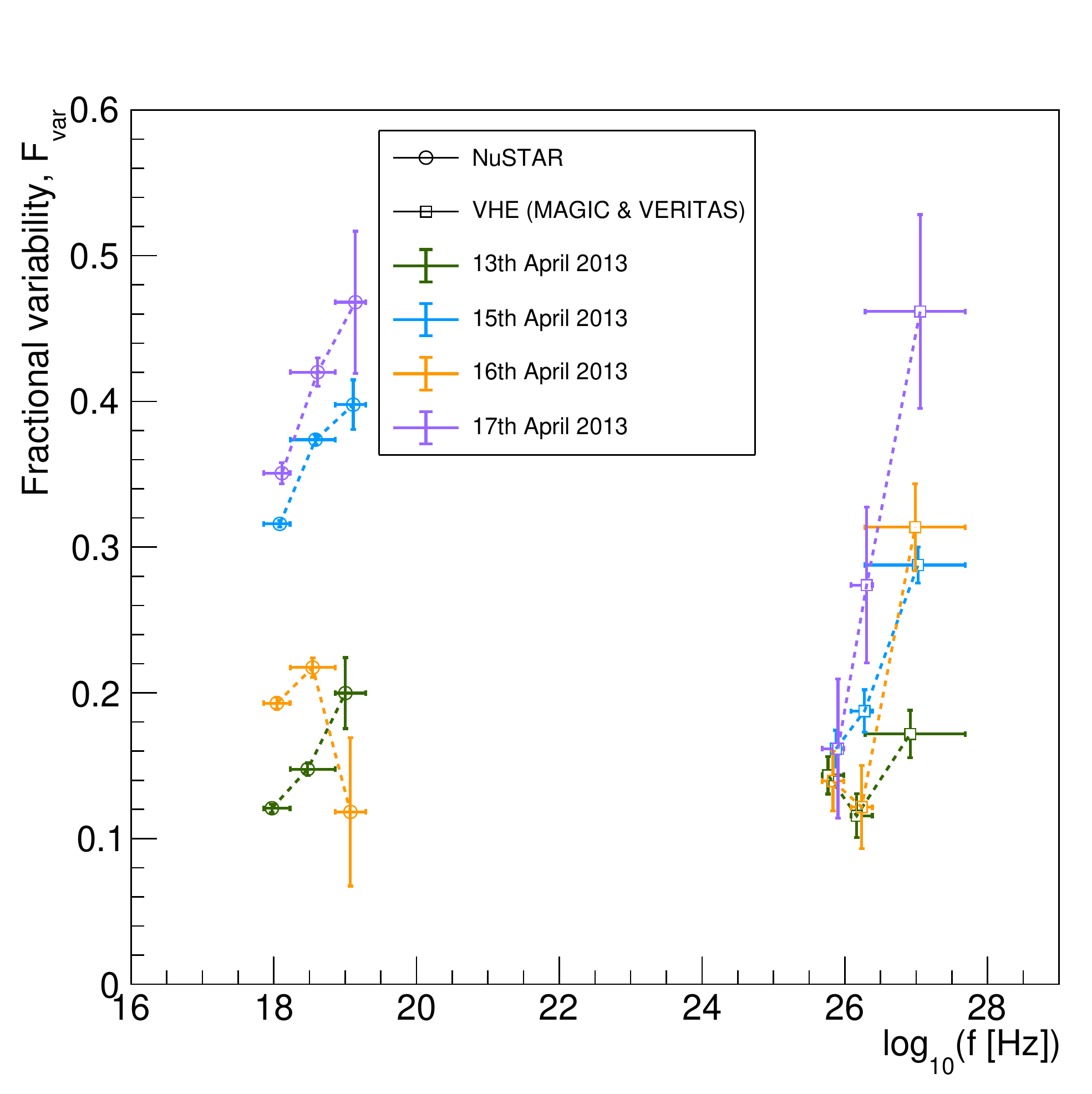}
\includegraphics[width=0.49\linewidth]{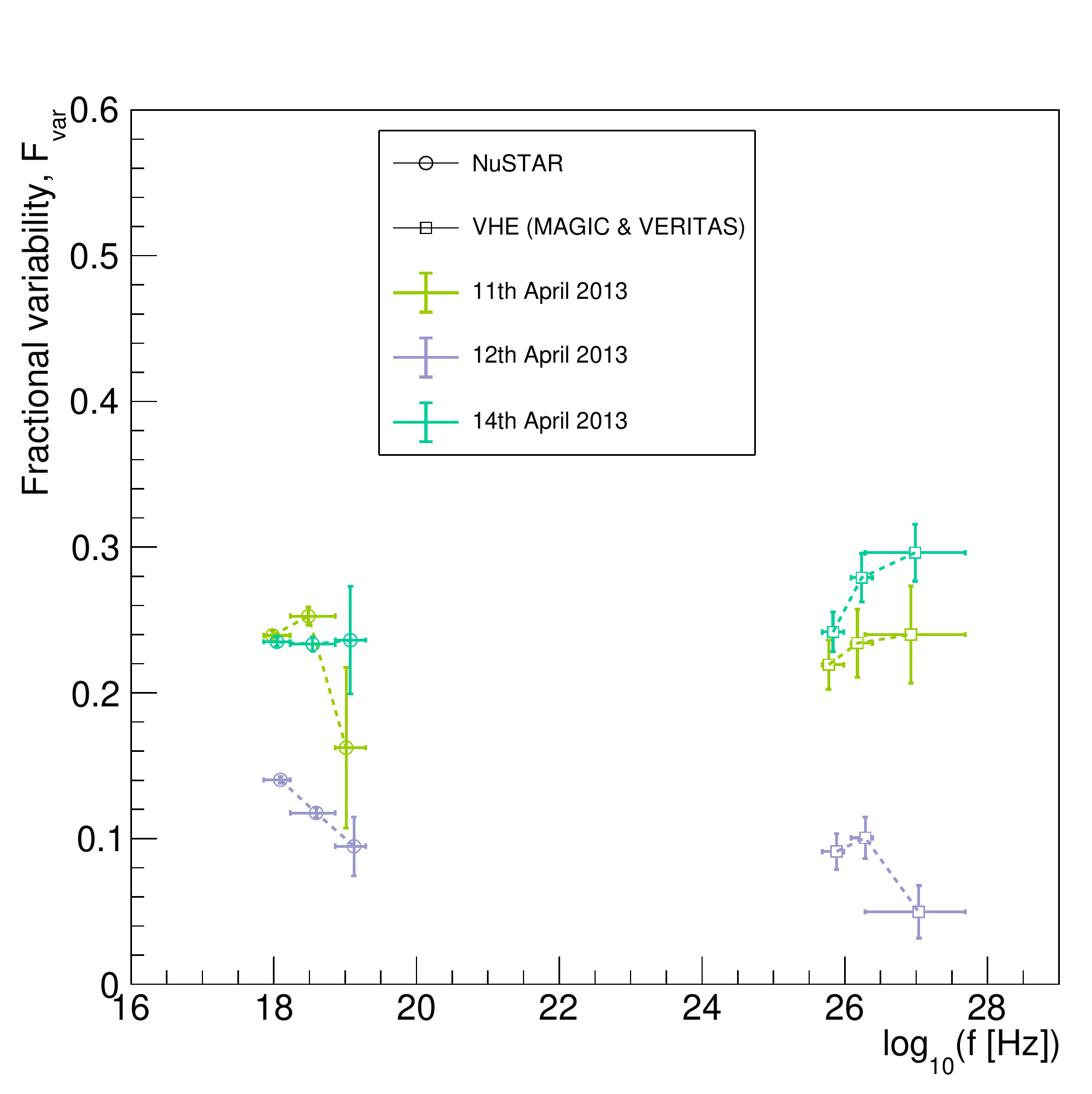}\\

\caption{Upper panel: Fractional variability $F_\mathrm{var}$ vs. energy band for the 9-day interval April 11-19. The panel reports the variability obtained using all available data (open symbols), and using only data which are taken simultaneously (filled symbols).  For comparison, the  $F_\mathrm{var}$ vs. energy  obtained with data from January-March 2013 \citep[see][]{2016ApJ...819..156B} are also depicted with grey markers. The grey shadowed regions depict the range of $F_\mathrm{var}$ obtained with data from single-night light curves, as shown in the lower panels.  Bottom panels: $F_\mathrm{var}$ vs. energy for the three X-ray and VHE bands, using simultaneous observations, and calculated for each night separately. In all plots, the vertical error bars depict the 1$\sigma$ uncertainty, while the horizontal error bars indicate the energy range covered. In order to improve the visibility of the data points, some markers have been slightly shifted horizontally (but always well within the horizontal bar). 
}
\label{fig:fvar1}
\end{figure*}

The fractional variability parameter $F_{\mathrm{var}}$ was computed using the flux values and
uncertainties reported in the light curves from Section
\ref{sec:MWLCs} (see Fig.~\ref{fig:mwl1} and Fig.~\ref{fig:mwl2}), hence providing a quantification of variability
amplitude for this nine-day long flaring activity from radio to VHE
gamma-ray energies. The results are depicted in the upper panel of
Fig.~\ref{fig:fvar1}, where open markers are used for the variability computed with all the available data, and the filled markers are used for simultaneous observations. Given the slightly different temporal coverage for different VHE bands, as described in the previous section, we decided to use the 0.2--0.4~TeV band to define the time slots for simultaneous X-ray/VHE observations. This ensures that the same temporal bins are being used for the 3$\times$3 X-ray and VHE bands. For comparison purposes, we added the $F_{\mathrm{var}}$  values obtained 
for the period January to March 2013, when Mrk\,421 showed a very low activity
\citep[see][]{2016ApJ...819..156B}.

The fractional variability plot shows the typical
double-bump structure, which is anlogous to the broadband SED. This
plot shows that most of the flux variations occur in the X-ray and VHE
bands, which correspond to the falling segments of the SED. Additionally, it
also shows that, during the nine-day  flaring activity in April 2013,
the amplitude variability in the hard X-ray band was substantially
larger than that measured during the low activity from January to March 2013. 
The higher the X-ray energy, the larger the difference between the
$F_{\mathrm{var}}$ values from the low and the high activity.

In addition to the study of nine-day behavior, the high photon fluxes and the deep exposures allow us to compute the $F_{\mathrm{var}}$ with the single-night light curves from six consecutive nights (from April 11 to April 17)\footnote{The light curves from April 18 and April 19 contain little data ($\sim$2 hours) and little variability, which prevents the calculation of significant ($>$3 sigma) variability for most of the energy bands.}, hence allowing
us to study the fractional variability on hour timescales for three
X-ray bands and three VHE gamma-ray bands. For this study, only simultaneous data (using the time bins from the 0.2--0.4~TeV band) were used, which means that the 3$\times$3 X-ray/VHE bands sample exactly the same source activity. The results are depicted in
the bottom panels of Fig.~\ref{fig:fvar1}. In general, all $F_{\mathrm{var}}$ values computed with the single-night light curves are lower than those derived with the 9-day light curve for the corresponding energy band. This is clearly visible when comparing the data points with the grey shadowed regions in the upper panel of Fig.~\ref{fig:fvar1}. Despite the X-ray and VHE flux varying on sub-hour timescales, the resulting intra-night
fractional variability is significantly lower than the overall fractional variability in the
nine-day time interval.  This result is expected because, while for single days the light
curves show flux variations within a factor of about 2, the nine-day light
curve shows flux variations larger than a factor of about 10. 
Unexpectedly, we find a large diversity in the variability versus energy patterns observed for the different days. For the days April 13, 15, 16, 17, one finds the typical pattern of higher fractional variability at higher photon energy within each of the two SED bumps. On the other hand, one finds that the fractional variability is approximately constant with energy for the days April 11 and April 14, and that the fractional variability decreases with energy for April 12. The decrease in $F_{\mathrm{var}}$ with increasing energy is only marginally significant in the VHE bands ($\sim$2$\sigma$), but very prominent in the X-ray emission, which within the synchrotron self-Compton (SSC) scenario, provides a direct mapping to the energy of the radiating electrons. These different variability versus energy  patterns suggest the existence of diverse causes (or regions) responsible for the variability in the broadband blazar emission on timescales as short as days and hours. This is the first time that the variability of Mrk\,421 can be studied with this level of detail, and the implications will be discussed in Section \ref{sec:discussion2}.

\subsection{Flux variations on multi-hours and sub-hours timescales}
\label{subsec:fastvariability}

This section focuses on the flux variations observed in the hard X-ray and VHE gamma-ray bands, which
are the ones with the largest temporal coverage and highest
variability (see section \ref{subsec:fvar}). The light curves for all
nights for these 3x3 energy bands are reported in Appendix \ref{app:NormLCs}.
There is clear intra-night variability in all the light curves, which can be significantly detected because of the high fluxes and the good temporal coverage, as described in the previous section (e.g. see Fig.~\ref{fig:fvar1}). The single-night light curves show a large diversity of temporal structures that relate to different timescales, from sub-hours (i.e. fast variation) to multi-hours (trends). We note that some of these fast components are present in both X-rays and VHE gamma rays, while some others are visible only at X-rays, or only at VHE gamma rays, and, in some cases, the features are present only in specific bands (either X-rays or VHE gamma rays) and not in the others. 
As it occurred with the study of the $F_{var}$ vs energy, the evaluation of the single-night multi-band light curves also suggests that there are different mechanisms responsible for the variability, some of them being achromatic (affecting all energies in similar way) and others chromatic (affecting the different energy bands in substantially different manner).

In this section we attempt to quantify the main trends and fast features, as well as their evolution across the various energy bands. We do that by fitting with a function formed by a \em{slow} trend $F_s (t)$ and \em{fast} feature $F_f (t)$ components.

\begin{equation}
 F(t) = F_s  (t) + F_f (t)
 \label{eq:flux_t}
\end{equation}

where 

\begin{equation}
  F_s (t) = Offset \cdot (1+ Slope \cdot t)
 \label{eq:slow_t}
\end{equation}

and 

\begin{equation}
  F_f (t) = \frac{2}{2^{-\frac{t-t_0}{t_\mathrm{rise}}} + 2^{\frac{t-t_0}{t_\mathrm{fall}}}}  \cdot A \cdot F_s(t_0)
 \label{eq:fast_t}
\end{equation}

Here $A$ is the flare amplitude, $t$ is the time since midnight for the chosen night, $t_0$ is the time of the peak flux of the flare, and $t_\mathrm{rise}$ and
$t_\mathrm{fall}$ are the flux-doubling timescales for the rising and falling part of the flare. 
This formulation, with the slope of the slow component normalized to the offset, and the flare amplitude (of the fast component) normalized to the slow component at  $t_0$, enables a direct comparison of the parameter values among the different energy bands, for which the overall measured flux may differ by factors of a few. 

In general, we find that, whenever fast flares occur, they appear to be quite symmetric and, given the relatively short duration (sub-hour timescales) and the flux measurement uncertainties, we do not have the ability to distinguish (in a statistically meaningful way) between different rise- and fall-doubling times. For the sake of simplicity, we decided to fit the light curves with a function given by Eq.~\ref{eq:fast_t} where $t_\mathrm{rise}=t_\mathrm{fall}=$~flux-doubling time. This fit function provides a fair representation of the intra-night rapid flux variations from all days but for April 16, where the flux variations have much longer (multi-hour) timescales.

\begin{figure*}
\plottwo{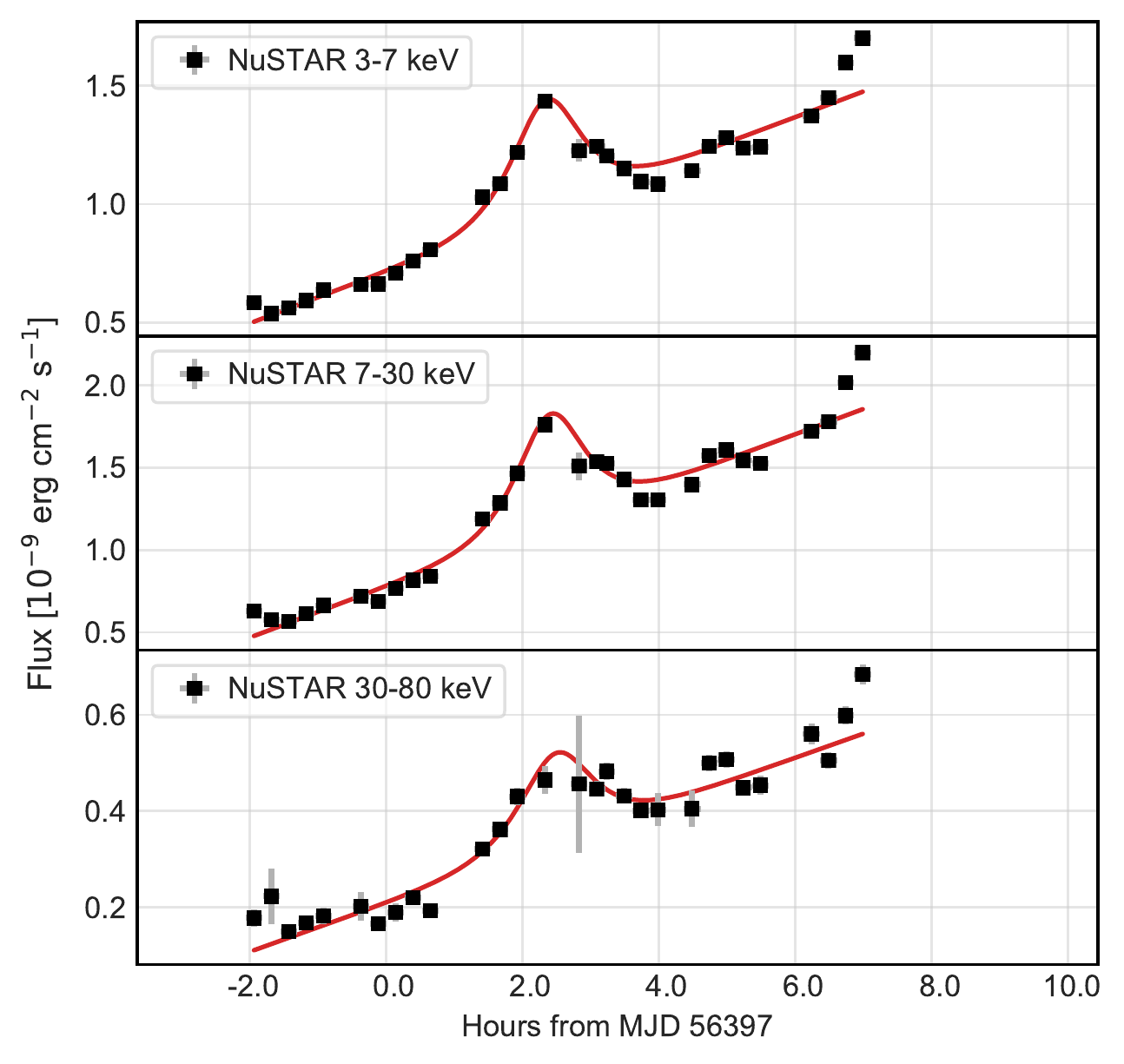}{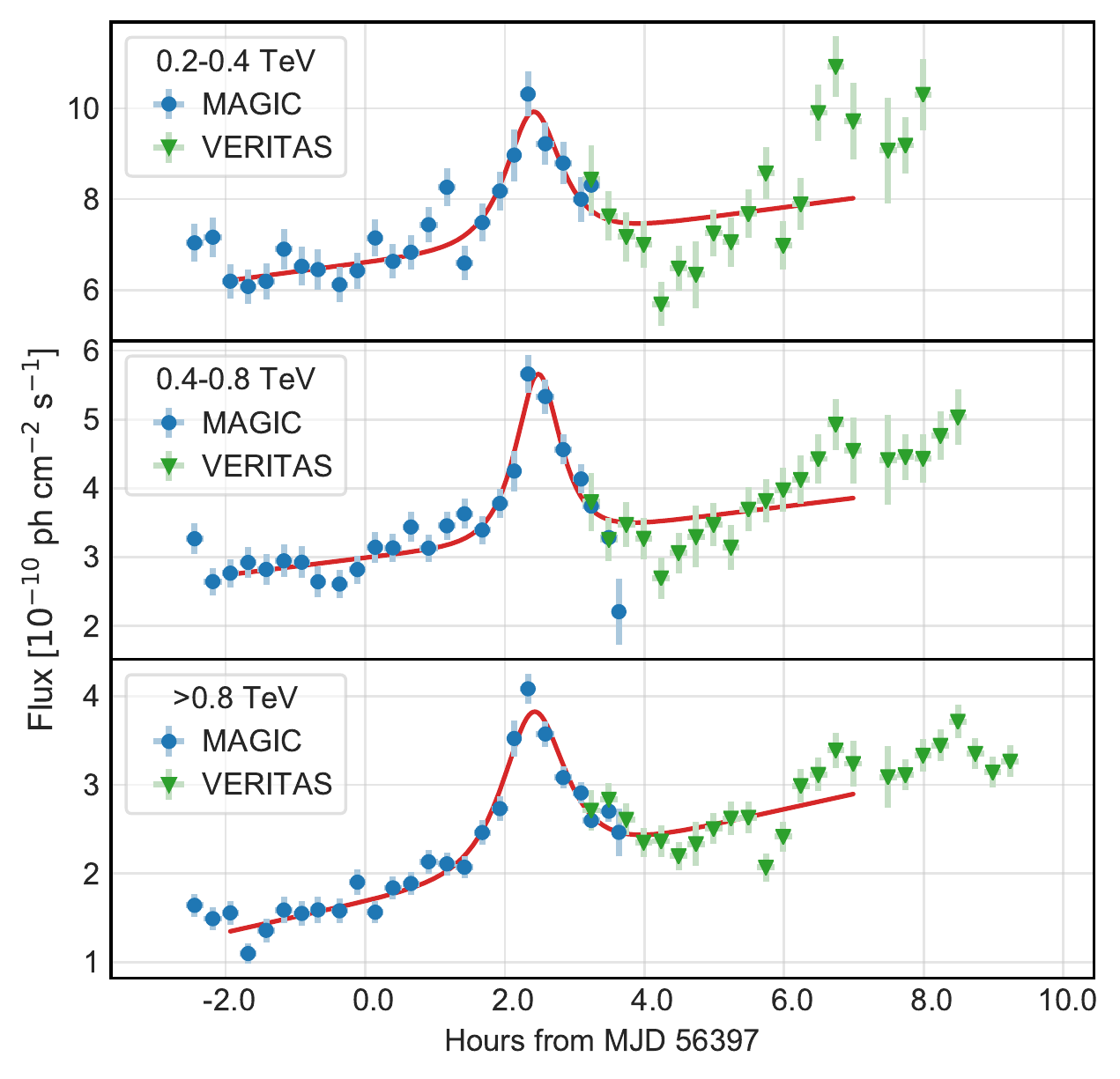}
\caption{Light curves from 2013 April 15 in three X-ray bands (left panel) and three VHE gamma-ray bands (right panel). 
The red curve is the result of a fit with the function in Eq.~\ref{eq:flux_t},  applied to the time interval with simultaneous X-ray and VHE observations. The resulting model parameters from the fit are reported in Table~\ref{tab:FitParametersDailyApril15}.}
\label{fig:LC15}
\end{figure*}  
  
This relatively simple function provides a rough description of the energy-dependent light curves, and may not describe perfectly well all the data points. For instance, in the low energy X-ray bands, the statistical uncertainties are very small and one can appreciate significant and complex substructure that is not reproduced by the above-described (and relatively simple) fitting function. We do not intend to find a model that describes accurately all the data points. Rather we look for a model that provides a description of the main flux-variability trends, and how they evolve with the X-ray and VHE energies.

\begin{deluxetable*}{c|c|c|c|c|c|c}
\tablecaption{Parameters resulting from the fit with Eq.~\ref{eq:flux_t} to the X-ray and VHE multi-band light curves from 2013 April 15.  \label{tab:FitParametersDailyApril15}}
\tablewidth{0pt}
\tablehead{
Band & {\it Offset}\tablenotemark{a}  & $Slope$      &   Flare   &  Flare  & Flare  & $\chi^2$/d.o.f \\
     &                          & [h$^{-1}$] & Amplitude $A$ & flux-doubling time\tablenotemark{b}  [h] & $t_0$ [h] & 
}
\startdata
 \multicolumn7l{ 15 April 2013 }  \\ \cline{1-7}
3-7 keV &  0.71 $\pm$ 0.01 &  0.153 $\pm$ 0.006 &  0.49 $\pm$ 0.07 &  0.30 $\pm$ 0.04 &  2.35 $\pm$ 0.06 &    836/24  \\
7-30 keV &  0.78 $\pm$ 0.02 &  0.199 $\pm$ 0.009 &  0.59 $\pm$ 0.11 &  0.30 $\pm$ 0.04 &  2.41 $\pm$ 0.06 &    889/24  \\
30-80 keV &  0.21 $\pm$ 0.01 &  0.241 $\pm$ 0.018 &  0.56 $\pm$ 0.18 &  0.32 $\pm$ 0.09 &  2.50 $\pm$ 0.10 &     111/24  \\
0.2-0.4 TeV   &  6.60 $\pm$ 0.17 &  0.031 $\pm$ 0.008 &  0.40 $\pm$ 0.09 &  0.23 $\pm$ 0.07 &  2.41 $\pm$ 0.09 &     96.9/38  \\
0.4-0.8 TeV   &  2.99 $\pm$ 0.07 &  0.042 $\pm$ 0.008 &  0.72 $\pm$ 0.09 &  0.19 $\pm$ 0.03 &  2.47 $\pm$ 0.04 &     68.1/42  \\
$>$0.8 TeV  &  1.68 $\pm$ 0.05 &  0.103 $\pm$ 0.010 &  0.82 $\pm$ 0.08 &  0.27 $\pm$ 0.03 &  2.41 $\pm$ 0.04 &     90.0/45  \\
\enddata
\tablenotetext{a}{For VHE bands in $10^{-10}$ ph cm$^{-2}$~s$^{-1}$, for X-ray bands in $10^{-9}$~erg cm$^{-2}$ s$^{-1}$.}
\tablenotetext{b}{Parameters $t_\mathrm{rise}$ and $t_\mathrm{fall}$ in Eq.~\ref{eq:flux_t} are set to be equal, and correspond to the Flare flux-doubling time in the Table.}
\end{deluxetable*}

\begin{figure*}[htb!]
\plottwo{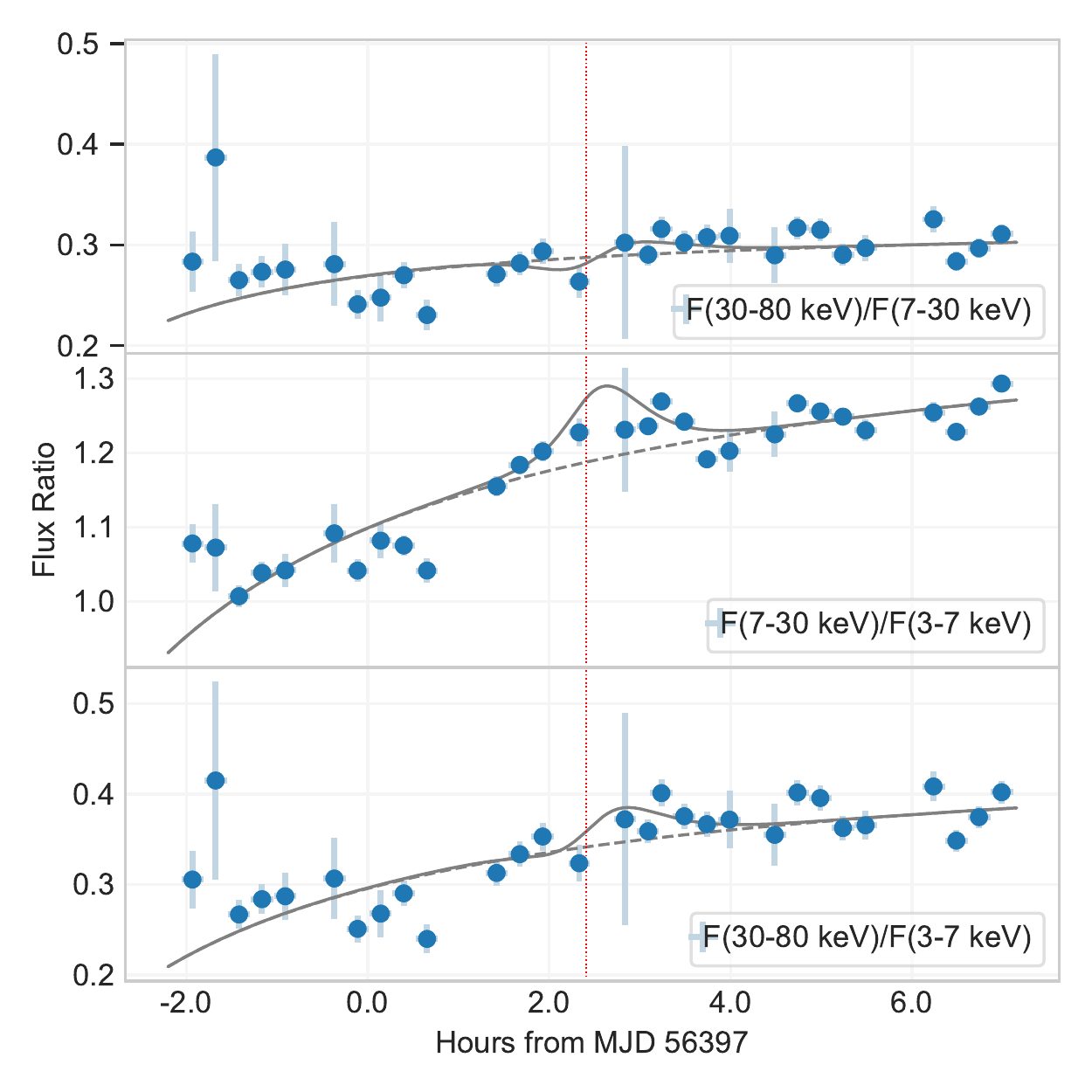}{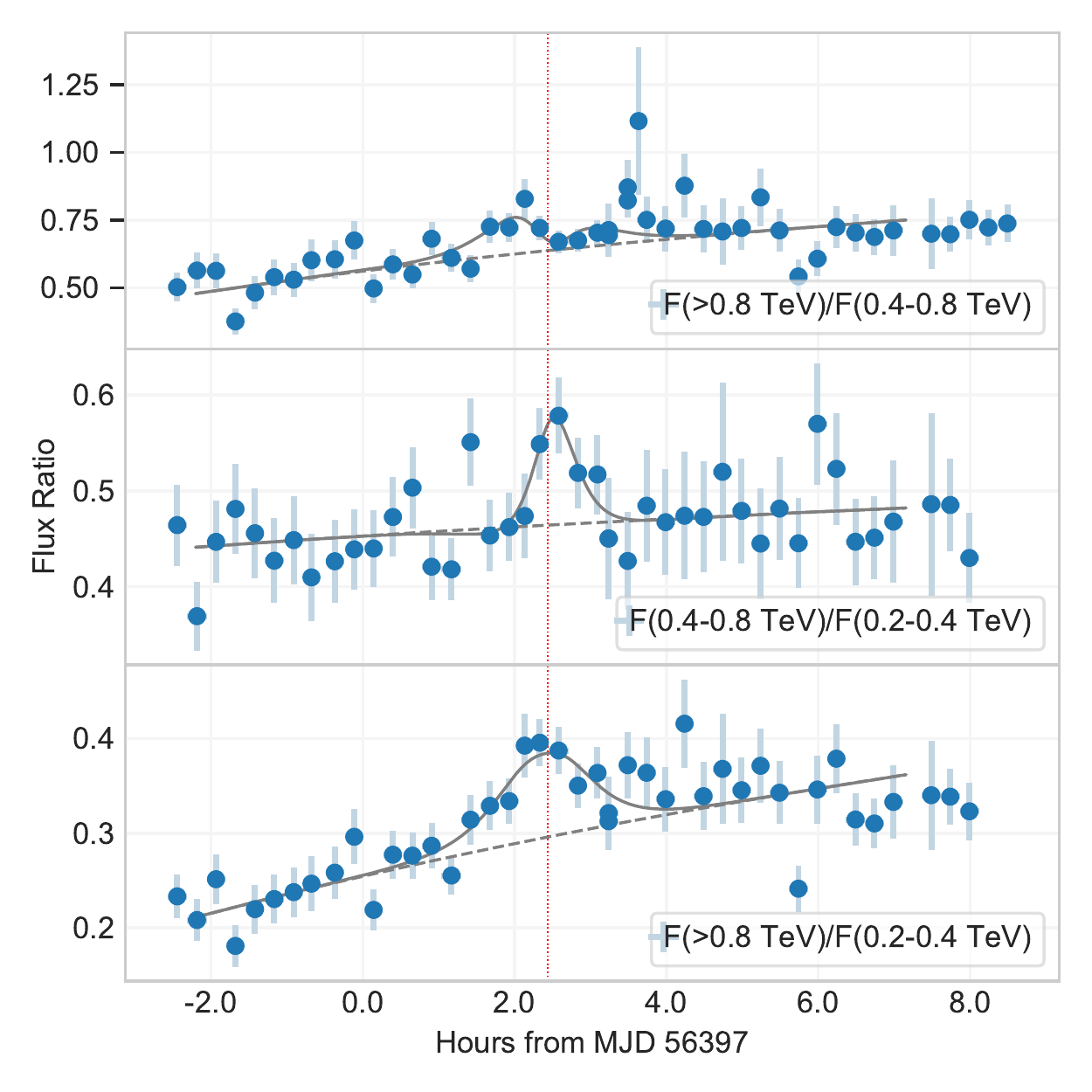}
\caption{The X-ray hardness (flux) ratios for several X-ray ({\em NuSTAR}) bands (left panel) and VHE (MAGIC+VERITAS) bands (right panel) for April 15. In both panels, the dashed red vertical line indicates the average time of the peak of the flare in VHE $t_{0,\text{VHE}} = 2.44 \pm 0.03$~hr, and X-rays $t_{0,\text{X-ray}} = 2.41 \pm 0.04$~hr,
where the average is calculated for the three bands (Table~\ref{tab:FitParametersDailyApril15}). The solid grey curve is  the ratio of the fitted functions with parameters reported in Table~\ref{tab:FitParametersDailyApril15}, and the dashed grey line is the ratio of the same fitted functions, but this time excluding the fast component. \label{fig:HR_April15}} 
\end{figure*}

The multi-band flux variations during April 15 and its related quantification using Eq.~\ref{eq:flux_t} are depicted in Fig.~\ref{fig:LC15}, with the parameters resulting from the fits reported in Table~\ref{tab:FitParametersDailyApril15}. 
The main multi-band emission varies on timescales of several hours, and hence it is dominated by the \textquotedblleft slow component" in equation~\ref{eq:flux_t}. The $Slope$ of this variation (quantified relative to the {\it Offset} in each band for better comparison among all bands) has a strong energy dependence, with the parameter value for the highest energies being around a factor of 2--3 times larger than that for the lowest energies for both the X-ray and VHE gamma-ray bands (e.g. $Slope_{>0.8~\mathrm{TeV}} \simeq 3 \cdot Slope_{0.2-0.4~\mathrm{TeV}}$). The second most important feature of this multi-band light curve is the existence of a short flare, on the top of the slowly varying flux, in all the energy bands for both X-ray and VHE gamma rays. The location of the flare $t_0$ is the same (within uncertainties) in all the three X-ray bands and VHE bands. In order to quantify better the location of the short flare at X-ray and VHE gamma-ray energies, using the information from Table~\ref{tab:FitParametersDailyApril15}, we computed the  weighted average separately for the three VHE bands, $t_{0,\text{VHE}} = 2.44 \pm 0.03$~hr and the three X-ray bands, $t_{0,\text{X-ray}} = 2.41 \pm 0.04$~hr past midnight. This indicates that, for this fast feature in the light curve, for all the energies probed, there is no delay in between X-ray and VHE gamma-ray emission down to the resolution of the measurement, which, adding the errors in quadrature, corresponds to 3 minutes. The flux-doubling time is comparable among all the energy bands, with about 0.3~hours for all the X-ray bands and the highest VHE band ($>$0.8~TeV), and about 0.2~hours for the lowest and middle VHE band. The characteristic by which the fast X-ray flare differs from the fast VHE flare is in the normalized flare amplitude $A$ (see Table~\ref{tab:FitParametersDailyApril15}): it is energy-independent (achromatic) for the X-ray fast flare, while it increases its value (chromatic) for the VHE fast flare (with amplitude \mbox{$A_{>0.8~\mathrm{TeV}} \simeq 2 \cdot A_{0.2-0.4~\mathrm{TeV}}$).}


In order to evaluate potential spectral variability throughout the $\sim$10-hour light curves measured on 2013 April 15, we computed the flux hardness ratios $HR$ (= $F_\mathrm{high-energy}/F_\mathrm{low-energy}$) for several energy bands in both X-ray and VHE gamma-ray domains. Fig.~\ref{fig:HR_April15} depicts the  $HR$ computed with the data flux measurements (in time bins of 15 minutes) and the $HR$ expected from the fitted functions reported in Fig.~\ref{fig:LC15} and Table~\ref{tab:FitParametersDailyApril15}. For comparison purposes, we also included the $HR$ from the fitted functions from Table~\ref{tab:FitParametersDailyApril15} excluding the fast component given by Eq.~\ref{eq:fast_t} (dotted line in Fig.~\ref{fig:HR_April15}). The dashed  vertical red line indicates the weighted average time of the peak of the flare $t_0$, calculated separately for the three X-ray bands and VHE gamma-ray bands (see above). One can see that the overall impact of the fast component in the $HR$ temporal evolution is small, and only noticeable in some panels (e.g. $F_{>0.8~\mathrm{TeV}}/F_{0.4-0.2~\mathrm{TeV}}$ or $F_{7-30~\mathrm{keV}}/F_{3-7~\mathrm{keV}}$). This is due to the relatively short duration of the fast component, and the relatively small magnitude of the flare amplitude, in comparison to the overall flux. Therefore, the temporal evolution of the spectral shape in both bands, X-ray and VHE, is dominated by the slow component, i.e. by the variations with timescales of several hours.


Besides April 15, we also performed the fit with Eq.~\ref{eq:flux_t} to the other five consecutive nights with large \mbox{X-ray} and VHE gamma-ray simultaneous datasets, namely all nights from April 11 to April 16 (both included). The results from these fits are reported in Appendix~\ref{app:lc_fits} (see Table~\ref{tab:FitParametersDailyOthers} and Fig.~\ref{fig:LC11}-\ref{fig:LC16}).
It is worth stating that, when comparing the quantification of the various light curves with the function Eq.~\ref{eq:flux_t}, we found diversity among the fit parameter values and their energy dependencies. For April 11, we did not find any fast component, and the flux decreases monotonically through the observation with energy-independent $Slope$ for both X-ray and VHE gamma rays (fully achromatic flux variations). 
On the other hand, during April 12, the emission increased throughout the observation, but with a $Slope$ that decreases with increasing energy in both X-ray and VHE gamma rays. This trend is also observed, from a different perspective, in the bottom-right panel of Fig.~\ref{fig:fvar1}, which displays a decreasing $F_{var}$  with increasing energy for both the X-ray and VHE gamma-ray emission from April 12. This is a very interesting behavior because it is opposite to the trend reported in most datasets from Mrk\,421, where the variability increases with energy. For this night, we can also see a fast X-ray flare (flux-doubling time of about 0.3~hours) whose amplitude increases with energy. Unfortunately, this fast X-ray flare occurred during a time window without VHE observations. 

For April 13, we also observed a slow flux variation with an energy-independent $Slope$, as in April 11, but this time with a flux increase, instead of a decrease. Additionally, we did observe a super-fast X-ray flare (flux-doubling time of 5$\pm$1 minutes) without any counterpart in the VHE light curve, i.e. an 'orphan' X-ray flare (see Fig.~\ref{fig:LC13}). 
As shown in Table~\ref{tab:FitParametersDailyOthers}, the 
X-ray {\it NusTAR} flare amplitude relative to the overall baseline is only about 11\%, but it is significant (3--4$\sigma$ depending on the energy band) and there is no correlated flux variation in the simultaneous VHE MAGIC fluxes, which have flux uncertainties of about 5\%. 

During April 14th we see again a monotonically decreasing flux with an energy-independent $Slope$ for both \mbox{X-ray} and VHE gamma rays, with another fast X-ray flare (flux-doubling time~$\sim 0.5$~hours) without counterpart in the VHE light curve. 


The night that differs most is April 16, which does not show any monotonic increase or decrease, and a largely non-symmetric flare with flux variation timescales of hours. In order to quantify the temporal multi-band evolution of the flux during April 16, we used Eq.~\ref{eq:fast_t} (i.e. the fitting function without the slow component), with $t_\mathrm{rise} \neq t_\mathrm{fall}$. See Appendix~\ref{app:lc_fits} for further details about the quantification of the multi-band flux variations during the six consecutive nights, from April 11 to April 16.

In summary, during these six consecutive nights with enhanced activity and with multi-hour long X-ray/VHE simultaneous exposures in April 2013, we found achromatic and chromatic flux variability with timescales spanning from multi-hours to sub-hours, and several X-ray fast flares without VHE gamma-ray counterparts. We did not see any VHE gamma-ray orphan fast flare (whenever we had simultaneous X-ray coverage). However, we did observe fast flares in some specific energy bands which are not detected in the other nearby energy bands (X-ray or VHE); which suggests the presence of flaring mechanisms affecting relatively narrow energy bands.

The temporal evolution of the X-ray and VHE emission, and the particularity of being able to approximately describe it with a two-component function with a fast (sub-hour variability timescale) and slow (multi-hour variability), will be discussed in section \ref{sec:discussion2}.

\begin{deluxetable*}{c|r|c|c|c|c|r}
\tablecaption{Correlation coefficients and slopes of the linear fit to the VHE vs. X-ray flux (in log scale) derived with the 9-day flaring episode of Mrk\,421 in April 2013. \label{tab:CorrelationCoefficients}}
\tablewidth{0pt}
\tablehead{
\colhead{VHE band} & \colhead{X-ray band} & \colhead{Pearson coeff.$^a$}  & \colhead{$N\sigma$ Pearson$^a$}  & \colhead{DCF} & Linear fit slope & $\chi^2$/d.o.f }
\startdata
0.2-0.4 TeV & 3-7 keV & $0.92 \pm 0.01$ & 20.2 & $0.93 \pm 0.12 $& $0.61 \pm 0.02$ &1183 / 162 \\ 
     & 7-30 keV & $0.87 \pm  0.02 $ & 17.0 & $0.88 \pm 0.11 $ & $0.45 \pm 0.03$ &1891 / 162 \\ 
     & 30-80 keV & $0.79 \pm  0.03$ & 13.6 & $0.81 \pm 0.11$ & $0.35 \pm 0.02$ & 2277 / 162 \\  \cline{1-7}  
0.4-0.8 TeV & 3-7 keV & $0.946^{ + 0.007}_{ - 0.009}$ & 23.4 & $0.96 \pm 0.11$ & $0.79 \pm 0.03 $ & 1038 / 170 \\ 
     & 7-30 keV & $0.91 \pm  0.01 $ & 19.8 & $0.92 \pm 0.11 $ & $ 0.58 \pm 0.03$ & 1725 / 170 \\ 
     & 30-80 keV & $0.84 \pm  0.02 $ & 15.8 & $0.86 \pm 0.11$ & $0.45 \pm 0.03$ & 2160 / 170 \\ \cline{1-7}
$>$0.8 TeV & 3-7 keV & $0.964^{ + 0.005}_{ - 0.006} $ & 26.0 & $0.97 \pm 0.11$ & $1.11 \pm 0.03$ & 704 / 170 \\ 
    & 7-30 keV & $0.947^{ + 0.007}_{ - 0.008}$ & 23.5 & $0.96 \pm 0.11 $ & $0.81 \pm 0.03$ & 1245 / 170 \\ 
     & 30-80 keV & $0.89 \pm  0.02$ & 18.6 & $0.91 \pm 0.10$ & $0.61 \pm 0.03$ & 1736 / 170 \\         
\enddata
\tablenotetext{a}{The Pearson correlation function 1$\sigma$ errors and the significance of the correlation are calculated following \citet{Press2002}. }
\end{deluxetable*}

\section{Unprecedented study of the multi-band X-ray and
  VHE gamma-ray correlations}
\label{sec:correlations}

We evaluated the correlations among all the frequencies covered during the April 2013 flare, and found that the largest flux variations and the largest degree of flux correlation occurs in the X-ray and VHE gamma-ray bands.  No correlation was found among the radio, optical and gamma-ray bands, a result that was expected  because of the lower activity and longer variability timescales at these energies. 
Apart from some variability in the GeV flux around April 15, which is the day with the highest X-ray activity, the GeV emission appears constant for the 12-hour time intervals related to flux variations by factors of a few at keV and TeV energies. If the GeV and TeV fluxes were correlated on 12-hour time scales, {\it Fermi}-LAT should have detected large flux variations, and hence we can exclude this correlation.

\begin{figure*}[htp]
\begin{center}
\includegraphics[width=\linewidth]{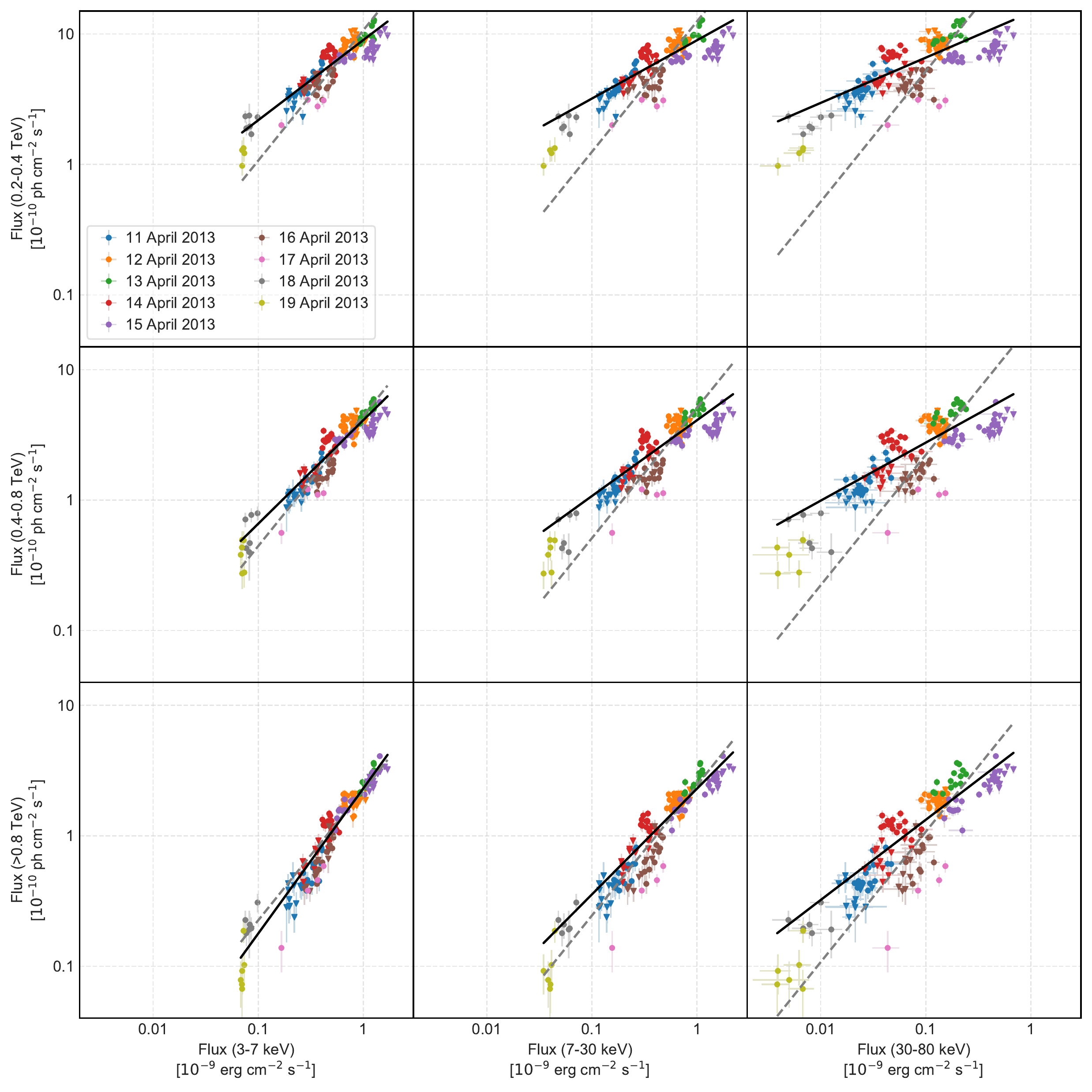}
\caption{VHE flux vs. X-ray flux in three X-ray and three VHE energy bands. Data from all nine days is
shown, with colors denoting fluxes from the different days.  The grey dashed line is a fit with slope fixed to 1, and black line is the best fit line to the data, with the slope quoted in Table \ref{tab:CorrelationCoefficients}. 
\label{fig:logFlogF1}}
\end{center}
\end{figure*}

\begin{figure*}[htp]
\begin{center}
\includegraphics[width=\linewidth]{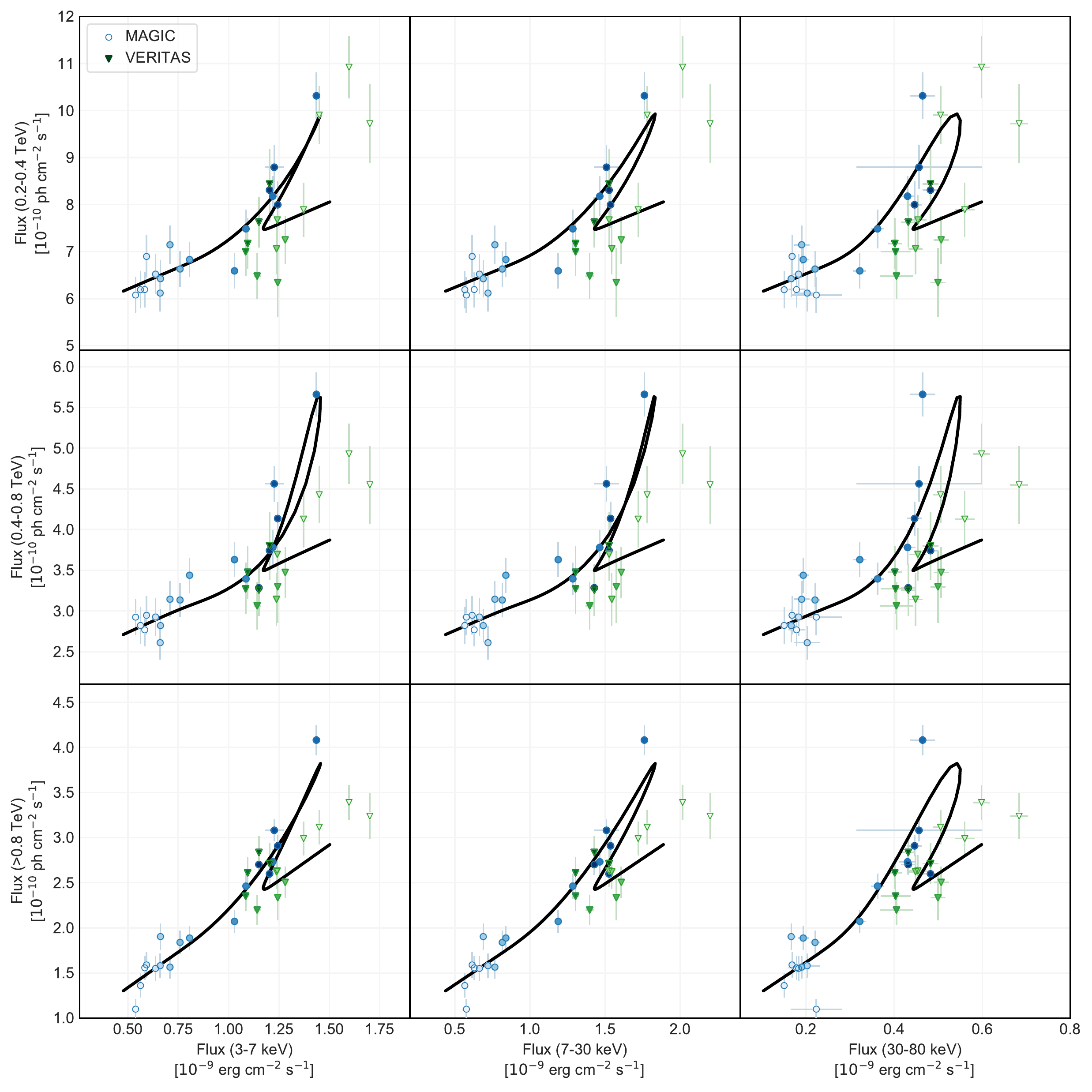}
\caption{VHE flux vs. X-ray flux in three X-ray and three VHE energy bands for April 15. The black line is the track predicted by Slow+Fast component fit from Eq.~\ref{eq:flux_t}. The lightness of symbols follows time: for MAGIC data lightness decreases with time, and for VERITAS data it increases in time, so that the central part of the night, where MAGIC and VERITAS observations overlap, is plot using darker symbols.}
 \label{fig:logFlogFtrack15}
\end{center}
\end{figure*}

The quality and extent of this dataset, both in time and energy, allows for a X-ray/VHE correlation study that is unprecedented among all datasets collected from Mrk\,421, and any other TeV blazar. The relation between the VHE gamma-ray and the \mbox{X-ray} fluxes in the 3$\times$3 energy bands is shown in Fig.~\ref{fig:logFlogF1} for the 9-day flaring activity, and in Fig.~\ref{fig:logFlogFtrack15} for April 15th. The Discrete Correlation Function (DCF) and Pearson correlation coefficients, as well as the slope of the VHE versus X-ray flux are reported in Table \ref{tab:CorrelationCoefficients}. There is a clear pattern: the strength of the correlation increases for higher VHE bands and lower X-ray bands. The strongest correlation is observed between the 3--7 keV and $>$0.8 TeV bands.
This combination of bands also shows a slope (from the fit in Table~\ref{tab:CorrelationCoefficients}) closest to 1, among all the 3$\times$3 bands reported. Moreover, the scatter in the plots is smaller as we increase the VHE band and decrease the X-ray energy band. The smallest scatter, which can be quantified with the $\chi^2$ of the fit (lower values of $\chi^2$ relate to smaller scatter in the data points), occurs for the combination $>$0.8~TeV and \mbox{3-7~keV.}

Fig.~\ref{fig:logFlogF1} reveals that the different days occupy (roughly) different regions in the VHE versus X-ray flux plots (for all the 3$\times$3 bands). This is expected because the largest flux changes occur on day-long timescales. In addition, individual days appear to show different patterns. In order to better characterize these different patterns (observed for the different days), we also computed the same quantities (DCF, Pearson and linear fit) to the simultaneous data points from the single nights with multi-hour light curves (namely April 11-16). The results are reported in Table \ref{tab:CorrelationCoefficientsDaily}, in Appendix~\ref{app:CorrFitOtherDays}.

   
The main conclusions from this study performed on data from April 11 to April 16 are the following\footnote{On the nights April 17-19, both the level of activity of Mrk\,421 and the amount of data collected was substantially smaller, which prevents us from making detailed studies of the multi-band correlations.}:
\begin{itemize}
\item In some nights, namely on April 15 and \mbox{April 16}, Mrk\,421 shows the ``general trend'' that is observed for the full 9-day flaring activity, with the highest magnitude and significance in the correlation occurring for the $>$0.8~TeV versus the 3--7~keV bands. 

\item  In other nights, namely on April 11, 13-14, the general trend from the 9-day data set is less visible: there is a larger similarity in the magnitude and significance of the correlation among the various energy bands. 
\item  In one night, April 12, we found no correlation between the X-ray and VHE gamma-ray bands, despite the significant variability in both bands. The lack of correlation between X-ray and VHE gamma rays (being both highly variable and characterized with simultaneous observations) has not been observed to date in Mrk\,421, although it has been observed in another HBL \citep[PKS~2155-304, in August-September 2008,][]{2009ApJ...696L.150A}.
\end{itemize}

The correlation study was also performed splitting the data set into two subsets: {\em (a)} April 15, 16, and 17 (which appear somewhat away from the main trend in Fig.~\ref{fig:logFlogF1}); and {\em (b)} April 11, 12, 13, 14, 18, and 19. In comparison to the nine-day data set, the two subsets analyzed separately yields a reduction of the scatter of the flux points around the main trends (which show up as a noticeable reduction in the $\chi^2$ values from the fit), and  also a 
smaller dependence of the magnitude and significance of the correlations with the specific combination of VHE and X-ray energy bands. The largest linear fit slope occurs for the combination $>$0.8~TeV and 3-7~keV  in both subsets (as with the nine-day data set),  but while for subset {\em (a)} we continue having the largest significance and magnitude of the correlation for $>$0.8~TeV and 3-7~keV, in subset {\em (b)} the change in magnitude and correlation with energy bands is much smaller, and the highest values occur for $>$0.8~TeV and 7-30~keV. This indicates a somewhat different physical state of the source during those three consecutive days (April 15,16 and 17) with respect to the others.

Besides the VHE gamma-ray and the X-ray fluxes in the 3$\times$3 energy bands for April 15,  Fig.~\ref{fig:logFlogFtrack15}  depicts also the flux-flux values from the fitted functions reported in Section~\ref{subsec:fastvariability}. This figure shows that multiple components in the flux evolution (e.g. fast component on the top of the slow component) appear as ``different trends'' in the flux-flux plots  with the flaring component having sharper VHE flux rise (with increasing X-ray flux) than the slow component. Because of the statistical uncertainties in the flux measurements, as well as the fact that one component has a much smaller flux and shorter duration, even for very good datasets such as this one, it is not easy to recognize and separate the contribution of different components
in the flux-flux plots. However, these different patterns can produce collective deviations (when considering many of these different single-trends) that are statistically significant when fitting the data points in the flux-flux plots with simple trends, such as the linear or quadratic functions in the log-log scale.

The discussion of these observational results is given in Section \ref{sec:discussion2}.





\section{Discussion of the results} 
\label{sec:discussion2}

Although detailed spectral energy distribution and light curve
modeling are beyond the scope of this paper, we discuss the main results of our analysis
and provide possible interpretations.

\subsection{Minimum Doppler factor}
\label{sec:dopplerfactor}
In general, VHE gamma rays can interact with low-energy synchrotron photons in order to produce electron-positron pairs.  
If both VHE gamma-ray and low-energy photons are produced in the same region, then the criterion that this attenuation is avoided so that the  VHE gamma rays may escape from the source and be detected leads to a lower limit on the Doppler factor \citep{dondi95,tavecchio98,finke08}.  Owing to the detection of 10 TeV photons from Mrk\,421 during this flaring activity, the relevant observed synchrotron frequency for their attenuation is close to $6\times10^{12}$\ Hz. In the $R$ band ($\nu \sim 4.5\times10^{14}$\ Hz), the observed flux is close to 50~mJy on 2013 April 15. Using 12~min as the shortest variability timescale (see Table~\ref{tab:FitParametersDailyApril15}), and extrapolating the $R$ band flux to $6\times10^{12}$\ Hz assuming the same spectral shape as the one obtained from the long-term SED \citep[i.e., photon index $\sim 1.6$,][]{2011ApJ...736..131A}, one finds $\delta\ga 35$, which lies to the high-end of values derived from SED modeling of \fermi-LAT detected blazars \citep{ghisellini10,tavecchio11,2017ApJ...851...33P}. The derived lower limit can be relaxed if the gamma-ray and optical emitting regions are decoupled, but the actual value would depend on the details of the theoretical model.

\vspace{2cm}

\subsection{Flux-flux correlations}
\label{discussion:flux-flux}

Flux-flux correlations have been the focus of many multi-wavelength campaigns during active and low states of blazar emission \citep[for Mrk\,421, see e.g.][]{maraschi99, fossati08, 2015A&A...576A.126A}, because their study may differentiate among emission models \citep[see e.g.][]{kraw02}. \cite{fossati08}, in particular, studied the  correlations on a daily basis between TeV fluxes ($E>0.4$~TeV) and X-ray fluxes, mostly in the 2--10~keV band, but also in the 2--4~keV, 9--15~keV and 20--60~keV bands. The work we show here goes one step further, as it allows for the first time, to study flux-flux correlations between multiple VHE gamma-ray and  X-ray energy bands on a daily basis and down to 15-min time bins.

The overall strong correlation found between the \mbox{$>$0.8~TeV}  band and the lowest energy X-ray band \mbox{(3--7~keV;} Fig.~\ref{fig:logFlogF1} and Table~\ref{tab:CorrelationCoefficientsDaily}) implies that the emissions are most likely co-spatial and they are produced by electrons with approximately the same energy via synchrotron in {\mbox{X-rays} and SSC processes in gamma rays. In TeV blazars, like Mrk\,421, the peak of the SSC spectrum is typically produced by inverse Compton scatterings in the Klein-Nishina regime \citep[e.g., ][]{tavecchio16}.

The results presented in Section~\ref{sec:correlations} reveal, however, a more complicated picture, as the strong correlation mentioned above weakens or even disappears on certain days (e.g., 2013 April 12). A weak correlation between X-rays and gamma rays can be produced if the emissions are produced by different components \citep[e.g.][]{petro14, chen16} or by different particle populations, as in lepto-hadronic models \citep{mast13}. Different strengths of the correlation can be predicted by adjusting the temporal variations of the model parameters (e.g., injection rate of accelerated particles) and/or by having more than one emitting component in X-rays and gamma rays.

The analysis presented in Section~\ref{sec:correlations} also reveals that the slope of the correlation between the X-ray bands and the gamma-ray bands is
generally sub-linear, i.e. $F_{\gamma}\propto F_X^m$ with $m\lesssim 1$, and changes with time
(Table~\ref{tab:CorrelationCoefficientsDaily}). In the standard (one-zone) SSC scenario, a value $m=2$ is
expected if only the electron distribution normalization is varying with time. Even in this scenario though, different values of $m$ can be obtained when looking at correlations between different energy bands in X-rays and gamma rays \citep{kat05}. 
Moreover, values $m<2$ are possible if several parameters changes with time (e.g., magnetic field strength and Doppler factor). \citet{kat05} explored in detail these flux-flux correlations for an SSC model for high-peaked BL Lac objects like Mrk\,421. They showed that $m\le 1$ is expected between energy bands close to the peaks of the synchrotron and SSC components, if the blob is expanding and the magnetic field is decreasing \citep[see Fig.~4 in][]{kat05}. \cite{petro14} showed that in a two-component SSC scenario the slope of the correlation between 2-10~keV and \mbox{0.4-10~TeV} may vary strongly from $m\sim 0$ to $\sim 1$ on day-long timescales, with a pattern that depends on the varying model parameter (i.e., injection rate or maximum electron energy). A quadratic relation between \mbox{X-rays} and gamma rays is also expected in lepto-hadronic models, where the former act as targets for the photo-hadronic interactions of accelerated protons that result in the production of gamma rays \citep{dim12, mast13}. A slope of $m \lesssim 1$ can also be produced in proton synchrotron models, where variations in the electron and proton injection rates are directly mapped to variations in the X-ray and gamma-ray flux, respectively \citep{mast13}. The detailed modeling of the flux-flux correlations will be the topic of a future study.

\subsection{Temporal variability}

One of the main results of this work is the detection of fast-evolving flares on top of a slower evolving emission in both X-ray and VHE gamma-ray bands. Although this temporal behavior was qualitatively discussed for some of the X-ray {\it NuSTAR} light curves in \citet{2015ApJ...811..143P}, here we present a quantitative study of these characteristics in the X-ray light curves and also the VHE gamma-ray light curves (see Fig.~\ref{fig:LC15} in Section~\ref{subsec:fastvariability} and Figs.~\ref{fig:LC11}-\ref{fig:LC16} in Appendix~\ref{app:lc_fits}). In the next paragraphs, we discuss possible interpretations for the origin of the multi-band temporal variability. 

\subsubsection{Acceleration and cooling processes}

The rise and decay timescale of a flare may be associated with the acceleration and cooling timescales of the radiating electrons. In this case, one can use the fact
that the acceleration and cooling timescales are found to be equal to estimate the
magnetic field of the emitting region, as follows.  If electrons undergo Fermi-1 (or Fermi-2) acceleration, then the acceleration timescale, in the co-moving frame, can be  written as:
\begin{flalign}
t\p_{acc} = \frac{2\pi m_e c\gp N_a}{e B^\prime}
\end{flalign}
\citep[e.g.,][]{finke08} where $B^\prime$ is the tangled (co-moving) magnetic field
strength, $\gp$ is the electron Lorentz factor in the co-moving frame,
and $N_a\ge 1$ is the number of gyrations an electron makes to double
its energy.  The synchrotron cooling timescale is
\begin{flalign}
t\p_\mathrm{syn} = \frac{6\pi m_e c^2}{4c\sT B^{\prime 2} \gp}\ .
\end{flalign}
Setting $t\p_\mathrm{acc}=t\p_\mathrm{syn}$ results in
\begin{flalign}
B^\prime =  \frac{3e}{4\sT\g^{\prime 2}N_a} \ .
\end{flalign}
 The Lorentz factor of electrons producing the peak of the SSC mission can be estimated as:
\begin{flalign}
\label{gammab}
\gp_b \approx \frac{E_\mathrm{ssc}}{\delta m_e c^2} \approx 5\times 10^4 \left(\frac{\delta}{40}\right)^{-1} \left( \frac{E_\mathrm{ssc}}{1 \ {\rm TeV}}\right)
\end{flalign}
 where $E_\mathrm{ssc}$ is the observed energy of the peak of the  SSC component. Here we use a value of $\delta$ consistent with the lower limit from $\g\g$ pair production (Section \ref{sec:dopplerfactor}).  Here and from this point forward, we neglect factors of $1+z$ which will be quite small given the redshift of Mrk\,421 ($z=0.03$). 
To accelerate electrons to this peak requires
\begin{flalign}
B^\prime \approx 9 \, \left(\frac{\delta}{40}\right)^2 \left(\frac{N_a}{100}\right)^{-1}\ \Gauss
\end{flalign}

Large values of
the magnetic field or large values of $N_a$ are required to accelerate electrons to $\gp_b$ where $t\p_\mathrm{acc} = t\p_\mathrm{syn}$. Similar results  were found for the luminous, rapid flare from PKS 2155$-$304 in 2006 \citep{finke08}. 

Acceleration and cooling may control the light curve time scales, if $R\p_b/c \ll t\p_{\rm syn} = t\p_{\rm acc}$ or $R^\prime_b\ll 1.4\times10^{12}\ (\delta/40)^{-3}(N_a/100)^{-2}\ \cm$, where $R^\prime_b$ is the radius of a spherical emitting region in the comoving frame. The required upper limit on $R^\prime_b$ is 2--3 orders of magnitude smaller than typical values for the size of the emitting region \citep[e.g.,][]{2011ApJ...736..131A}, even for flaring episodes \citep[e.g.,][]{2015A&A...576A.126A}. 
Additionally, if the flare's rise and decay times were dominated by the acceleration and cooling timescales, one would expect the flaring timescales to be energy-dependent; however, they seem to have the same timescale across energy bands (see Table~\ref{tab:FitParametersDailyApril15}). Therefore, the timescales of the fast component of the light curve are likely controlled by the light-crossing time of a blob with fixed size. As a result, they should appear symmetric and with timescales independent of the energy.

\subsubsection{Plasmoids in magnetic reconnection}

Magnetic reconnection is invoked as an efficient particle acceleration process in a variety of astrophysical sources of non-thermal high-energy radiation including AGN jets \citep{romanova_92,giannios09,giannios_10b,giannios_13}. It has been proposed that plasmoids (i.e., blobs of magnetized plasma containing energetic particles) that are formed and accelerated in the reconnection regions of jets can serve as high-energy emission sites in both blazars and radio galaxies
\citep{giannios09, sironi15}. \citealt[][(hereafter, PGS16)]{pgs16} presented a
semi-analytic model of flares powered by plasmoids in a reconnection
layer, simplifying the results of detailed particle-in-cell (PIC)
simulations \citep[for a full numerical treatment, see][]{christie19}.

A single plasmoid produces a flare with peak luminosity and
flux-doubling timescale that depend on its size and Doppler
beaming. A unique feature of this model is that the flux-doubling timescale in
the rising part of the flare is mostly determined by the acceleration
of the plasmoid in the layer (i.e., by its bulk motion). As a result, similar rise timescales should be
observed at different energy bands of a flare powered by a single plasmoid 
(achromatic behavior). In contrast to the rising part of a flare, 
its decay is not constrained by PIC simulations. By setting the decay timescale to be approximately equal to the rise timescale of the fast flares, as observed, one can infer the declining rate at which accelerated particles are injected in the plasmoid or the decay rate of the magnetic field after the plasmoid has left the layer (PGS16).

No strong spectral evolution is expected during a flare
produced by a {\it single} plasmoid \citep[PGS16;][]{christie19}. At any given
time though, an observer receives radiation from a large number of plasmoids
in the layer, having different sizes and Doppler factors. Those
plasmoids that move with mildly relativistic speeds (in the jet frame)
and have intermediate sizes (in terms of layer's size) can
contribute and even dominate the overall emission. The superposition of their
emission could result in a slow varying and more luminous component of
the light curve \citep[][]{giannios_13, christie19}, which may exhibit spectral variations and drive the fractional variability on longer timescales than the fast varying component of the light curve (Christie et al. 2019b, in preparation).  

\textit{Energetics \& timescales} -- PGS16 provided simple formulae to estimate the flux-doubling
timescale ($t_{1/2}$) and peak luminosity ($L_{\rm pk}$) of
flares produced by individual plasmoids in a reconnection layer (with half-length
$L^\prime$\footnote{Primed quantities are measured in the jet's rest frame.}) of a blazar jet \citep[for an illustration, see Fig.~3 in][]{christie19}. The bolometric peak flare luminosity can be written as:
\eqb
\label{eq:Lbol}
L_{\rm pk,bol}=\frac{\pi}{2} \beta_{\rm g} c w^{\prime 2}_{\rm f}  \delta_{\rm p,f}^4 u^{\prime}_{\rm e}
\eqe
where $\beta_{\rm g}$ is the growth rate of a plasmoid, $w^\prime_{\rm f}$ and $\delta_{\rm p, f}$
are, respectively, the plasmoid transverse size and Doppler factor  at the end of its lifetime (i.e., when it is being advected from the layer or when it merges with a another bigger plasmoid). The Doppler factor takes into account the relativistic motion of the plasmoid in the layer and the relativistic motion of the layer itself (for definition, see Eq.~8 in PGS16). In the above equation, $u^{\prime}_{\rm e}\simeq f_{\rm rec} L_{\rm j}/ 4 \pi \varpi^2 c \beta_{\rm j} \Gamma_{\rm j}^2$, where $\Gamma_{\rm j}$ is the jet's bulk Lorentz factor,  $\beta_j\sim 1$, $L_{\rm j}$ is the absolute power of a two-sided jet, $\varpi/L^\prime=\epsilon_{\rm rec}/\Gamma_{\rm j} \theta_{\rm j}$ is the cross-sectional radius of the jet, $\epsilon_{\rm rec}\simeq 0.15$ is the reconnection rate, and $f_{\rm rec}\simeq 0.5$ is the fraction of dissipated magnetic energy transferred to relativistic pairs \citep[e.g.,][]{sironi15}.  The flux-doubling timescale can be estimated as: 
\eqb
\label{eq:double}
t_{1/2} \approx \frac{1}{c \beta_{\rm g} }\int_{w^\prime_{1/2}}^{w^\prime_{\rm f}} \!\!\!\frac{{\rm d}\tilde{w}}{\delta_{\rm p}(\tilde{w})}. 
\eqe
where $\delta_{\rm p}$ is the plasmoid's Doppler factor (see Eq.~8 in PGS16), which evolves during the plasmoid lifetime as this accelerates in the layer. Here, $w^\prime_{1/2}$ is the size of the plasmoid at the moment the flare luminosity reaches half of its peak value. For details about the derivation of Eqs.~\ref{eq:Lbol}-\ref{eq:double} and assumptions therein, we refer the reader to PGS16.  The free parameters of the model are: the plasma magnetization $\sigma$,  the orientation of the layer with respect to the jet axis $\theta^\prime$, the observer's angle $\theta_{\rm obs}$, $L^\prime$,  $L_{\rm j}$, and $\Gamma_{\rm j}$.  

\begin{figure}
\centering
 \includegraphics[width=0.45\textwidth]{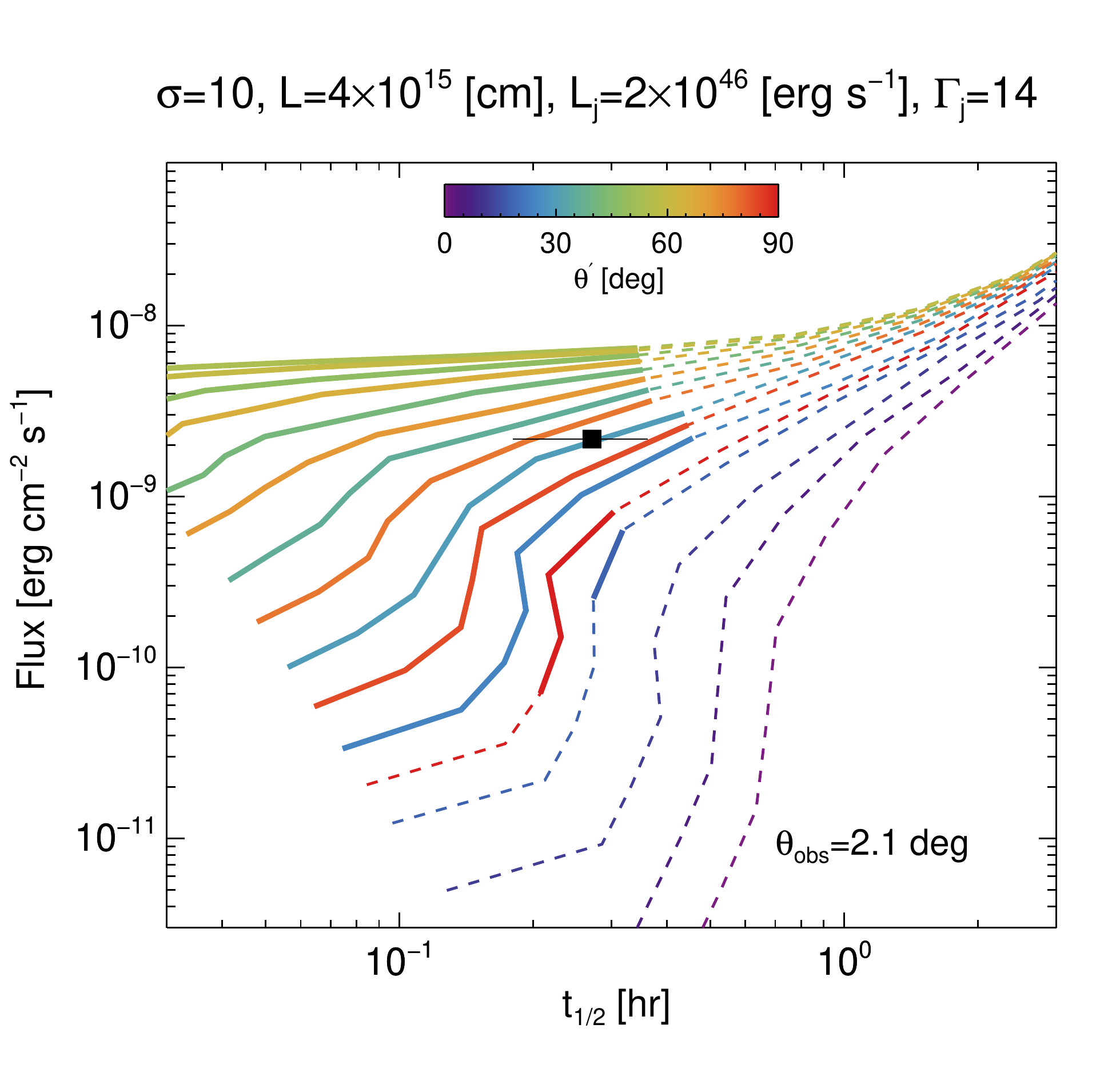}
   \caption{Bolometric peak flux of flares produced by plasmoids of different sizes as a function of the respective flux-doubling timescale for $\sigma=10$. Colored curves correspond to different orientations (see inset color bar) of the reconnecting layer with respect to the jet axis \citep[for an illustration, see Fig.~3 in][]{christie19}. Solid thick lines indicate that the plasmoid Doppler factor $\gtrsim 35$, thus satisfying the $\gamma \gamma$ opacity constraint (see Section~\ref{sec:dopplerfactor}).  
   The peak bolometric flux and  flux-doubling time of the fast flare observed in April 15 2013 is also shown (black symbol).  A choice of larger $L^\prime$ would shift the curves horizontally to longer timescales, while an increase of $\Gamma_{\rm j}$ would shift the curves diagonally towards the upper left corner of the plot. }
  \label{fig1}
\end{figure}

We apply the PGS16 model to the fast flaring activity
observed in Mrk\,421. We use as an illustrative example
the results for 2013 April 15, where fast flares with similar flux-doubling timescales  (Table~\ref{tab:FitParametersDailyApril15})
have been detected in
all energy bands (3$-$7 keV, 7$-$30 keV, 30$-$80 keV, 0.2$-$0.4~TeV, 0.4$-$0.8~TeV, $>0.8$~TeV). Using the peak flux of the fast flare in each energy band, we estimate the peak bolometric luminosity of the fast flare. Given the similar rise timescales in all bands, we use their mean value as the observed $t_{1/2}$ of the fast flare (see black symbol in Fig.~\ref{fig1}).

We consider a case with plasma magnetization $\sigma=10$. Besides the microphysical parameters, which are benchmarked with PIC simulations of
reconnection (see Table~1 in PGS16), we adopt the following values for the free parameters of the model: $L^\prime=4\times10^{15}$~cm, $L_{\rm j}=2\times10^{46}$~erg
s$^{-1}$, $\Gamma_{\rm j}=14$, and $\theta_{\rm
  obs}=\left(2\Gamma_j\right)^{-1}\simeq 2.1^{\rm o}$. Using Eqs.~\ref{eq:Lbol} and \ref{eq:double} we can then estimate the peak flux and flux-doubling
timescale of flares produced by plasmoids with different sizes, as
shown in Fig.~\ref{fig1}. 
We note that the curves showing the different orientations of the reconnecting layer do not depend monotonically on $\theta^\prime$ due to the kinematics of plasmoids in the jet (for details, see PGS16).
For the adopted parameter values, 
we find that a larger range of $\theta^\prime$ values results in fluxes and flux-doubling
timescales that are compatible  with the observed values. A choice of larger $L^\prime$ would just shift the curves horizontally to longer timescales, while an increase of $\Gamma_{\rm j}$
would shift the curves diagonally towards the upper left corner of the plot (for scalings, see equations~(33)--(37) in PSG16). A higher $\sigma$ value would have a similar effect as that of a higher $\Gamma_{\rm j}$, as it would result in stronger relativistic motions of the plasmoids in the layer. The predicted peak flux depends strongly on the angle at shorter variability timescales, whereas it is almost independent of the orientation of the layer at longer timescales. This merely reflects the fact that flares with longer timescales are produced by the largest plasmoids in the layer that move with non-relativistic speeds in the jet frame.  For these plasmoids, the beaming of the radiation is basically determined by $\theta_{\rm obs}$ and $\Gamma_j$. On the contrary, flares with short durations are produced by plasmoids that move relativistically in the jet frame and whose Doppler factor $\delta_p$ depends sensitively on both angles $\theta^\prime$ and $\theta_{\rm obs}$.

\begin{figure}
\centering
 \includegraphics[width=0.45\textwidth]{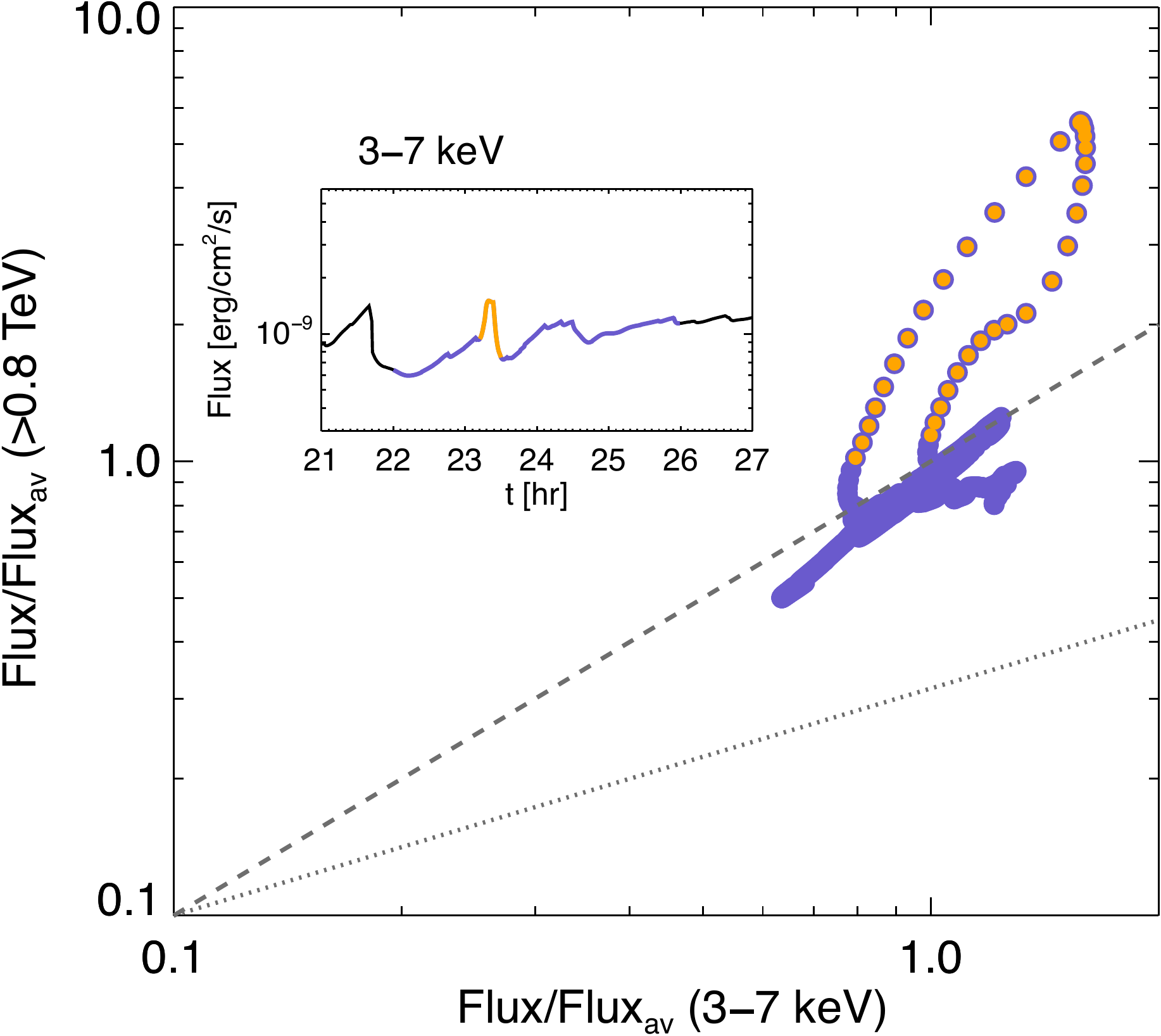}
 \caption{VHE flux ($>0.8$~TeV) versus X-ray flux (3$-$7 keV) of a plasmoid-powered light curve, computed for   a ``vanilla'' model of a BL Lac source  \citep[see model BL10 in][]{christie19}. The fluxes are extracted from a 4-hr time window of the total light curve (see purple line in the inset plot) and are normalized to their time-averaged values. The loop-like structure in the flux-flux plot is produced during a fast flare of duration $\sim 0.3$~hr (see orange points). Lines with slopes 1 (dashed) and 0.5 (dotted) are overplotted to guide the eye.} 
 \label{fig2}
 \end{figure}
 
\textit{Flux-flux correlations} -- Here  we examine the flux-flux correlations predicted by the model for the fast and slow components of a  plasmoid-powered light curve. To do so, we adopt the results of \cite{christie19} and, in particular, the light curves computed for a  ``vanilla'' model\footnote{Performing detailed radiative transfer calculations, as in \cite{christie19}, for parameters similar to those in Fig.~\ref{fig1} lies beyond the scope of this paper.} of a BL Lac-like source for $\sigma=10$ (BL10) with $\theta^\prime=30^\circ$ and $\theta_{\rm obs}=0^\circ$. The values for the angles correspond to the vanilla BL Lac model of \cite{christie19}. A choice of 2.1$^\circ$ instead of 0$^\circ$ would have no effect on the conclusions. We focus on a four hour-long segment of the total light curve that displays a fast flare (with duration $\sim 0.3$~hr) emerging on top of a less variable component (see inset plot in Fig.~\ref{fig2}). The fluxes computed in the 3-7 keV and $>0.8$~TeV energy bands during the selected time window are displayed in the main panel of Fig.~\ref{fig2} (blue points). The fluxes computed during the fast flare are highlighted for clarity (oranged colored points). Although our vanilla model cannot explain the details of the observed correlations shown in  Fig.~\ref{fig:logFlogFtrack15}, the predicted flux-flux correlations bear some similarities with the observed ones: a tight correlation with slope close to 1 is produced by the slow component of the light curve, whereas the fast flare leads to a looser correlation with a steeper slope. 

We note that the temporal resolution of our model light curve is about 30~s. The loop-like structure  produced by the fast component would likely be missed in real data due to the coarser sampling. For example, even with the 15-min sampling used in this work, the loop would consist of just two data points. Our results are also in agreement with those shown in Fig.~\ref{fig:logFlogFtrack15},  where the loop is not evident in the real data, but becomes visible only when the fitted fluxes of the slow and fast components are plotted. 

\textit{Caveats} -- We next discuss some caveats of the plasmoid-dominated reconnection model. The fast component of the optical light curve is predicted to be wider than the one seen in X-rays and VHE gamma rays, due to the longer cooling timescales of electrons radiating in the optical band. Nevertheless, some degree of correlation between the optical and other high-energy bands is expected in this model.
Unfortunately, this cannot be tested with our current dataset because we do not have a proper optical coverage around the peak times of the fast (sub-hour) flares in X-rays and VHE gamma rays (see Fig.~\ref{fig:mwl1} and Fig.~\ref{fig:LC15}, \ref{fig:LC12}, \ref{fig:LC13}, \ref{fig:LC14}). 
On longer (multi-hour) timescales, the observed optical flux appears to roughly follow the temporal trends in X-rays and VHE gamma rays for some days (e.g., April 13 and April 17), while it appears to be anti-correlated for some other days (e.g., April 11, April 12, April 15, and April 18). Such variability patterns cannot be explained by the vanilla model we described above. If magnetic reconnection can be triggered in different jet locations, then it is possible to have the formation of multiple layers in the jet with different properties 
\citep[e.g., sizes, magnetizations, and orientations;][]{giannios09, giannios_19}.
It is then possible that the observed optical emission is dominated by a different layer (e.g., larger and with different orientation) than the one producing the high-energy emissions (e.g., smaller and more aligned to the observer).

To fully account for all the observed properties of the flares (i.e., absolute flux and spectra), one would have to adjust some of the parameters entering the vanilla model, such as   the properties of the injected particle distribution within each plasmoid (e.g., slope, minimum and maximum Lorentz factor). The vanilla model described above could be improved through the inclusion of continuous slow acceleration of particles within plasmoids due to plasmoid compression \citep{ps18} and/or impulsive particle injection with a harder energy spectrum during plasmoid mergers \citep{christie19}. 

PGS16 and \cite{christie19} developed a radiative model based on the PIC simulations of \cite{sironi16}. Although the radiative model takes into account the plasmoid kinematics and dynamics, it treats each plasmoid as a homogeneous source. Thus, this model cannot be used to predict the polarization signatures of a reconnection event. This issue has been recently tackled in \cite{zhang18}, where the polarized emission was computed directly from 2D PIC simulations of reconnection, demonstrating that the latter can produce variable optical polarization. Due to the different simulation setups used in \cite{sironi16} and \cite{zhang18}, it is not clear how the polarization results found in the latter study can be applied to the plasmoid-dominated model discussed here.



\section{Summary and Conclusions}
\label{sec:conclusions}

This manuscript presents results from an extensive multiwavelength campaign in 2013 on the blazar Mrk\,421, which included, for the first time, highly sensitive hard X-ray observations with {\it NuSTAR}. The article focuses on the exceptional flaring activity observed during nine consecutive days, from April 11 to April 19, and complements the results obtained with the low VHE gamma-ray activity from January to March 2013, which was reported in \citet{2016ApJ...819..156B}. The 2013-April flaring activity is the brightest detected with MAGIC to date, and the second brightest observed in Mrk\,421 to date, after the 2010-February flaring activity reported in VERITAS-MAGIC-2019 (ApJ publication in press). More important than the bright blazar activity, is the comprehensive multi-instrument dataset collected during these nine days, which includes 42 hours with MAGIC, 27 hours with VERITAS, 18 hours with {\it Swift}, and 71 hours with {\it NuSTAR}, out of which 43 hours were taken simultaneously to the VHE observations from MAGIC and VERITAS.

Using simultaneous MAGIC-VERITAS observations of Mrk\,421 during this high activity, we noticed that the VHE gamma-ray flux values from VERITAS are systematically lower than those from MAGIC throughout the flaring activity.  The flux difference is energy dependent, being about 10\% in the 0.2--0.4~TeV band, and about 30\% in the $>$0.8~TeV band. These differences are within the quoted systematic uncertainties from both instruments, and probably related to the absolute energy reconstruction. In order to perform variability and correlation studies during single days, the VERITAS flux values were scaled up to match those from MAGIC (see Appendix~\ref{app:scalingfactors}). The physics results reported in this manuscript relate to the variations in the VHE gamma-ray flux and its correlation to the X-ray flux, and hence do not depend on the absolute value of the VHE gamma-ray flux. The results reported here would essentially be the same if we had scaled down the MAGIC fluxes to match the VERITAS fluxes.

Owing to the large fluxes and the unprecedented coverage provided by the simultaneous {\it NuSTAR} and MAGIC/VERITAS observations during these nine consecutive days, this dataset allowed us to evaluate the variability and correlations over three \mbox{X-ray} bands (3--7~keV, 7--30~keV and 30--80~keV) and three \mbox{VHE gamma-ray} bands (0.2--0.4~TeV, 0.4--0.8~TeV and $>$0.8~TeV) on timescales of 15 minutes, producing the most detailed X-ray/VHE variability and correlation study of Mrk\,421 to date. This study yielded a number of results, which we summarise below:

\begin{itemize}
\item The fractional variability $F_{var}$ vs. energy for the 9-day flaring activity shows a similar pattern to that during the low  X-ray and VHE gamma-ray activity from \citet{2016ApJ...819..156B}. The $F_{var}$ vs energy shows a double-bump structure with the highest variability occurring in the X-ray and VHE gamma-ray bands, while the variability in radio and optical bands is very low.  Additionally, we find that {\it a)} the $F_{var}$ values for the highest X-ray and VHE energies during this 9-day flaring activity are much higher than those from   the low activity reported in \citet{2016ApJ...819..156B}; 
{\it b)} the $F_{var}$ values obtained with data from the single nights are much smaller than those obtained with the 9-day flaring activity, indicating that the processes with timescales larger than half day dominate the variability at X-ray and VHE over the processes that have timescales of hours; and {\it c)} $F_{var}$ typically increases with energy for single nights, in the same way as observed for multi-day or multi-month timescales; but we also find nights where $F_{var}$ does not increase (or even decreases) with energy, hence suggesting the existence of distinct mechanisms (or distinct particle populations and/or regions) responsible for the variability of Mrk\,421.	

\item There is significant variability in the optical emission, as well as in the polarization degree and EVPA (see Fig.~\ref{fig:mwl1}), but they are not correlated to the X-ray and VHE gamma-ray emission. The optical polarization variations, while statistically significant, relate to very low polarization degree (typically less than 5\%) and appear to be random and without any obvious coherent structure. This is consistent with a multi-zone scenario such as the one proposed in \citet{2014ApJ...780...87M}, where the sum of polarization vectors from many zones would result in a low level of polarization with random fluctuations in both the polarization degree and the EVPA. At 86~GHz, the polarization degree is also low (about 3\%), as it happens with the optical data, but the polarization angle differs by about 70$^{\circ}$, which indicates that the radio and optical emissions are produced, or at least dominated, by different regions along the jet of Mrk\,421.  

\item The flux variability measured at GeV is dominated by the large flux variation around April 15, which is the day with the highest X-ray activity. Despite the lower sensitivity of {\it Fermi}-LAT, in comparison to that of the X-ray and VHE gamma-ray instruments, the GeV flux of Mrk\,421 can be measured significantly on 12-hour timescales, and it does not show the large flux variations (by factors of a few) that are reported for the X-ray and VHE gamma-ray energies. Therefore, this observation indicates that, when considering timescales of days, there is no correlation between the GeV band and the keV and TeV bands. 

\item Differently to the January-to-March 2013 results reported in \citet{2016ApJ...819..156B}, the light curves from the April 2013 flaring activity show substantial intra-day variability in both the X-ray and VHE gamma-ray bands. Moreover, one can clearly distinguish variability on sub-hour timescales on the top of flux variations occurring on multi-hour timescales.
The intra-night flux variations were quantified with a function consisting of an exponential increase and decay ({\it fast component}) on the top of a monotonically increasing or decreasing flux ({\it slow component}). We found out that, within the X-ray and VHE gamma-ray bands, the parameters describing the {\it fast component} do not depend on energy, while the ones describing the {\it slow component} can depend strongly on the energy.
We also found out that the {\it fast component} is symmetric (rise time $\sim$ decay time).
This suggests that the mechanisms that dominate the production of the sub-hour flux variability appear to be \textquotedblleft achromatic", while those responsible for the production of the multi-hour flux variability can be \textquotedblleft chromatic", at least in some cases.

\item A lower limit to Doppler factor of 35 was derived by requiring that the emitting region producing the fastest VHE variability  is optically thin to $\gamma\gamma$ pair production (see Section~\ref{sec:dopplerfactor}). This limit implicitly assumes that the low-energy (optical/IR) and VHE gamma-ray photons are produced in the same region. However, the multi-band variability and correlations derived with the 9-day April 2013 flaring activity indicates that these two emissions are likely originated in different regions, which would relax the derived lower limit.

\item Using the parameters from the {\it fast component}, we find that there are no delays between the X-ray and VHE gamma-ray emission down to the resolution of our measurement (3 minutes during April 15th). This is the strongest constraint on the correlated behavior between X-ray and gamma rays in Mrk\,421, and among the strongest constraints  derived with TeV blazar data until now.

\item The correlation between VHE gamma rays and \mbox{X-rays} is positive (i.e., the VHE gamma-ray emission increases when the X-ray emission increases); but there are multiple flavors in the strength and characteristics of this correlated behavior that change both across energy (even for nearby energy bands) and over time (on day timescales). The strongest correlation occurs between the lowest X-ray band (3$-$7~keV) and the highest VHE gamma-ray band ($>$0.8~TeV), where one finds an approximately linear change in the VHE flux with the X-ray flux. On the other hand, the weakest correlation occurs between the highest X-ray band (30$-$80~keV) and the lowest VHE gamma-ray band (0.2$-$0.4~TeV), where the VHE flux changes with the X-ray flux with a slope of $\sim$0.3 in log-log scale. This indicates that the particle population dominating the emission in the 3$-$7~keV and $>$0.8~TeV bands are closely related, while this does not occur for the bands 30$-$80~keV and 0.2$-$0.4~TeV. The decrease in the magnitude and significance of the correlation (with respect to the maximum) when increasing the X-ray energy and decreasing the VHE gamma-ray energy is expected if the variability is dominated by the highest energy electrons; but  the rapid change in the correlation pattern with energy is remarkable, and it has never been observed to date neither for Mrk\,421 nor for any other TeV blazar. 

\item  Within the context of a synchrotron/SSC model, the  approximately linear correlation between VHE and X-ray  could be an indication of an expanding blob with a decreasing magnetic field \citep{kat05}.  It could also be an indication of more than one emitting region \citep{fossati08}.

\item The temporal and spectral properties of the multi-band flares  disfavor a single-zone interpretation of the results (see Section \ref{sec:discussion2}). A scenario with multiple zones (and possibly with narrow electron energy distributions) is likely needed to explain the achromatic (chromatic) variability of the fast (slow) component of the light curves, as well as the changes of the flux-flux correlation within day-long timescales. We showed that plasmoids forming in the jet of Mrk\,421 due to magnetic reconnection might explain some of the main observational results of this campaign.
In this scenario, the multi-hour flux variations, which are found to be chromatic for some nights, would be dominated by the {\sl combined} emission from various plasmoids of different sizes and velocities, while the sub-hour flux variations, which appear to be achromatic, would be dominated by the emission from a {\sl single} small plasmoid moving with relativistic speeds along the magnetic reconnection layer. Due to the different origin of the fast and slow components of plasmoid-powered light curves, the flux correlations between different energy bands are also expected to differ. In particular, the reconnection model predicts tight correlations (with linear or sub-linear relation between energy bands) for the slow component of the light curve, and weaker correlations (with steeper than linear relation between fluxes) during fast flares. However, the characteristic loop-like structure in the VHE flux vs. X-ray flux plot of predicted for fast flares is likely to be missed with the realistic temporal binning.
\end{itemize}

The accuracy and the level of detail of this study, which is unmatched among all VHE gamma-ray emitting blazars (including past observations of Mrk\,421), shows a large degree of complexity in the variability and correlation patterns. This complexity may be present in other blazars, but it may be difficult to observe owing to insufficient temporal and energy coverage in the observations, especially with the instruments that can resolve the VHE gamma-ray fluxes with high precision. The study presented here on Mrk\,421 sheds
some light into this complex behavior, and represents a pathfinder to the studies that may be done with the next generation of ground-based VHE gamma-ray instruments like the Cherenkov Telescope Array (CTA), which is expected to resolve many VHE blazars with a level of accuracy comparable to that of MAGIC and VERITAS to resolve Mrk\,421 during this period of outstanding activity.

\acknowledgments

%
The MAGIC collaboration would like to thank the Instituto de Astrof\'{\i}sica de Canarias for the excellent working conditions at the Observatorio del Roque de los Muchachos in La Palma. The financial support of the German BMBF and MPG, the Italian INFN and INAF, the Swiss National Fund SNF, the ERDF under the Spanish MINECO (FPA2017-87859-P, FPA2017-85668-P, FPA2017-82729-C6-2-R, FPA2017-82729-C6-6-R, FPA2017-82729-C6-5-R, AYA2015-71042-P, AYA2016-76012-C3-1-P, ESP2017-87055-C2-2-P, 
FPA2017‐90566‐REDC), the Indian Department of Atomic Energy, the Japanese JSPS and MEXT, the Bulgarian Ministry of Education and Science, National RI Roadmap Project DO1-153/28.08.2018 and the Academy of Finland grant nr. 320045 is gratefully acknowledged. This work was also supported by the Spanish Centro de Excelencia ``Severo Ochoa'' SEV-2016-0588 and SEV-2015-0548, and Unidad de Excelencia ``Mar\'{\i}a de Maeztu'' MDM-2014-0369, by the Croatian Science Foundation (HrZZ) Project IP-2016-06-9782 and the University of Rijeka Project 13.12.1.3.02, by the DFG Collaborative Research Centers SFB823/C4 and SFB876/C3, the Polish National Research Centre grant UMO-2016/22/M/ST9/00382 and by the Brazilian MCTIC, CNPq and FAPERJ.

The {\em Fermi}\,-LAT\ Collaboration acknowledges generous ongoing support
from a number of agencies and institutes that have supported both the
development and the operation of the LAT as well as scientific data analysis.
These include the National Aeronautics and Space Administration and the
Department of Energy in the United States, the Commissariat \`a l'Energie Atomique
and the Centre National de la Recherche Scientifique / Institut National de Physique
Nucl\'eaire et de Physique des Particules in France, the Agenzia Spaziale Italiana
and the Istituto Nazionale di Fisica Nucleare in Italy, the Ministry of Education,
Culture, Sports, Science and Technology (MEXT), High Energy Accelerator Research
Organization (KEK) and Japan Aerospace Exploration Agency (JAXA) in Japan, and
the K.~A.~Wallenberg Foundation, the Swedish Research Council and the
Swedish National Space Board in Sweden. Additional support for science analysis
during the operations phase is gratefully acknowledged from the Istituto
Nazionale di Astrofisica in Italy and the Centre National d'\'Etudes Spatiales in France.

This work made use of data from the {\it NuSTAR} mission, a project led by the California Institute of Technology, managed by the Jet Propulsion Laboratory, and funded by the National Aeronautics and Space Administration. We thank the {\it NuSTAR} Operations, Software, and Calibration teams for support with the execution and analysis of these observations. This research has made use of the {\it NuSTAR} Data Analysis Software (NuSTARDAS) jointly developed by the ASI Science Data Center (ASDC, Italy) and the California Institute of Technology (USA).

This research has made use of the XRT Data Analysis Software (XRTDAS) developed under the responsibility of the ASI Science Data Center (ASDC), Italy.

D.P. and A.B. are grateful to Amy Furniss and Wystan Benbow for providing the VERITAS VHE gamma-ray fluxes and for useful discussions about them. M.P. acknowledges support from the Lyman Jr. Spitzer Postdoctoral Fellowship and NASA Fermi grant No. 80NSSC18K1745., and J.F. is partially supported by NASA under contract S-15633Y. M.B. acknowledges support from NASA Headquarters under the NASA Earth and Space Science Fellowship Program (grant NNX14AQ07H) and the Black Hole Initiative at Harvard University, which is funded in part by the Gordon and Betty Moore Foundation (grant GBMF8273) and in part by the John Templeton Foundation.

This research was partially supported by the Bulgarian National Science Fund of the Ministry of Education and Science under grants DN 18-13/2017, DN 18-10/2017, KP-06-H28/3 (2018) and KP-06-PN38/1 (2019), as well as for the Spanish MIMECO (AYA2016-80889-P, RYC-2013-14511) and the IAA-CSIC \textquotedblleft Severo Ochoa" program SEV-2017-0709. The St. Petersburg University team acknowledges support from Russian Science Foundation grant 17-12-01029.
The Abastumani team acknowledges financial support of the project FR/638/6-320/12 by the Shota Rustaveli National Science Foundation under contract 31/77.
T.G. acknowledges support from Istanbul University (Project numbers 49429 and 48285), Bilim Akademisi (BAGEP program), and TUBITAK (project numbers 13AT100-431, 13AT100-466, and 13AT60-430).
The Boston University effort was supported in part by NASA grants NNX12AO90G and NNX14AQ58G.
Data from the Steward Observatory spectropolarimetric monitoring project were used in this paper. This program is supported by Fermi Guest Investigator grants NNX08AW56G, NNX09AU10G, NNX12AO93G, and NNX15AU81G.
The OVRO 40 m monitoring program is supported in part by NASA grants NNX08AW31G and NNX11A043G and NSF grants AST-0808050 and AST-1109911.
The Metsähovi team acknowledges the support from the Academy of Finland to our observing projects (numbers 212656, 210338, 121148, and others).

%

\vspace{5mm}


          }

%

\appendix

\section{Quantification and correction of the offset between MAGIC and VERITAS flux measurements}
\label{app:scalingfactors}

The high accuracy in the VHE gamma-ray flux measurements presented in this work reveals an offset in the simultaneous MAGIC and VERITAS fluxes. This offset, which is energy-dependent and can reach up to 30\%, is within the flux systematic uncertainties reported by each instrument (on the order of 20--25\%). The factor that dominates the systematic uncertainties in an IACT is the uncertainty on the absolute energy scale calibration.
The energy reconstruction of VHE gamma-ray showers by IACTs  relies heavily on Monte Carlo simulations. The main source of systematic uncertainty on the energy scale is the uncertainty on the \textquotedblleft light yield" (i.e. the total light throughput) considered in these simulations, associated for instance to the average transparency of the atmosphere or the light collection efficiency of the telescopes. Each instrument tunes these parameters over long periods of time with stable performance (during years). For this reason, deviations with respect to the considered atmosphere and telescope models are expected for shorter time-scales on the order of 15\% (as discussed in \cite{Madhavan2013,2016APh....72...76A}).

\begin{figure}
\centering
\includegraphics[width=0.45\linewidth]{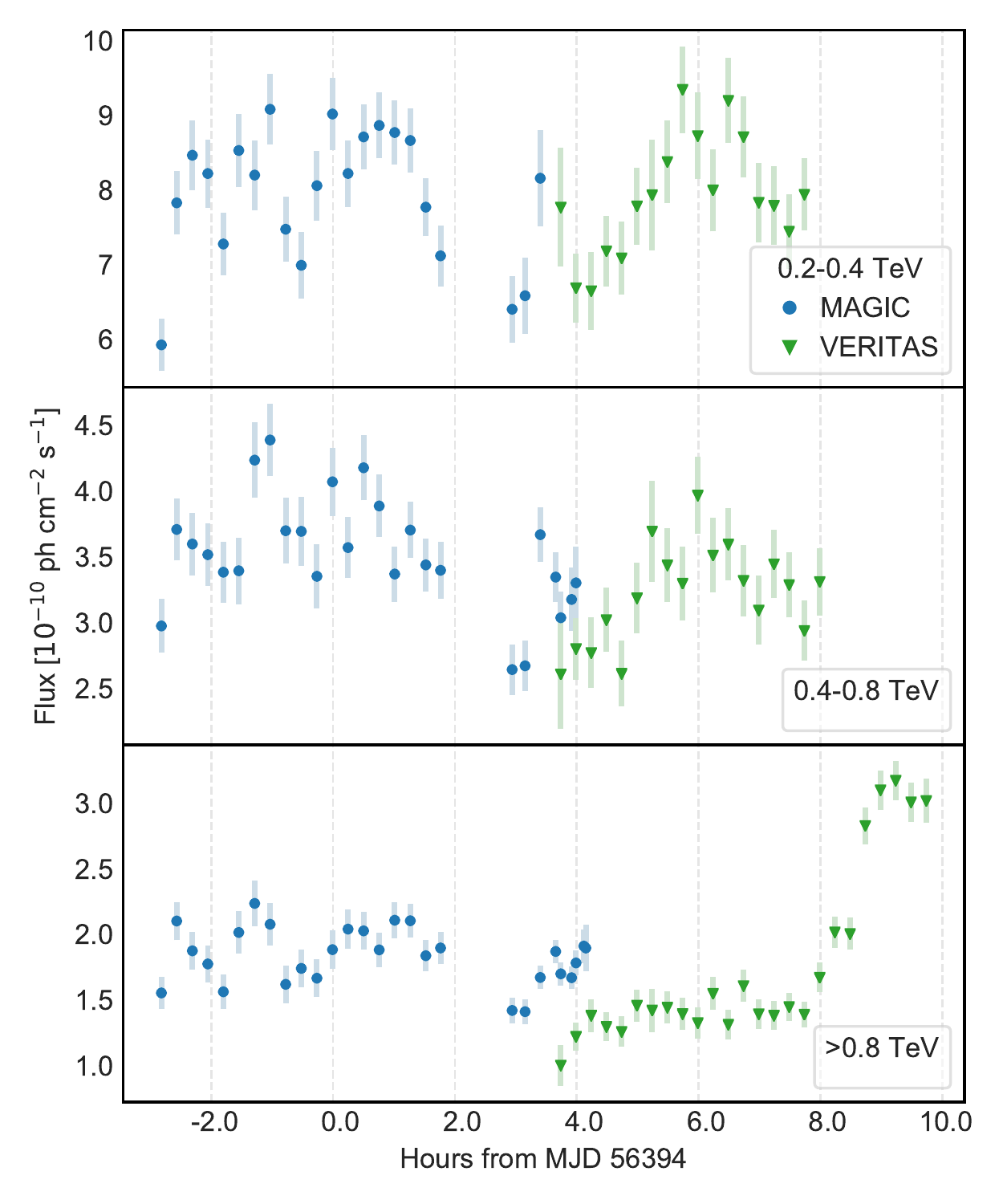}
\includegraphics[width=0.45\linewidth]{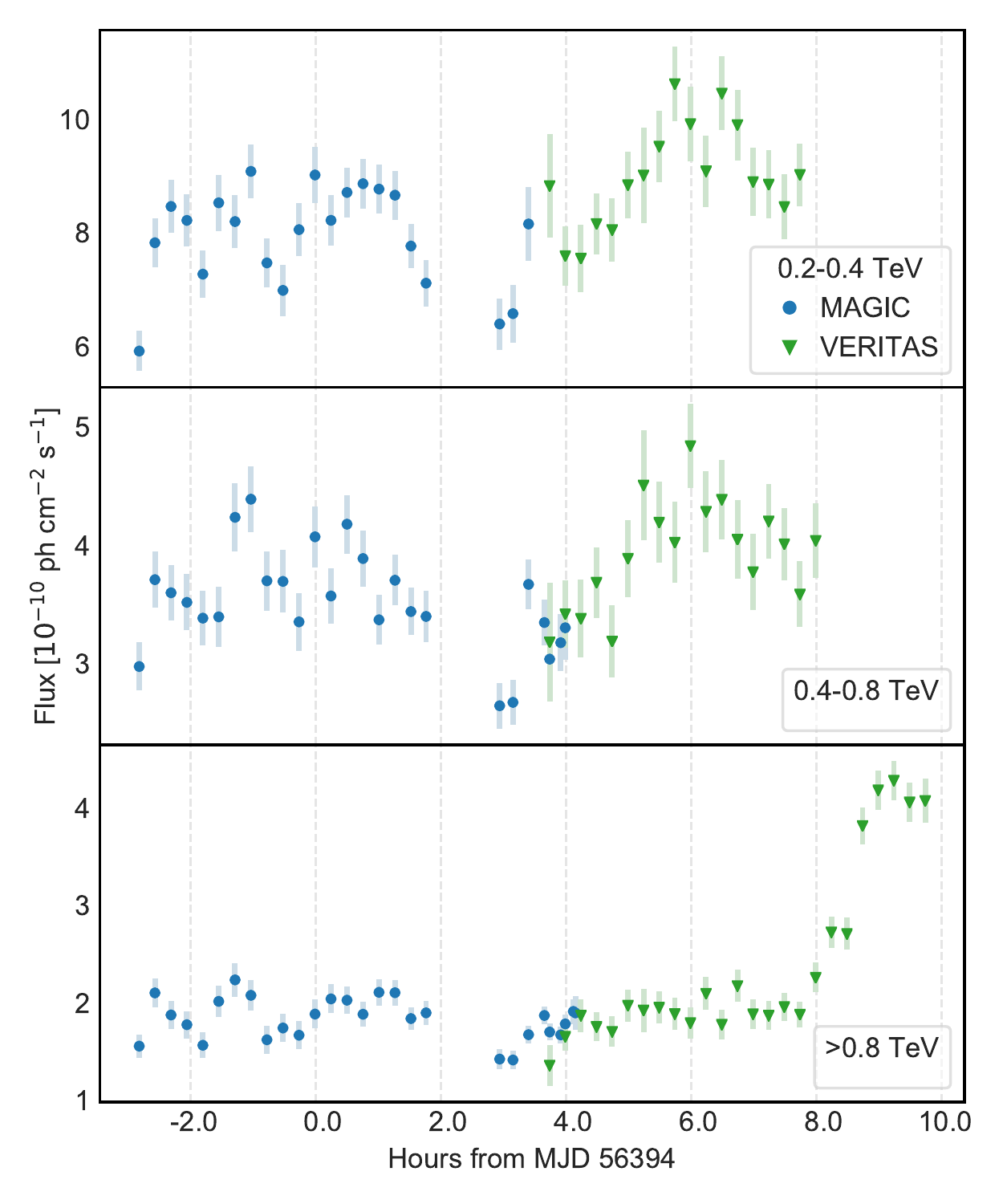}
\caption{Light curves from April 12th in three VHE gamma-ray energy bands: 0.2-0.4~TeV, 0.4-0.8~TeV and $>$0.8~TeV, as measured with MAGIC and VERITAS before (left) and after (right) applying the correction factors from Table~\ref{tab:CorrectionFactors}.  The temporal coverage (i.e. number of data points) varies with the energy band because of the increasing energy threshold with the increasing zenith angle of the observations (see Sect.~\ref{sec:obs} for further details). \label{fig:April12}}
\end{figure}

In order to exploit the excellent coverage of this dataset and perform intra-night VHE gamma-ray flux variability and X-ray/VHE flux correlation studies, these energy-dependent offsets need to be corrected. 
This mismatch shows up prominently when comparing single-night light curves. 
As an example, the left panel in Fig.~\ref{fig:April12} displays the multi-band VHE light curves during the night of April 12th, showing an increasing offset with energy reaching about 30\% for the VHE gamma-ray fluxes with E $>$ 0.8~TeV. We note that an increasing offset with energy is expected if the main source of this systematic difference between the MAGIC and VERITAS fluxes is the energy scale. 

The simplest approach to correct for this difference is to scale the lightcurves of one of the instruments 
by an energy-dependent factor. As we have been using throughout this work three different energy ranges (i.e. 0.2--0.4~TeV, 0.4--0.8~TeV, and $>$0.8~TeV), three different flux ratios need to be calculated to properly correct for this bias. To calculate these correction factors, two different methods were tested: a) obtain three scaling factors by normalising the MAGIC and VERITAS fluxes in the three energy bands; and b) calculate the energy-scale correction that makes MAGIC and VERITAS spectra compatible, and then obtain the three scaling factors for the three energy bands considered.
The first approach is purely agnostic, 
while the second one assumes that the small mismatch between MAGIC and VERITAS VHE gamma-ray measured fluxes is dominated by the systematic uncertainty on the energy scale of each instrument. 
Both strategies make use of the 2h~45min simultaneous MAGIC and VERITAS observations taken during four different nights (see Tab.~\ref{tab:XVHEoverlap}).

\begin{figure}
\centering
\includegraphics[width=0.8\linewidth]{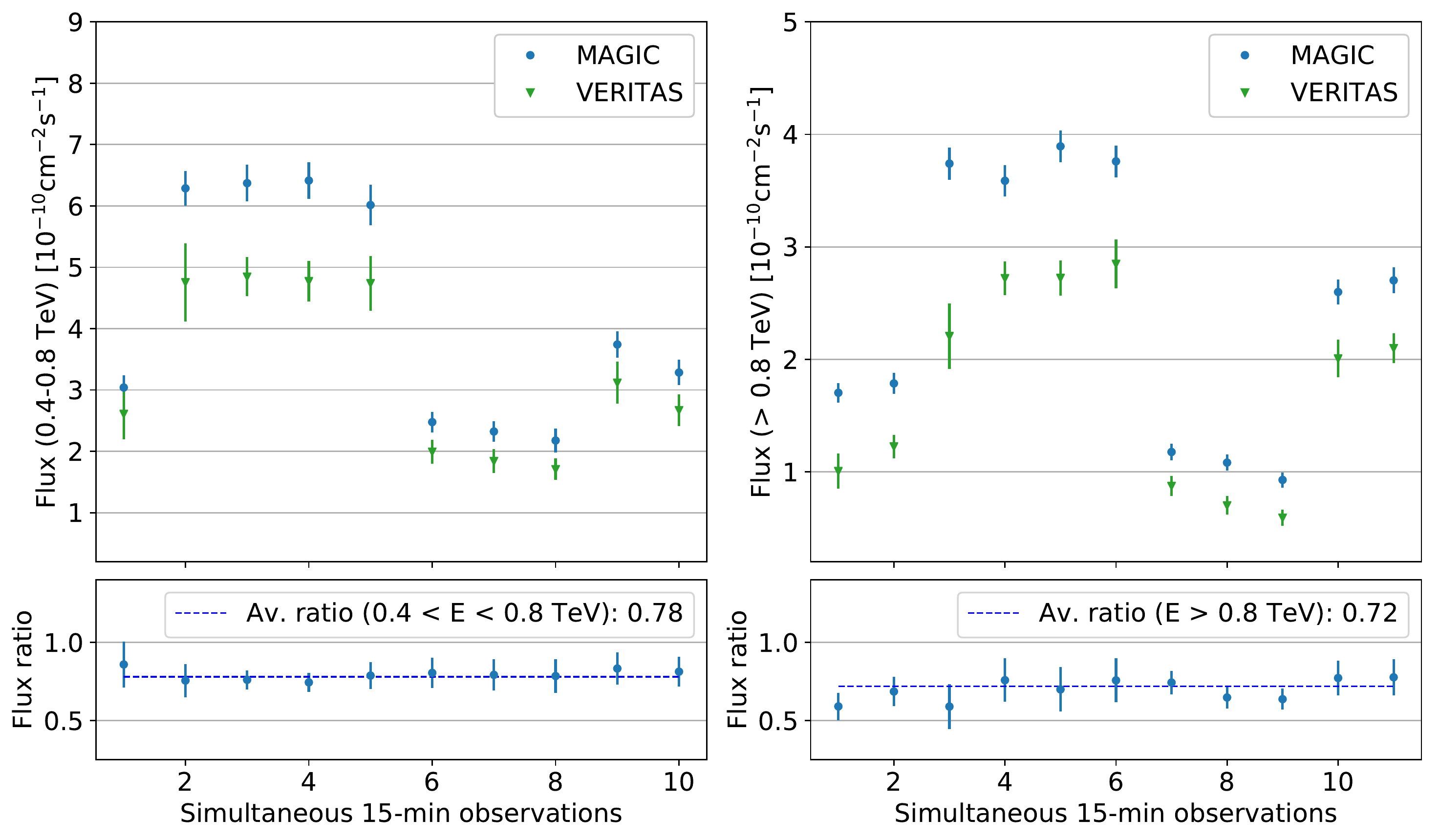}
\caption{Comparison of the VHE gamma-ray fluxes measured with MAGIC and VERITAS for the 15-min time intervals of simultaneous MAGIC-VERITAS observations. 
\label{fig:MAGIC-VERITAS_fluxratios}}
\end{figure}

Inferring the correction factor from available simultaneous data is the most direct method to calculate these factors; but it cannot be applied to the 0.2--0.4~TeV energy band because the simultaneous MAGIC-VERITAS data relate to MAGIC observations at zenith angle above 55$^{\circ}$, which have an analysis energy threshold above 0.3~TeV (see Sect.~\ref{sec:obs}).
 Fig.~\ref{fig:MAGIC-VERITAS_fluxratios} shows a comparison of the VHE gamma-ray fluxes (in the two available energy bands) derived with simultaneous 15-min time bins, which span throughout four consecutive nights: from April 12 to April 15.

The figure shows two important characteristics of this offset. First, the ratio at a given energy band is approximately the same for all the simultaneous observations, hence indicating a systematic effect between MAGIC and VERITAS, which is not related to the peculiarity of one single day (e.g. bad weather in one of the two telescope sites). 
Second, the ratio of MAGIC and VERITAS fluxes deviates from 1 by a larger amount in the highest energy band, hence confirming the energy dependence of the offset. The correction factors and uncertainties inferred with this method are reported in the first column of  Table~\ref{tab:CorrectionFactors}.

As described in \cite{2016APh....72...76A}, MAGIC analysis allows to test the impact of a modified light yield over a given reconstructed spectral energy distribution (SED). By applying relative light yield corrections to the MAGIC data simultaneous with VERITAS, we are able to test which value minimizes the MAGIC-VERITAS SED differences. This search determined that a decrease of 20\% in the MAGIC light makes the simultaneous MAGIC and VERITAS spectra compatible\footnote{A similar effect is expected while increasing the light yield of VERITAS, but these tests were only possible over MAGIC data.}. The flux correction factors for the three energy ranges considered here were then computed through the ratio of the pre- and post-light-yield-corrected MAGIC lightcurves in the three energy bands. In order to reduce the statistical uncertainties, we used the full MAGIC data set from these four nights (April 12-15), instead of using only the time intervals with simultaneous MAGIC-VERITAS observations.  The flux ratios derived with this method are reported in the second column of Table~\ref{tab:CorrectionFactors}.

\begin{table}
\centering
\caption{Scaling factors (defined as $F_\text{VERITAS}$ / $F_\text{MAGIC}$) inferred from two different methods (see text for further details) for the three VHE gamma-ray energy bands considered. The values reported correspond to the best fit value and the related statistical uncertainty. Both methods use data from simultaneous MAGIC-VERITAS observations.}
\label{tab:CorrectionFactors}
\medskip
\begin{tabular}{c|c|c}
\hline
 Energy band & Flux ratio & Energy-scale corr. \\
\hline
0.2--0.4~TeV & - & 0.88 $\pm$ 0.03 \\ 
0.4--0.8~TeV & 0.78 $\pm$ 0.03 & 0.82 $\pm$ 0.06\\ 
~$>$~0.8~TeV & 0.72 $\pm$ 0.02 & 0.74 $\pm$ 0.05 \\ 
\hline
\end{tabular}
\end{table}



As shown in Table \ref{tab:CorrectionFactors}, there is a very good agreement in the flux correction factors calculated with these two different methodologies. From here on, the VERITAS light curves are normalized to those of MAGIC using the method derived with the energy-scale correction test (second column in Table \ref{tab:CorrectionFactors}). We chose this method (instead of applying the flux ratio method) because this one provides a scaling factor for the lowest energy band, 0.2-0.4~TeV, for which we do not have simultaneous MAGIC-VERITAS data. 
An example of a light curve after the application of the scaling factors to the VERITAS fluxes is given in the right panel of Fig.~\ref{fig:April12}. 


In order to test the impact of the scaling factor uncertainty on the VHE/X-ray correlations reported in Section~\ref{sec:correlations} and Tables~\ref{tab:CorrelationCoefficients} and \ref{tab:CorrelationCoefficientsDaily}, we scaled the VERITAS fluxes by a value drawn from a normal distribution with mean and sigma given by the parameters in Table~\ref{tab:CorrectionFactors}. This was repeated 1000 times, and each time all the correlations were calculated. The standard deviation of the obtained set of correlation parameters was taken as a measure of the systematic uncertainty associated to the scaling of the VERITAS fluxes. For the nine-day correlations in Table~\ref{tab:CorrelationCoefficients}, the systematic uncertainty was found to be an order of magnitude smaller than the statistical uncertainty. This is expected because the flux-corrections are of the order of 20\%, while the nine-day flaring activity is dominated by flux changes of about one order of magnitude. On the other hand, the impact of this flux scaling is not negligible for quantities derived with the single-night light curves, which are dominated by flux changes of about a factor of 2. The uncertainties related to the flux-scaling are reported as the second uncertainty in Table~\ref{tab:CorrelationCoefficientsDaily}. Typically, these systematic uncertainties are found to be smaller, or at most comparable, to the statistical uncertainty.

\section{Daily normalized light curve sets}
\label{app:NormLCs}

Using the scaling factors from Appendix \ref{app:scalingfactors},
one can now up-scale all the VERITAS flux measurements and produce
single-night light curves with simultaneous VHE and X-ray data spanning about 8--10~hours. The VHE flux measurements can be best compared with the {\it NuSTAR} measurements on timescales of 15 minutes, through light curves where the X-ray and VHE gamma-ray fluxes are normalized so that one can compare the relative differences among them. 

\begin{figure*}
\includegraphics[width=\linewidth]{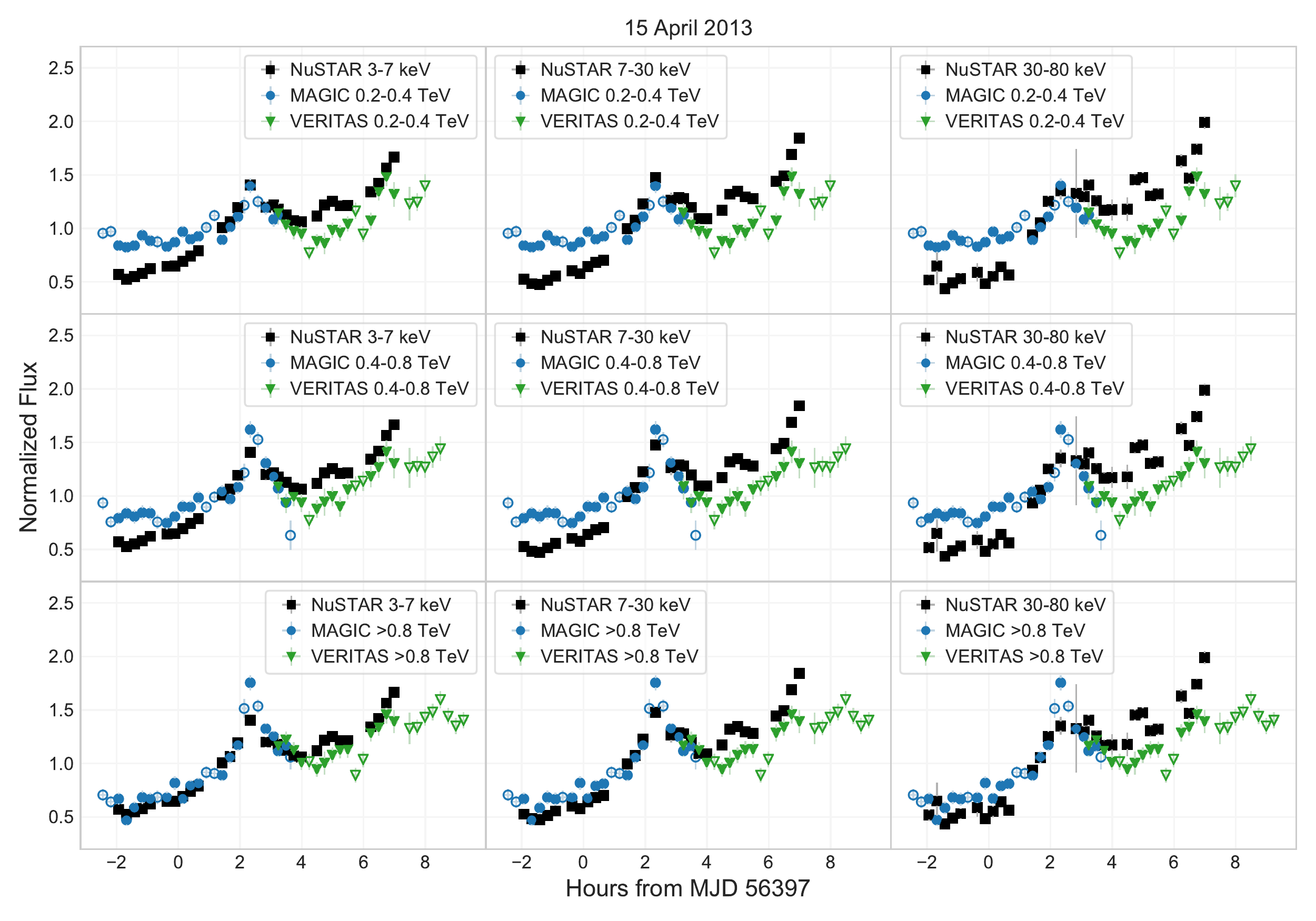}
\caption{Normalized VHE and X-ray light curves with 15-min time bins for 2013 April 11-19 (continuing on following pages). Successive panels show VHE light curves in three energy bands, namely $E>0.8$~TeV, $0.4-0.8$~TeV, and $0.2-0.4$~TeV energy bands, and three X-ray bands, namely $3-7$~keV, $7-30$~keV and $30-80$~keV. Filled markers are used for those 15-min temporal bins with simultaneous X-ray and VHE data. The fluxes are normalized with the average flux for the night, where the average flux is computed excluding the non-simultaneous X-ray and VHE observation.}
\label{fig:NormLC_April15}
\end{figure*}

In order to enable direct (normalized) flux comparisons between the X-ray and VHE bands, the fluxes are normalized by calculating the average flux for each night and each band using only data points that have simultaneous VHE and X-ray observations (filled markers in Fig.~\ref{fig:NormLC_April15}). The resulting normalization factors are reported in the Table \ref{tab:Normalizations}. Subsequently, all the light curve fluxes (including the non-simultaneous data marked with open symbols) are divided by this normalization factor. To be able to visually compare the intra-night flux changes, all VHE and X-ray band combinations for each day are shown in Fig.~\ref{fig:NormLC_April15}. Using the normalization factors from the Table \ref{tab:Normalizations}, the scaling factors from Table~\ref{tab:CorrectionFactors}, and the normalized light curves flux points from Fig.~\ref{fig:NormLC_April15}, one can retrieve the primal X-ray and VHE light curves from {\it NuSTAR}, MAGIC and VERITAS. 



Normalized light curves in Fig.~\ref{fig:NormLC_April15}  all show that, during these few-days-long activity, the X-ray and VHE emissions of Mrk\,421 show quite a number of structures on timescales from multiple hours, and down to timescales smaller than one hour. They also show remarkable correlations in some of the band combinations for selected nights (e.g. 3-7~keV and $>$0.8~TeV bands on 15 April, or 3-7~keV and 0.2-0.4~TeV bands on 11 April). Looking at single day band combinations plots, one can also explicitly see different trends for the fast and the slow components (e.g. 15 April slow change in 3-7~keV flux is not as perfectly matched in 0.2-0.4~TeV band as is in the highest VHE band, while the fast component shows comparable agreement).
 
\clearpage

\begin{center}
\includegraphics[width=0.85\linewidth]{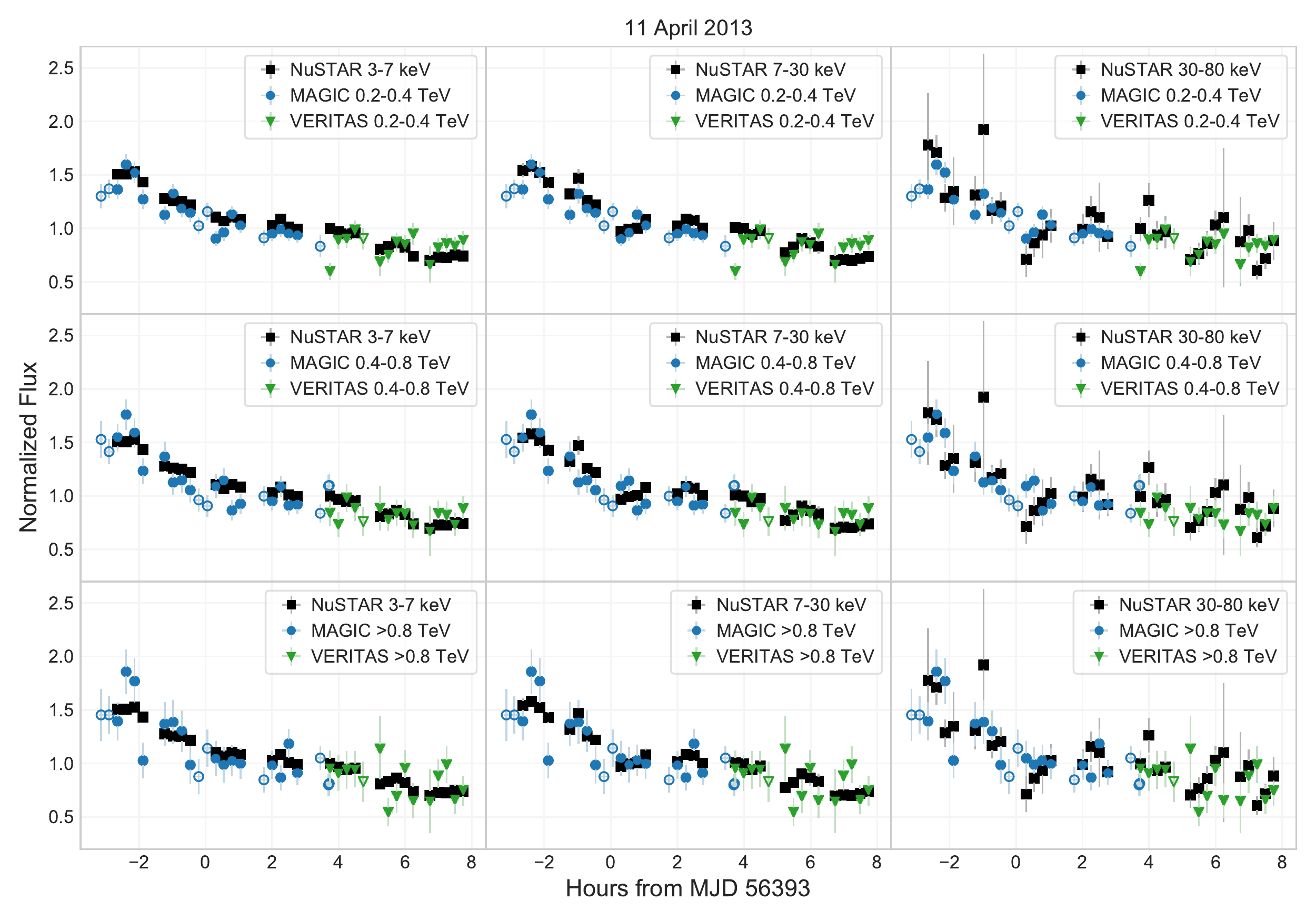}
\includegraphics[width=0.85\linewidth]{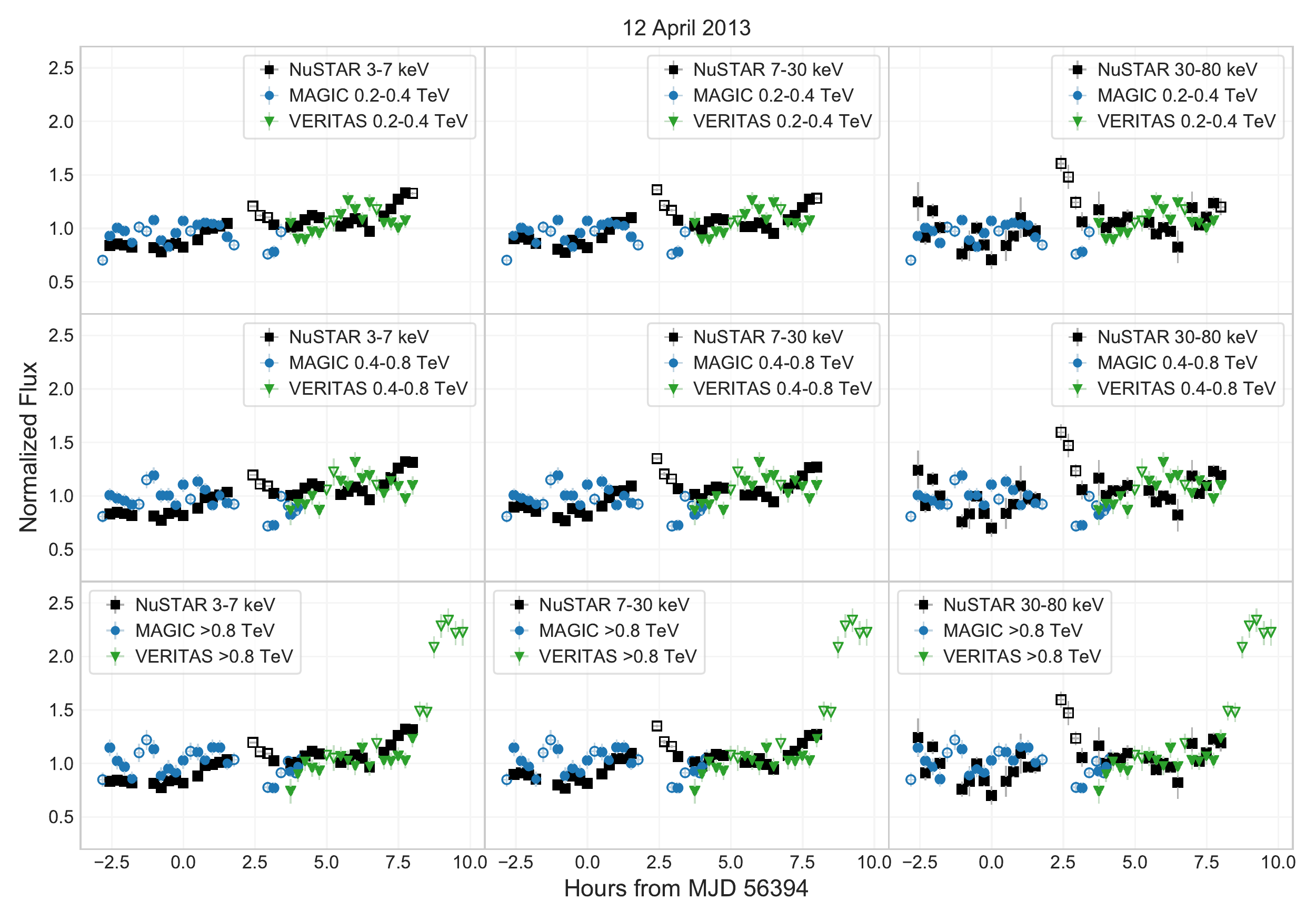}\\
\em{Figure 12 continued}
\end{center}
\clearpage
\begin{center}
\includegraphics[width=0.85\linewidth]{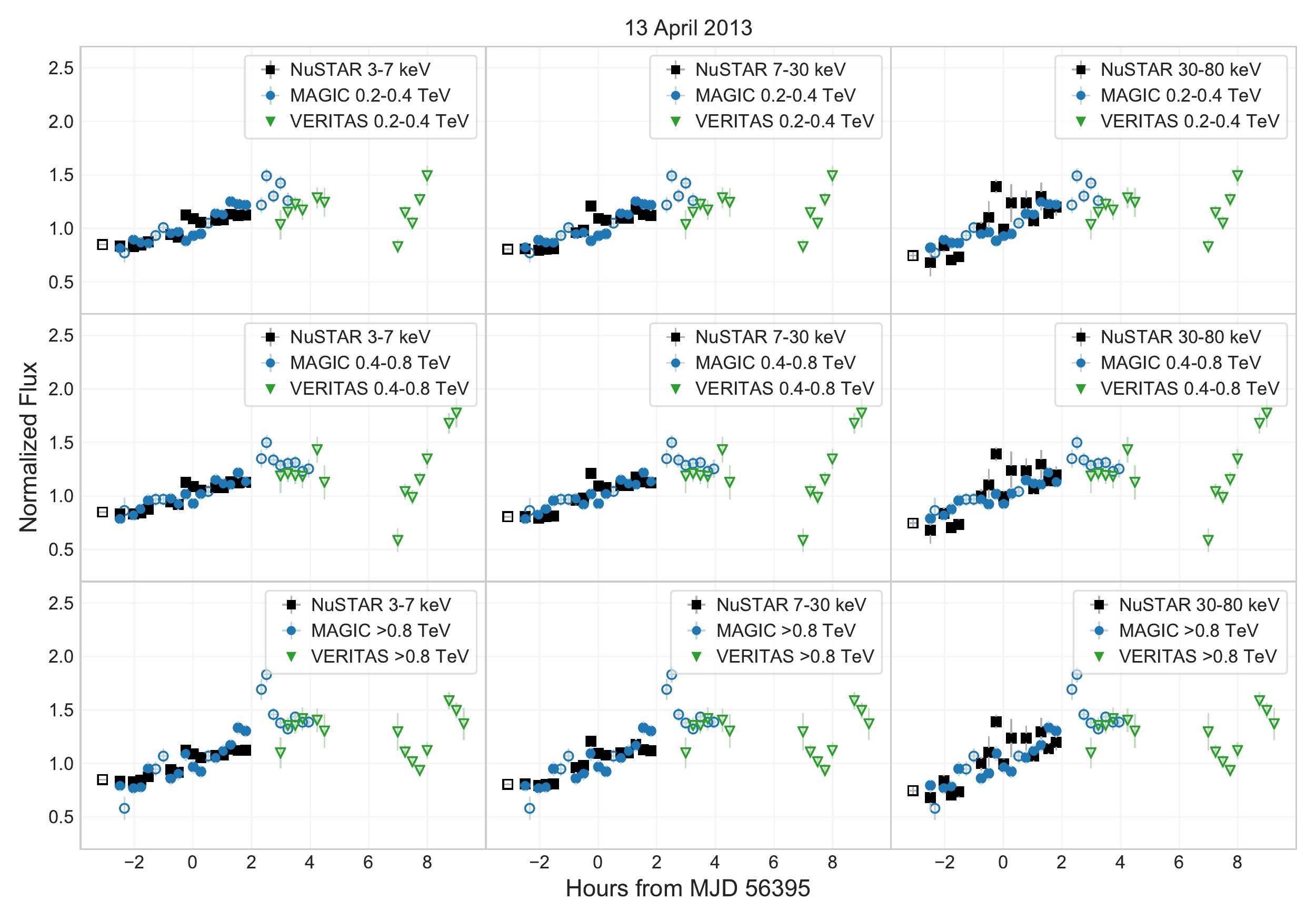}
\includegraphics[width=0.85\linewidth]{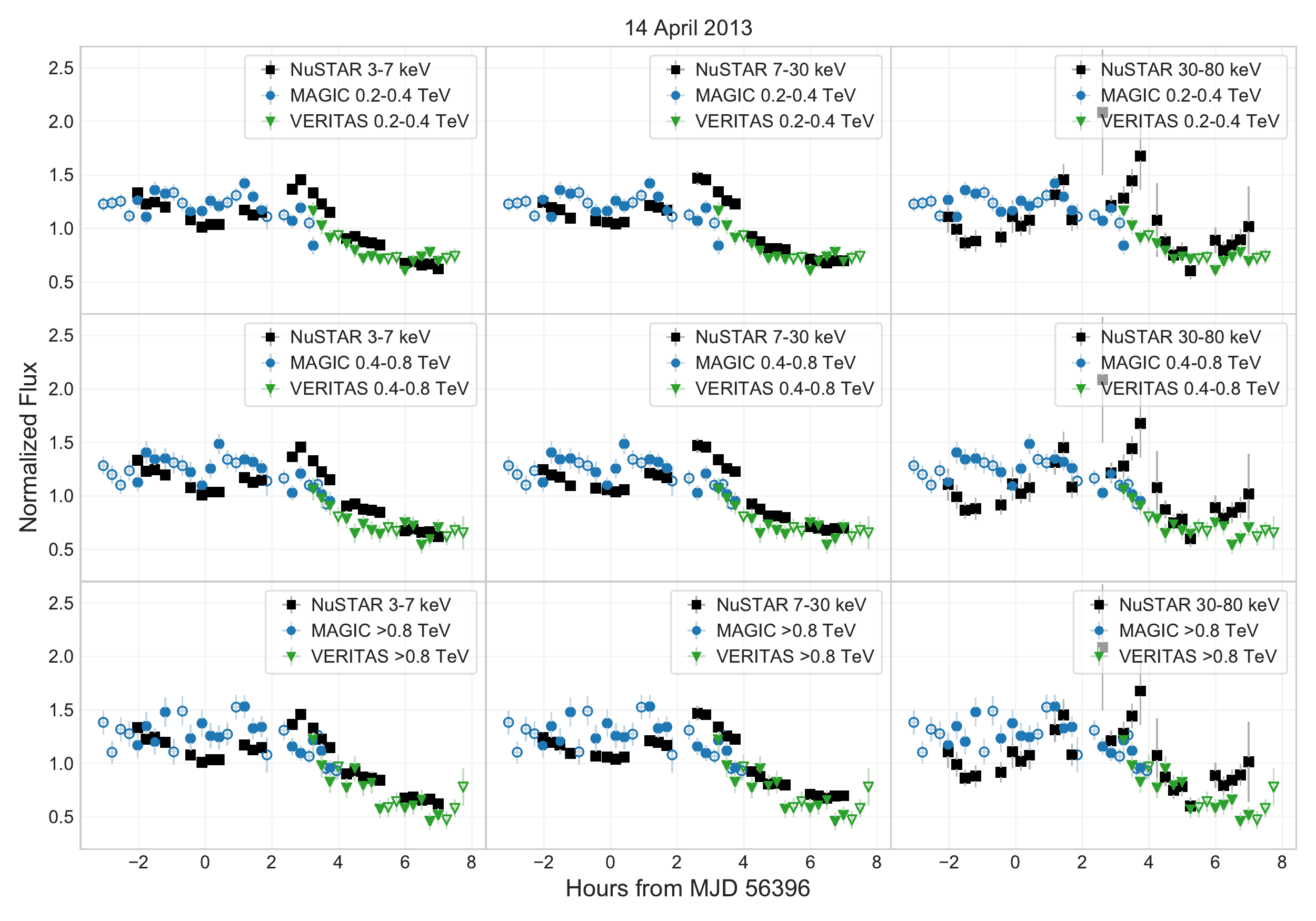}\\
\em{Figure \ref{fig:NormLC_April15} continued}
\end{center}

\begin{center}
\includegraphics[width=0.85\linewidth]{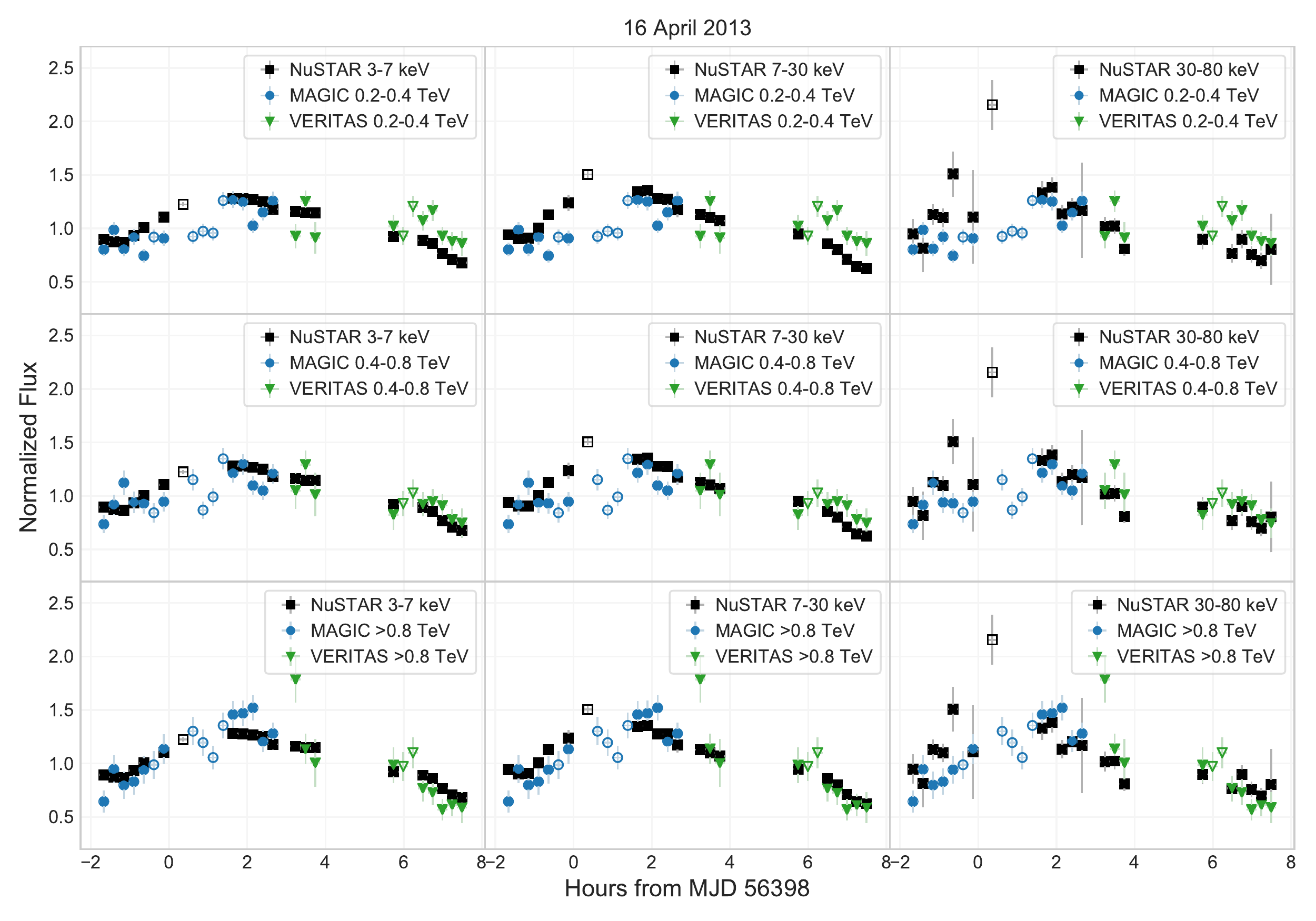}
\includegraphics[width=0.85\linewidth]{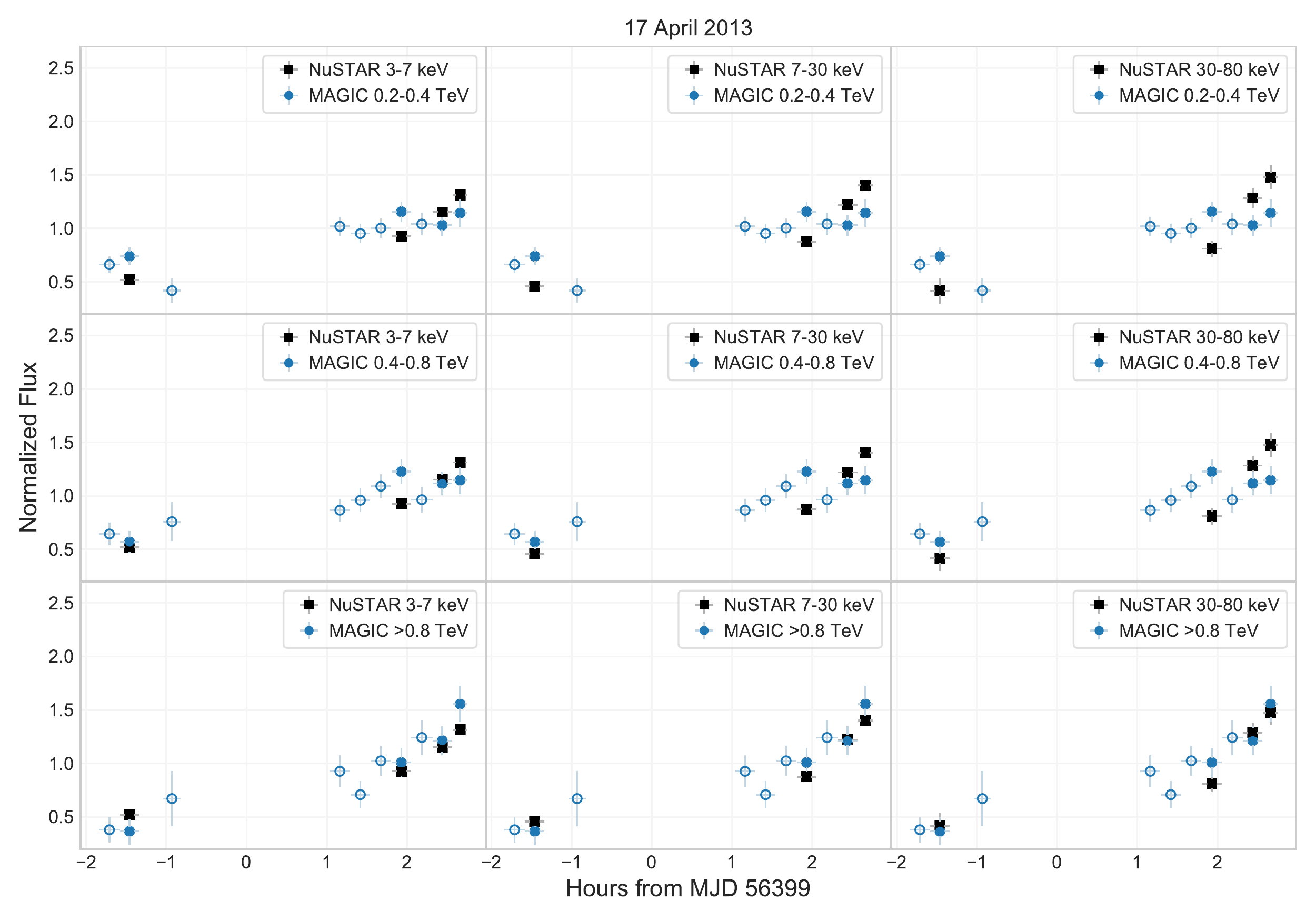}\\
\em{Figure \ref{fig:NormLC_April15} continued}
\end{center}

\begin{center}
\includegraphics[width=0.85\linewidth]{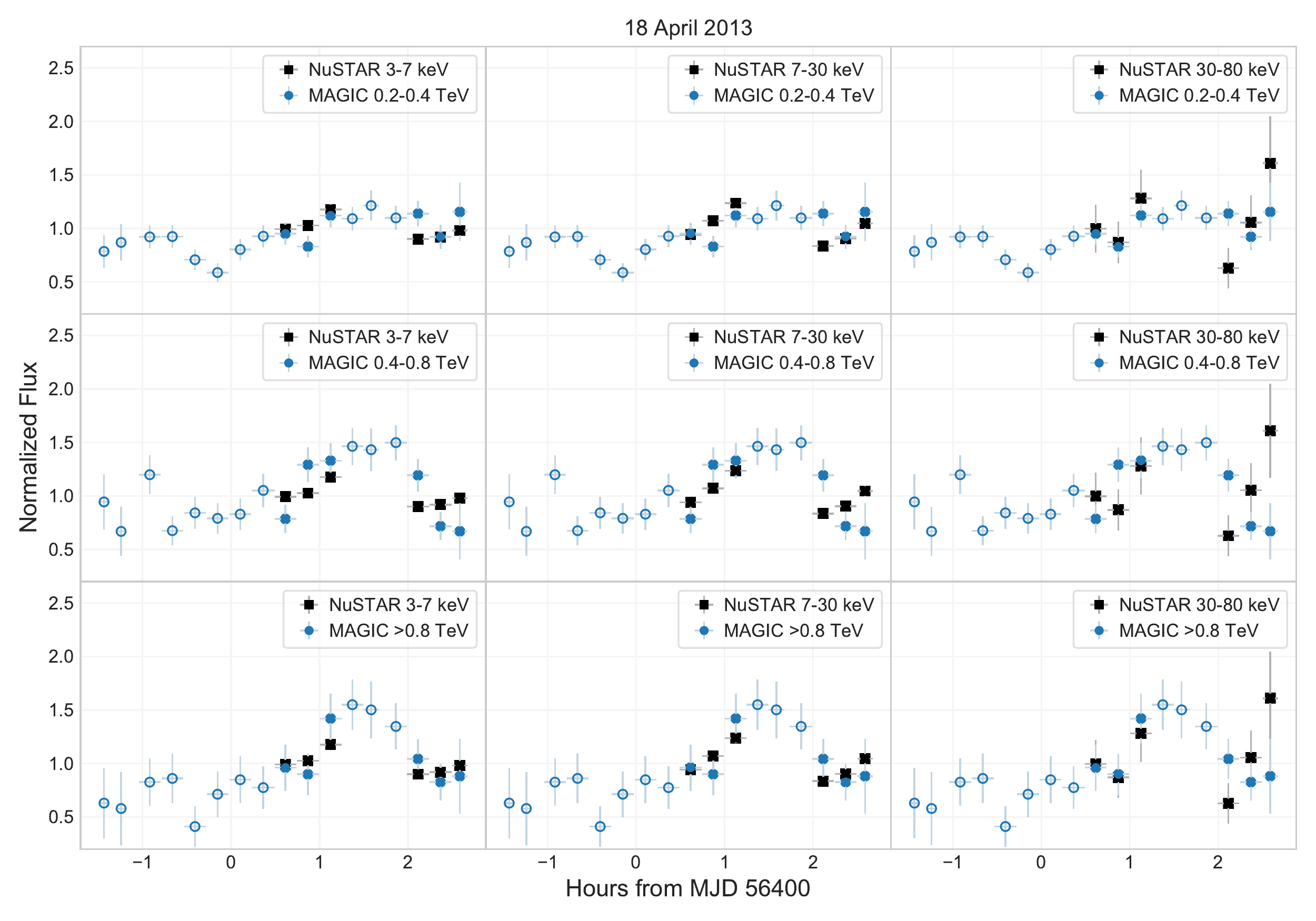}
\includegraphics[width=0.85\linewidth]{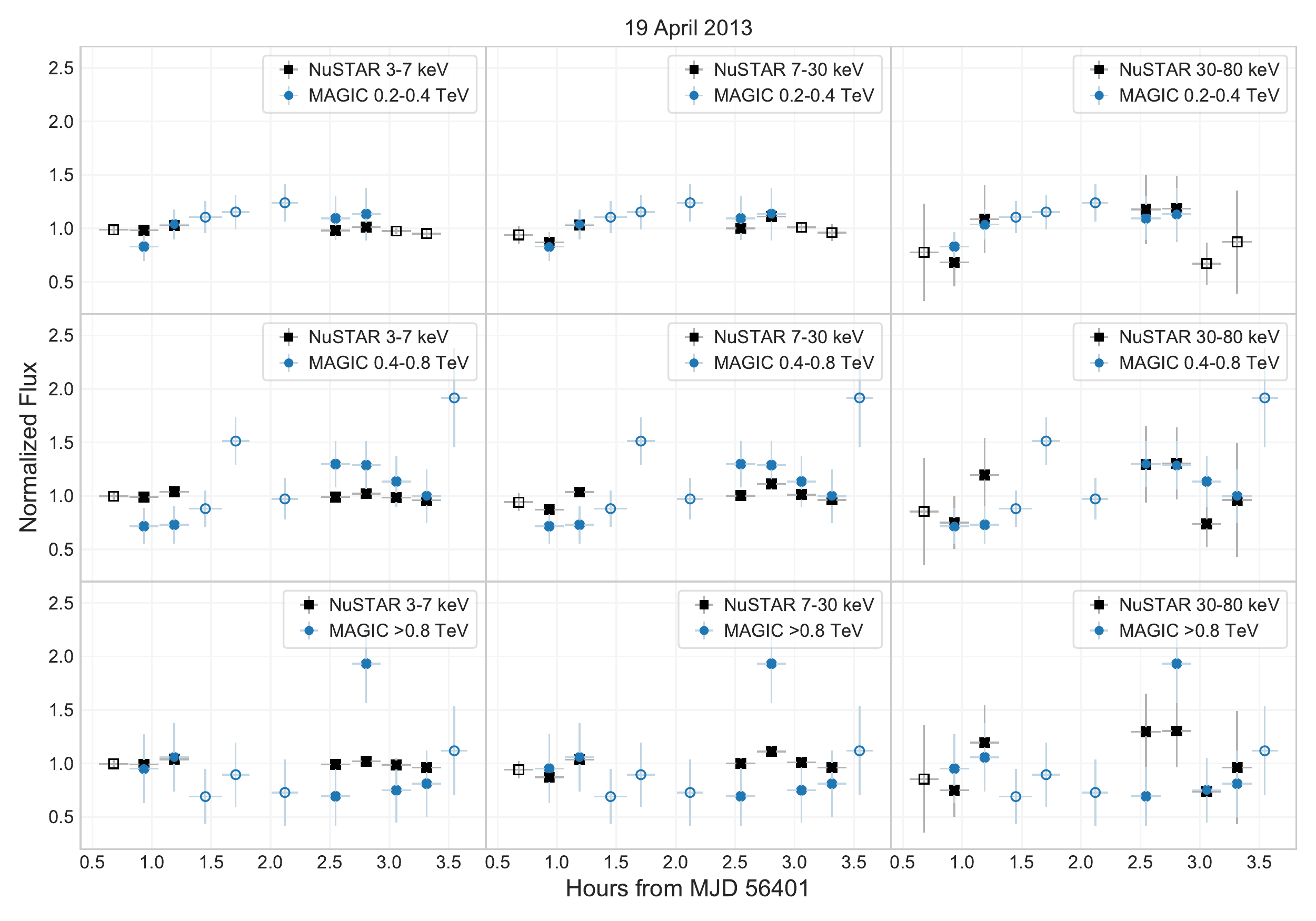}\\
\em{Figure \ref{fig:NormLC_April15} continued}
\end{center}
\clearpage

\begin{deluxetable*}{c|c|c|c|c|c|c}
\tablecaption{Normalization factors\tablenotemark{a} used in the daily light curves reported in Fig.~\ref{fig:NormLC_April15}. \label{tab:Normalizations}}
\tablewidth{0pt}
\tablehead{
Day & $3-7$ keV & $7-30$ keV  & $30-80$ keV & $0.2-0.4$ TeV  & $0.4-0.8$ TeV & $>$0.8 TeV  
}
\startdata
11 April 2013 & 0.2657 & 0.1670 &0.02437 & 3.872 &1.312 & 0.4375 \\ 
12 April 2013 & 0.7881 & 0.6725 &0.1291 & 8.431 &3.683 & 1.834 \\ 
13 April 2013 & 1.119 & 0.9506 &0.1737 & 10.23 &4.889 & 2.713 \\ 
14 April 2013 & 0.4043 & 0.2806 &0.04366 & 5.758 &2.286 & 0.9662 \\ 
15 April 2013 & 1.021 & 1.194 &0.3437 & 7.373 &3.493 & 2.326 \\ 
16 April 2013 & 0.4054 & 0.3498 &0.07931 & 4.211 &1.561 & 0.6639 \\ 
17 April 2013 & 0.3190 & 0.3397 &0.1042 & 2.706 &0.9845 & 0.37643 \\ 
18 April 2013 & 0.08387 & 0.05741 &0.007850 & 2.053 &0.5959 & 0.2168 \\ 
19 April 2013 & 0.07172 & 0.03977 &0.005220 & 1.175 &0.3813 & 0.09668 \\ 
\enddata
\tablenotetext{a}{Mean X-ray fluxes are given in units $10^{-9}$ erg cm$^{-2}$ s$^{-1}$, and VHE gamma-ray fluxes are given in \mbox{$10^{-10}$ ph cm$^{-2} $ s$^{-1}$.} }
\end{deluxetable*}

\section{Flux variations on multi-hours and sub-hours timescales}
\label{app:lc_fits}

This section reports the quantification of the X-ray and VHE gamma-ray multi-band  flux variability during single nights using the templates defined by Eq.~\ref{eq:flux_t} in Section~\ref{subsec:fastvariability}. The best-fit results obtained with the light curves from April 11-14 are reported in Table~\ref{tab:FitParametersDailyOthers}, while the results obtained for April 16 are reported in Table~\ref{tab:LCFitApril16}.
Below we provide a brief description of the used strategy and the obtained results for each of the nights. 
 
\subsection{11 April 2013}
\begin{figure*}
\plottwo{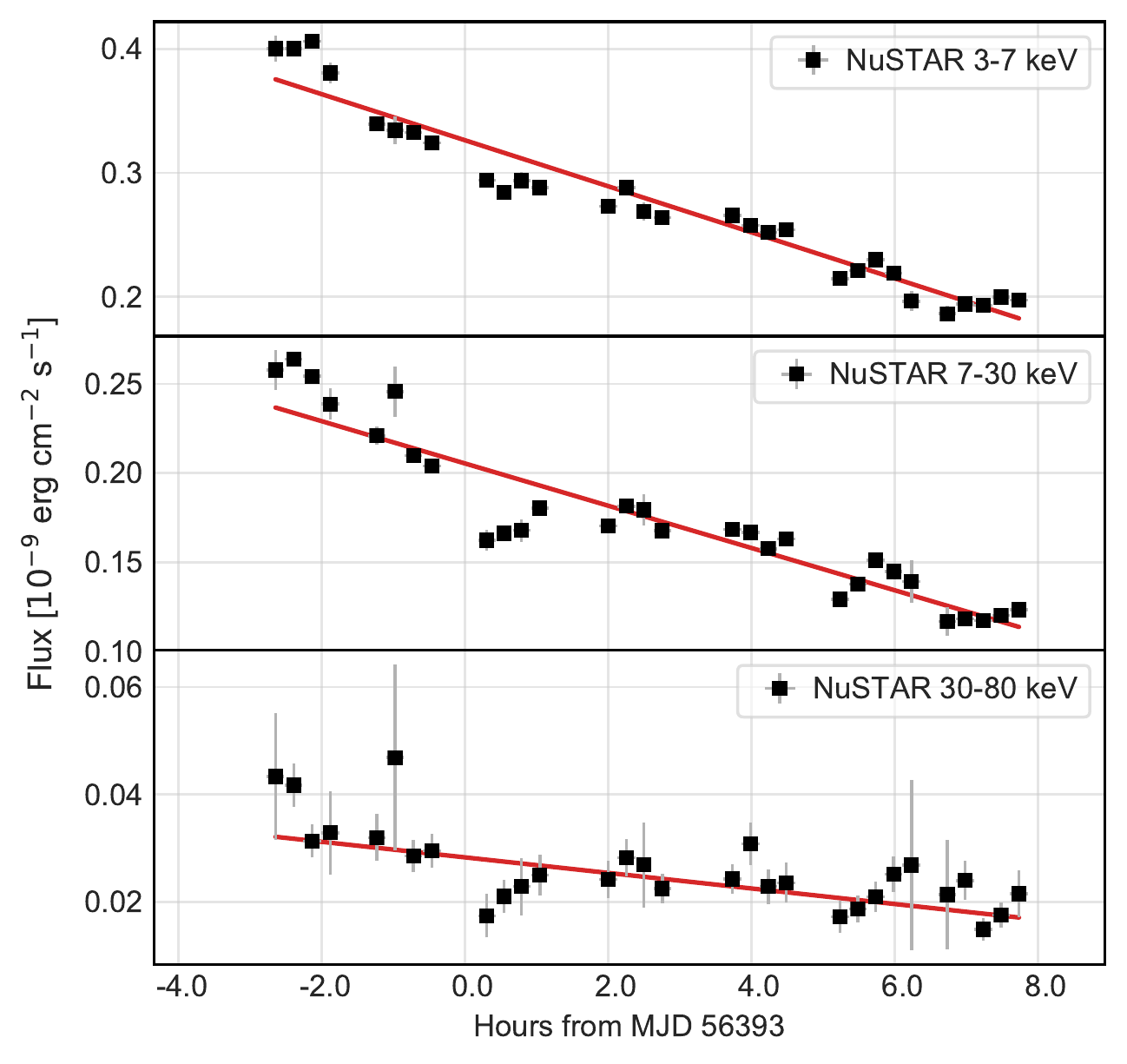}{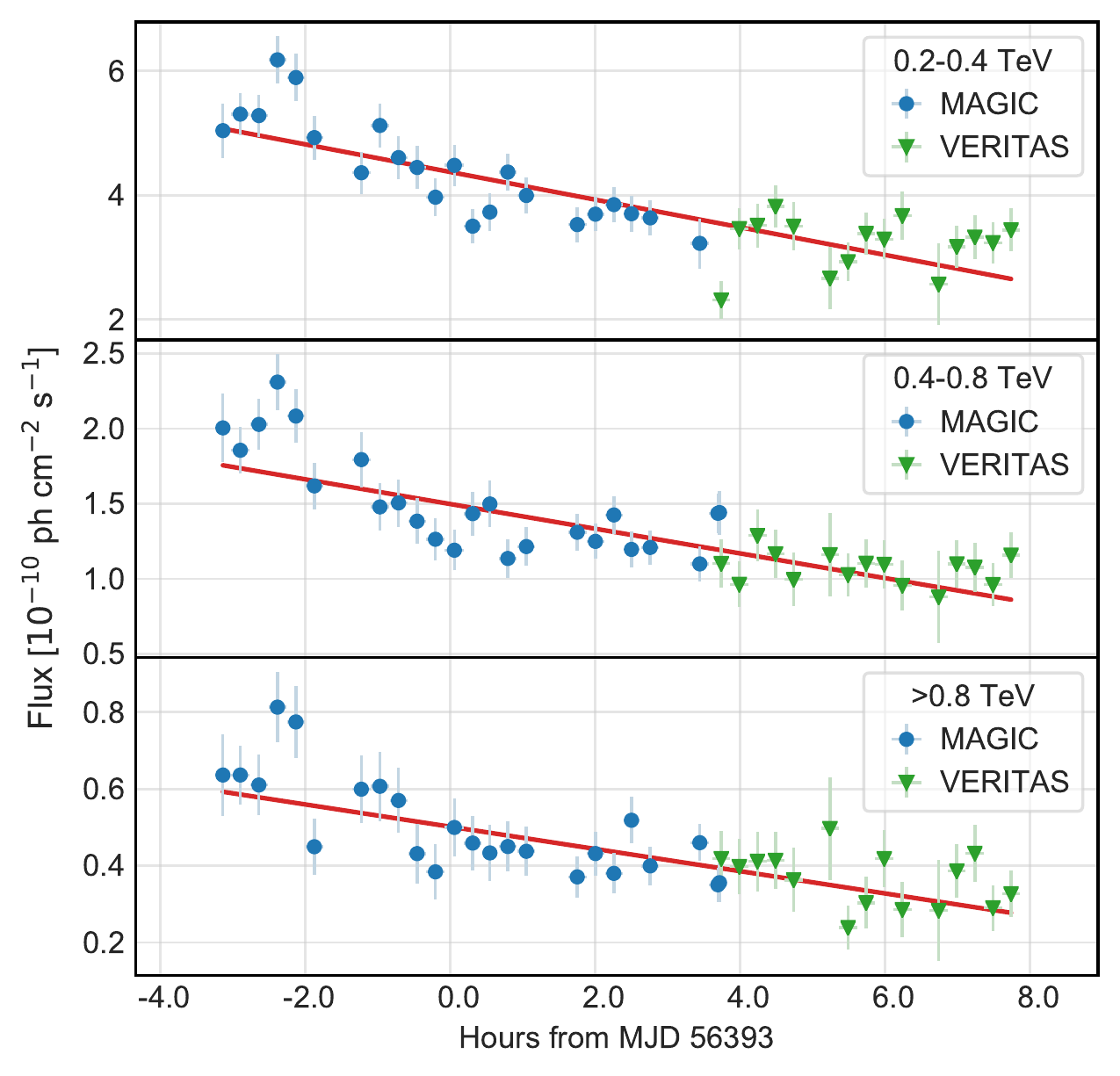}
\caption{Light curves from 2013 April 11 in three X-ray bands (left panel) and three VHE gamma-ray bands (right panel). The red curve is the resulting fit with the function defined by Eq.~\ref{eq:flux_t}, whose model parameters are reported in Table~\ref{tab:FitParametersDailyOthers}.}
\label{fig:LC11}
\end{figure*}

There are no flares visible in any of the VHE or X-ray bands. The light curves are not consistent with constant flux, and are fit with a linear model, i.e. the Slow component. The results are shown in Fig.~\ref{fig:LC11}. For all the X-ray and VHE gamma-ray bands from these 10-hour long light curves, the slopes are all consistent within the error bars.

\subsection{12 April 2013}
\begin{figure*}
\plottwo{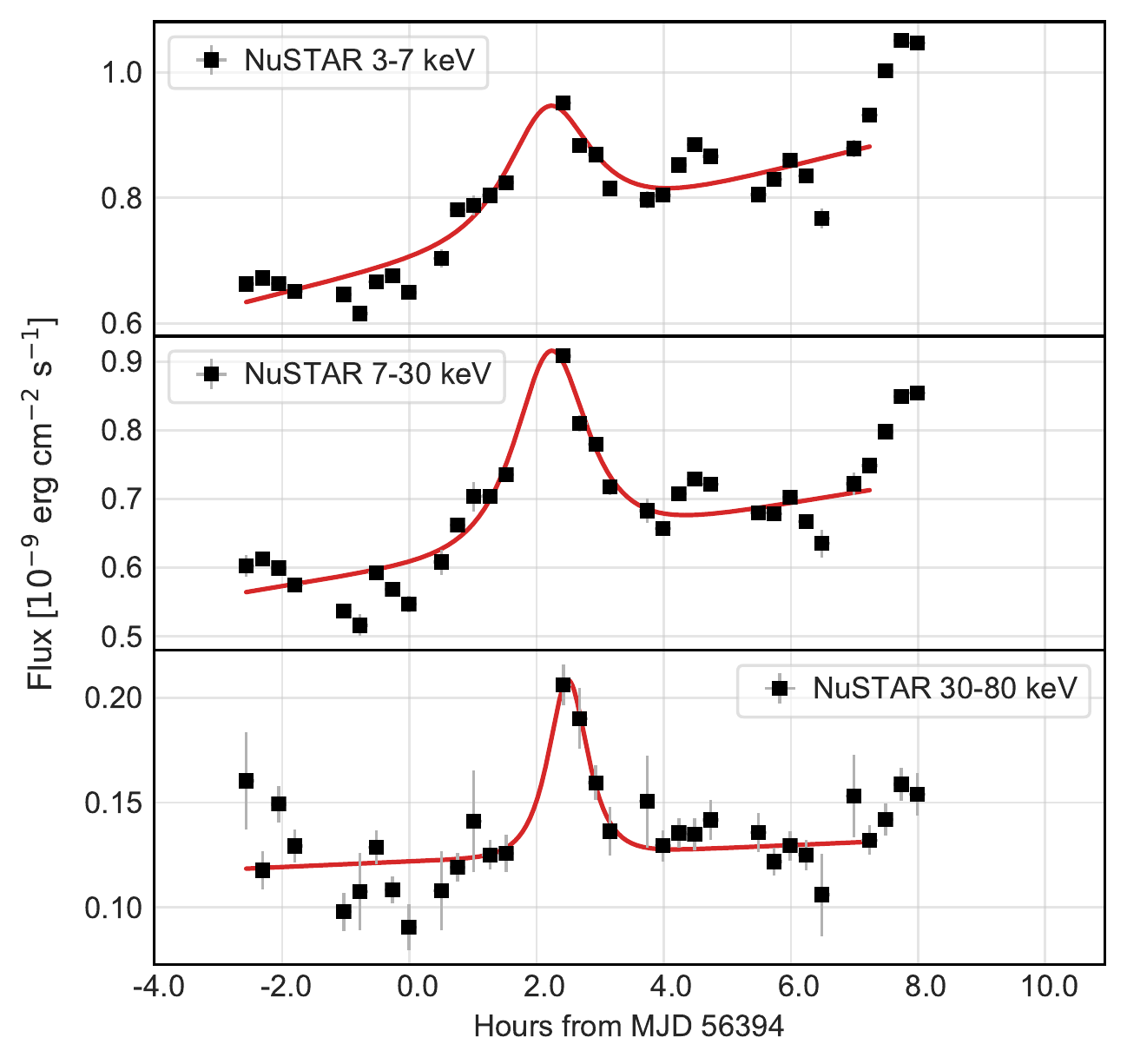}{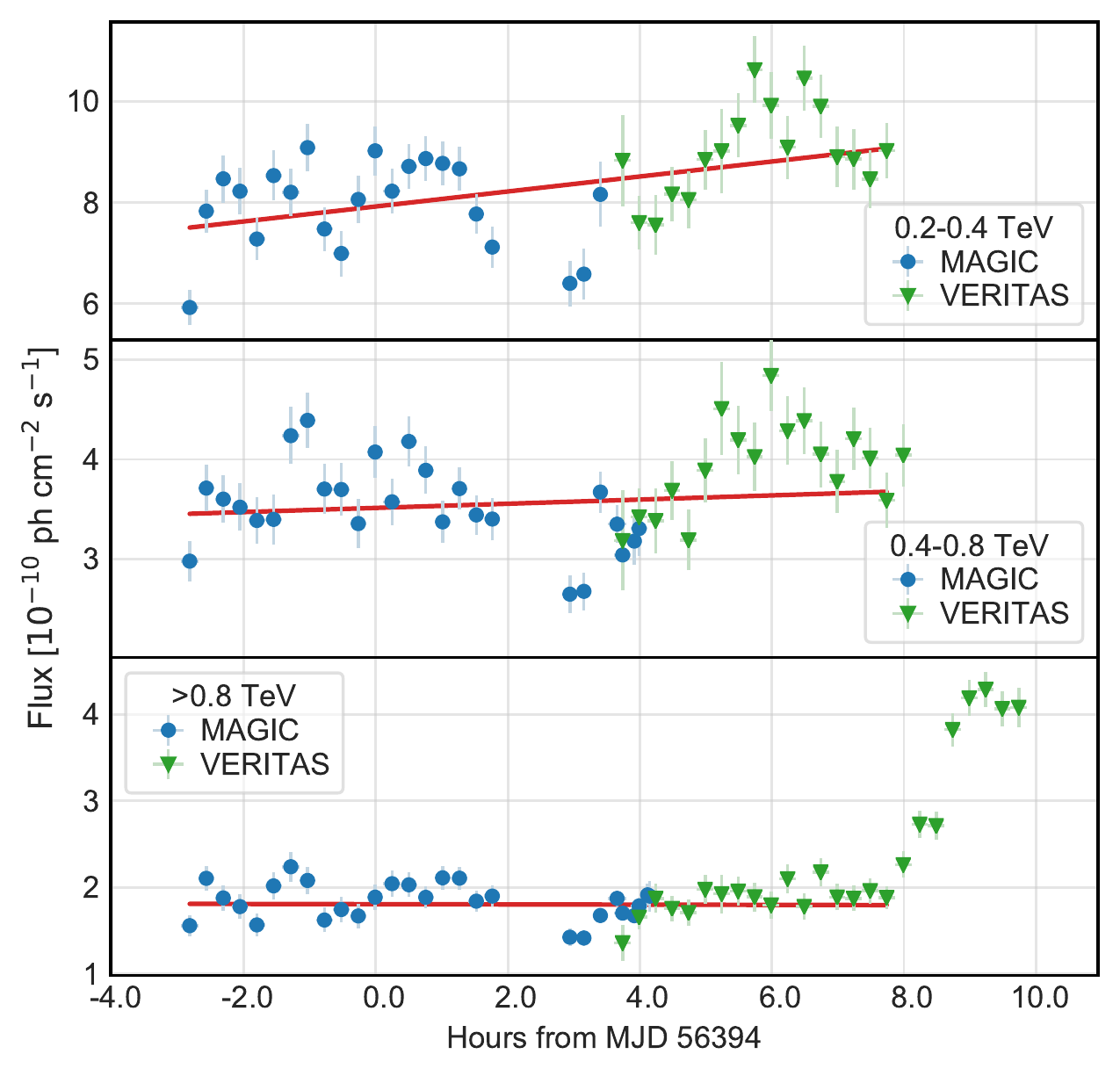}
\caption{Light curves from 2013 April 12 in three X-ray bands (left panel) and three VHE gamma-ray bands (right panel). The red curve is the resulting fit with the function defined by Eq.~\ref{eq:flux_t}, whose model parameters are reported in Table~\ref{tab:FitParametersDailyOthers}.}
\label{fig:LC12}
\end{figure*}

There is a clear flare visible in all X-ray bands around 2 hours past midnight; but unfortunately there is a gap in the VHE coverage for that time interval.  Therefore VHE light curves are fit with a linear function (Slow component) only. The results are shown in Fig.~\ref{fig:LC12}. All fits are performed only up to 7.7~hrs past midnight. There is flickering in VHE bands, with light curves  inconsistent with constant flux, and simple linear model or Slow+Fast component model from section \ref{subsec:fastvariability} does not describe data well either. Fits with Slow component give low slope values, consistent with zero for bands above 0.4~TeV. 

Slopes of the Slow component decrease with increasing energy in X-ray bands as well, while flare amplitude increases with energy.

\subsection{13 April 2013}
\begin{figure*}
\plottwo{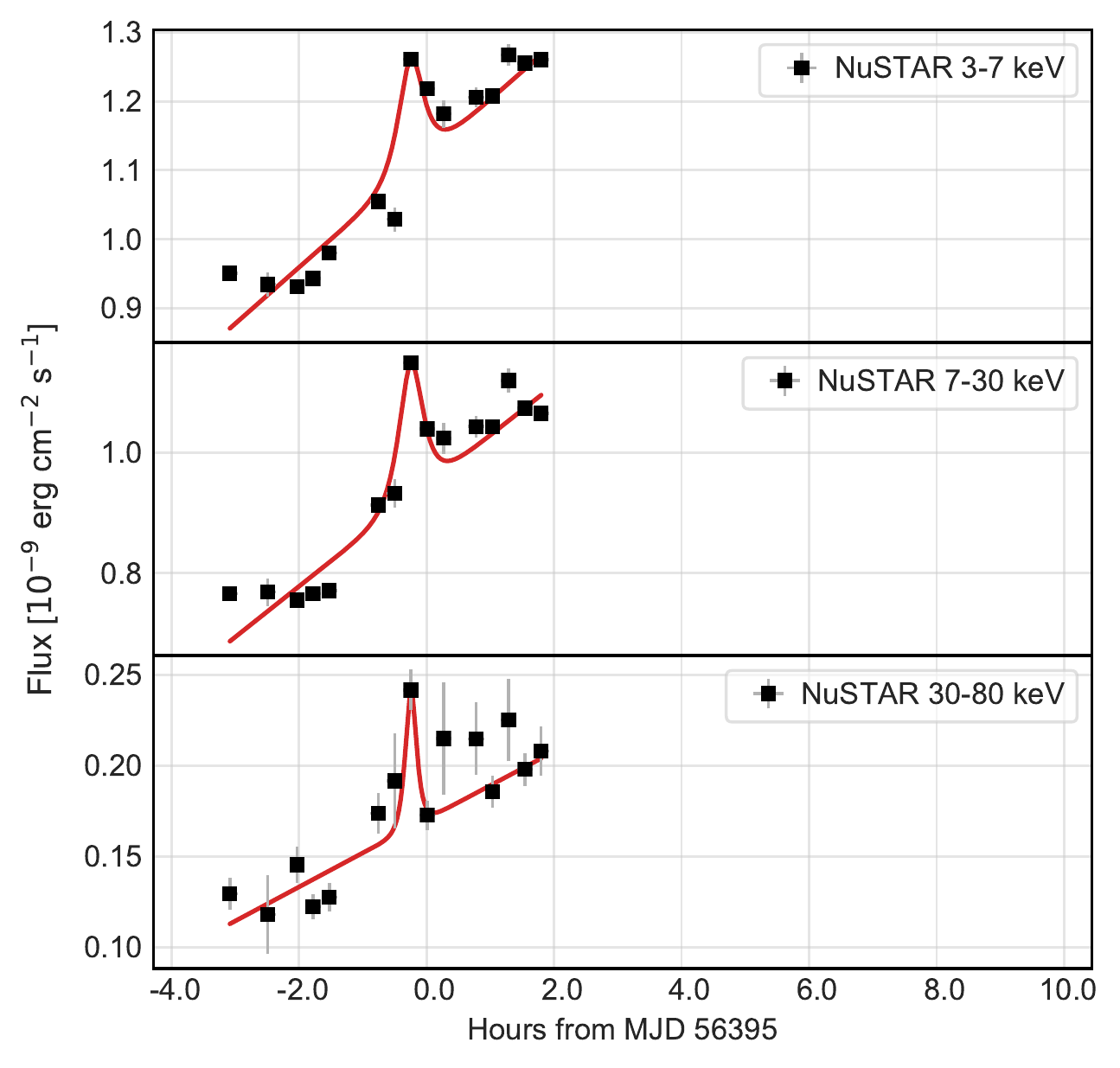}{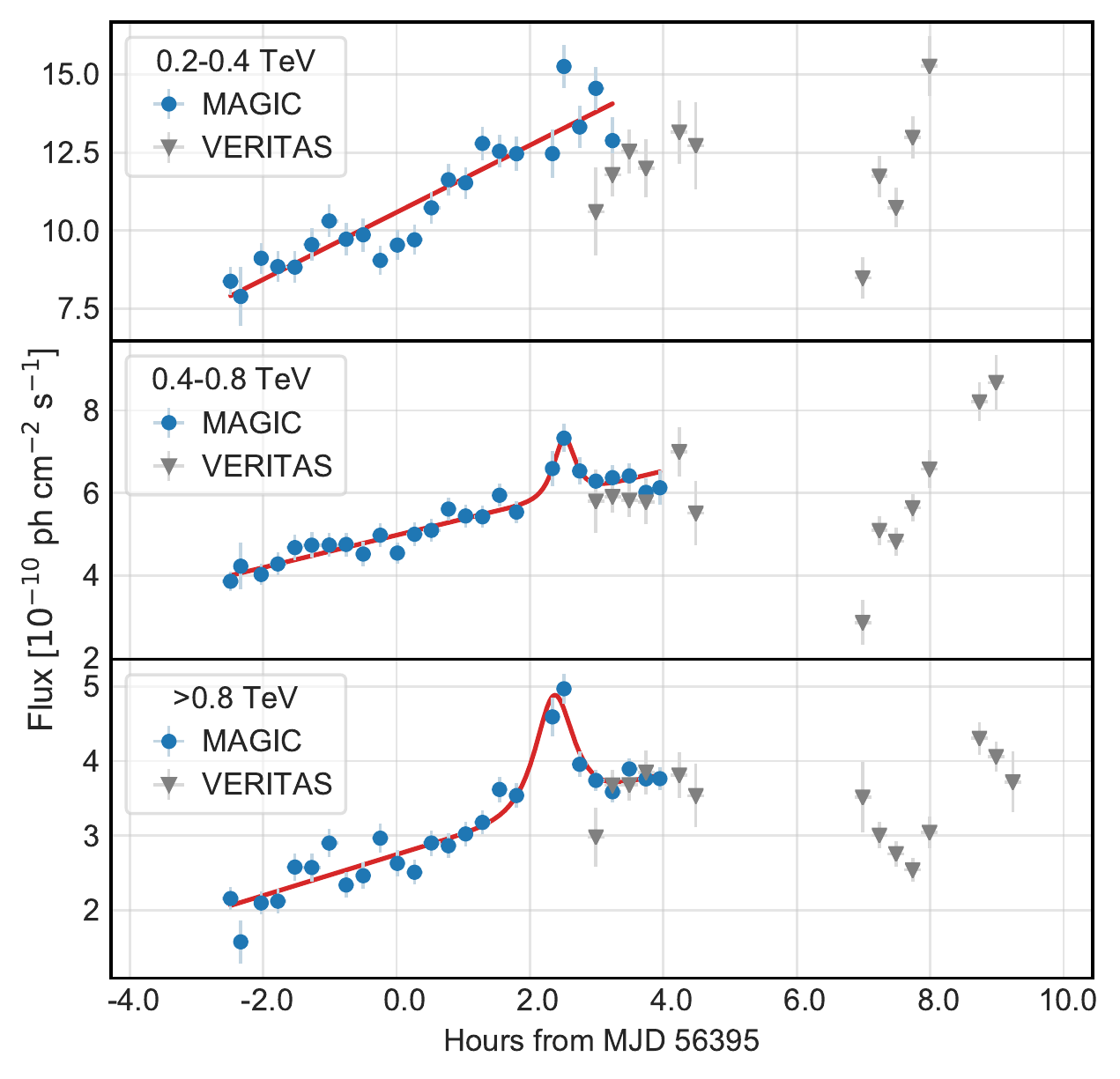}
\caption{Light curves from 2013 April 13 in three X-ray bands (left panel) and three VHE gamma-ray bands (right panel). The red curve is the resulting fit with the function defined by Eq.~\ref{eq:flux_t}, whose model parameters are reported in Table~\ref{tab:FitParametersDailyOthers}.}
\label{fig:LC13}
\end{figure*}

There is a flare visible in the two higher VHE bands, with the amplitude and the width of the flare both approximately halved in the 0.4-0.8~TeV band w.r.t. the $>0.8$~TeV band. In the lowest VHE band, there are flux measurements above the Slow component at the location of the flare, but no satisfactory Slow+Fast model fit (section \ref{subsec:fastvariability}) could be obtained. There is a $\sim 2$-hour gap between the MAGIC and VERITAS data, and the fit is performed only with MAGIC data point, denoted by using gray instead of green color for VERITAS data in Fig.~\ref{fig:LC13}.

There is no X-ray data covering the time of the VHE flare. On the other hand, there is a small-amplitude flare around 2.5~hrs earlier visible in X-ray bands, without obvious counterpart in VHE bands. Being relatively weak mini-flare, with the highest flux measured at the same time in all three X-ray bands, the fit is performed with Slow+Fast model, but with flare time (location) fixed to the time of the highest flux measurement, at midnight - 0.24~hrs.

\subsection{ 14 April 2013}
\begin{figure*}
\plottwo{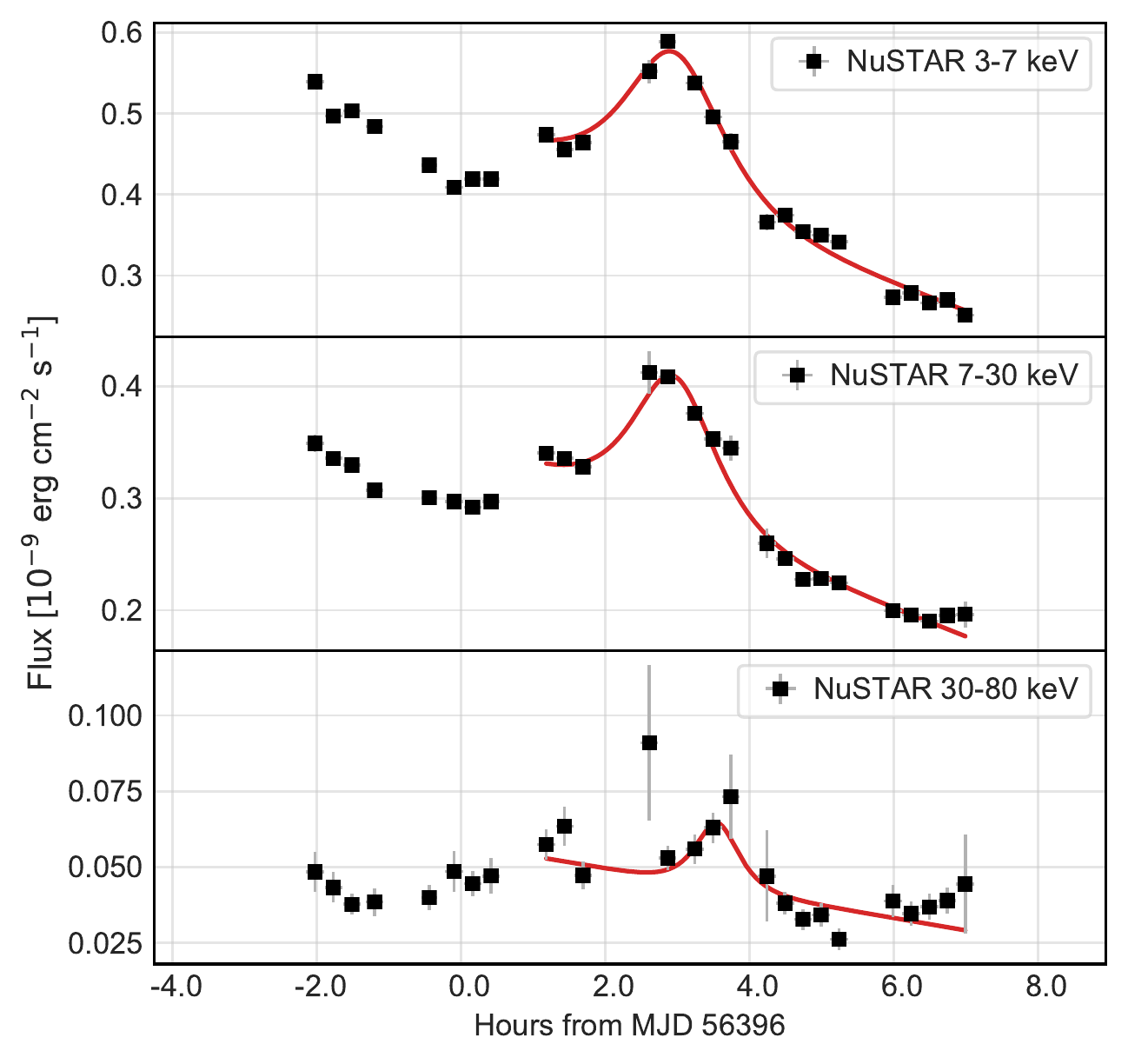}{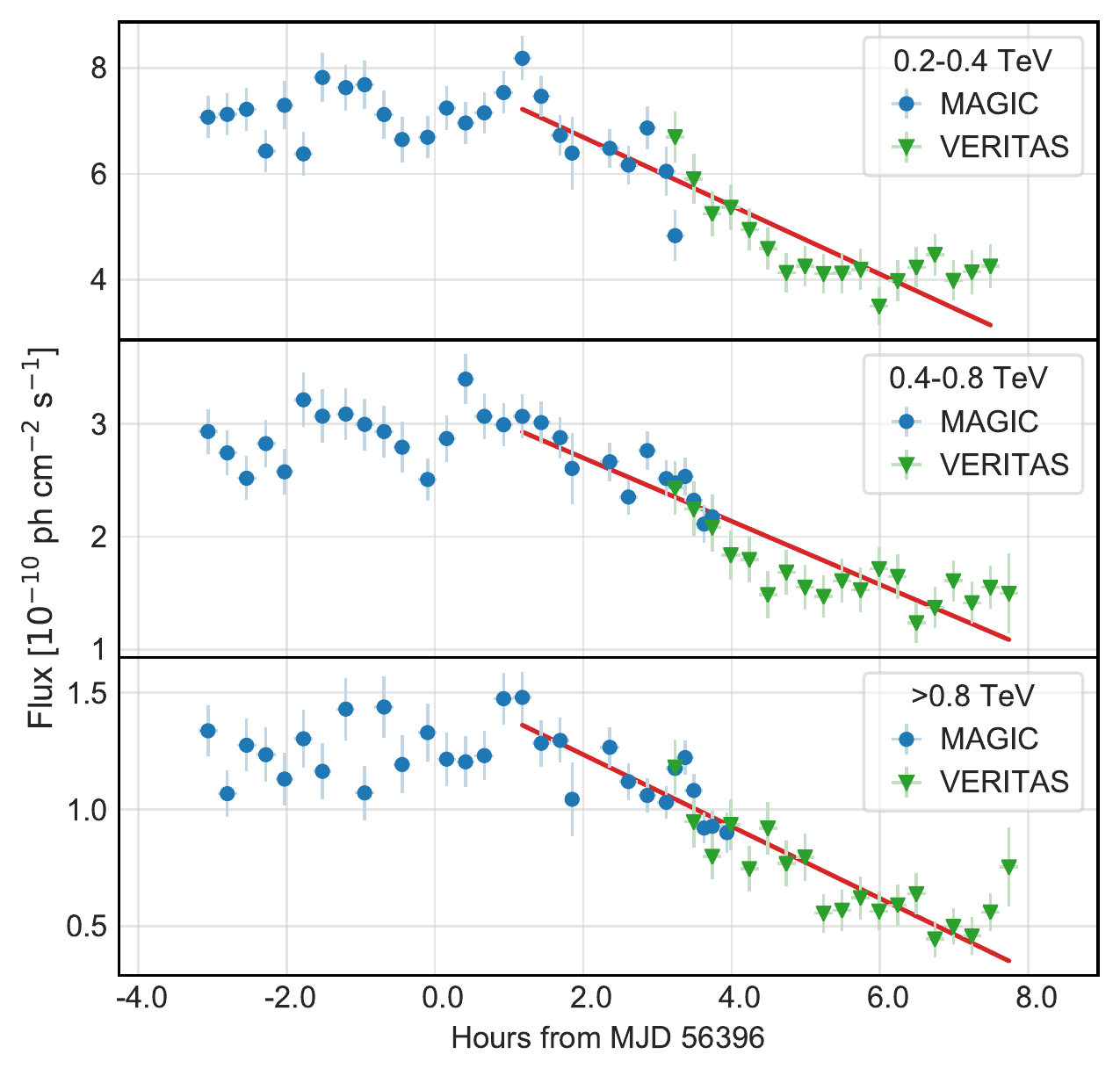}
\caption{Light curves from 2013 April 14 in three X-ray bands (left panel) and three VHE gamma-ray bands (right panel). The red curve is the resulting fit with the function defined by Eq.~\ref{eq:flux_t}, whose model parameters are reported in Table~\ref{tab:FitParametersDailyOthers}.}
\label{fig:LC14}
\end{figure*}

There is no obvious flare component in the VHE bands, therefore VHE bands are fit with Slow components only (Fig.~\ref{fig:LC14}). In X-ray bands the flux goes down, then increases in a broad flare visible in two lower X-ray bands, while no such behavior is seen in VHE bands.  Flare amplitudes in X-ray bands are consistent within the error bars, but there are some data missing during the time covered by the Fast component, therefore no definite conclusions about the flaring component can be made.  

The second part of the light curves (about 1~hour after midinght) has more simple structure, with a significant decrease in flux. Slow component slopes in all bands are similar.

\subsection{16 April 2013}
There are no characteristic mini-flares in the 10-hour long light curve from April 16. Instead, there is a prominent rise and then fall light curves from all energy bands. This behavior cannot be fit by the Slow+Fast model from section \ref{subsec:fastvariability}. Because of that, we modify the model to exclude the Slow component, and only have a broad 'bump' with different rise and fall times:
\begin{equation}
F_B\,(t) =   
\frac{2}{2^{-\frac{t-t_0}{t_\mathrm{rise}}} + 2^{\frac{t-t_0}{t_\mathrm{fall}}}} 
\cdot A
\label{eq:bump}
\end{equation}

The fit is performed to all three X-ray bands simultaneously, giving a unique rise time \mbox{$t_\mathrm{rise} = 1.38 \pm 0.23$~hrs,} fall time \mbox{$t_\mathrm{fall} = 4.96 \pm 0.47$~hrs,} and break time \\ 
\mbox{$t_0 = -0.42 \pm 0.34$~hrs} from midnight, with \\
\mbox{$\chi^2 / \mathrm{d.o.f.} = 582 / 57$. }
Only flare amplitude is allowed to vary in different bands, with results reported in Table \ref{tab:FitParameters16}. After fitting to X-ray data, the function shape is fit to each VHE band, again allowing only the flare amplitude to vary. The results, along with the $\chi^2$ values for each band are reported in  Table \ref{tab:FitParameters16}.

\begin{figure*}
\plottwo{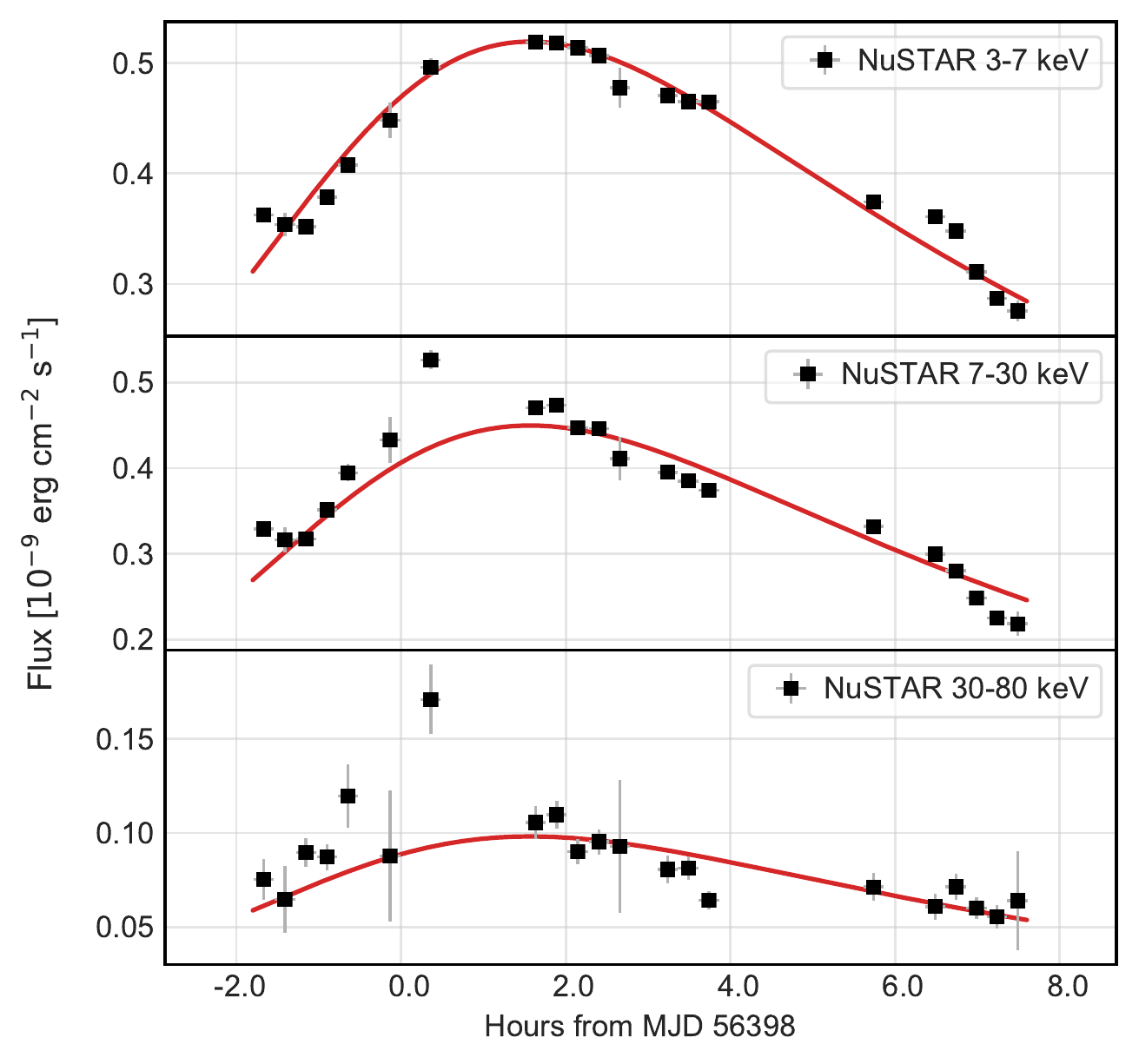}{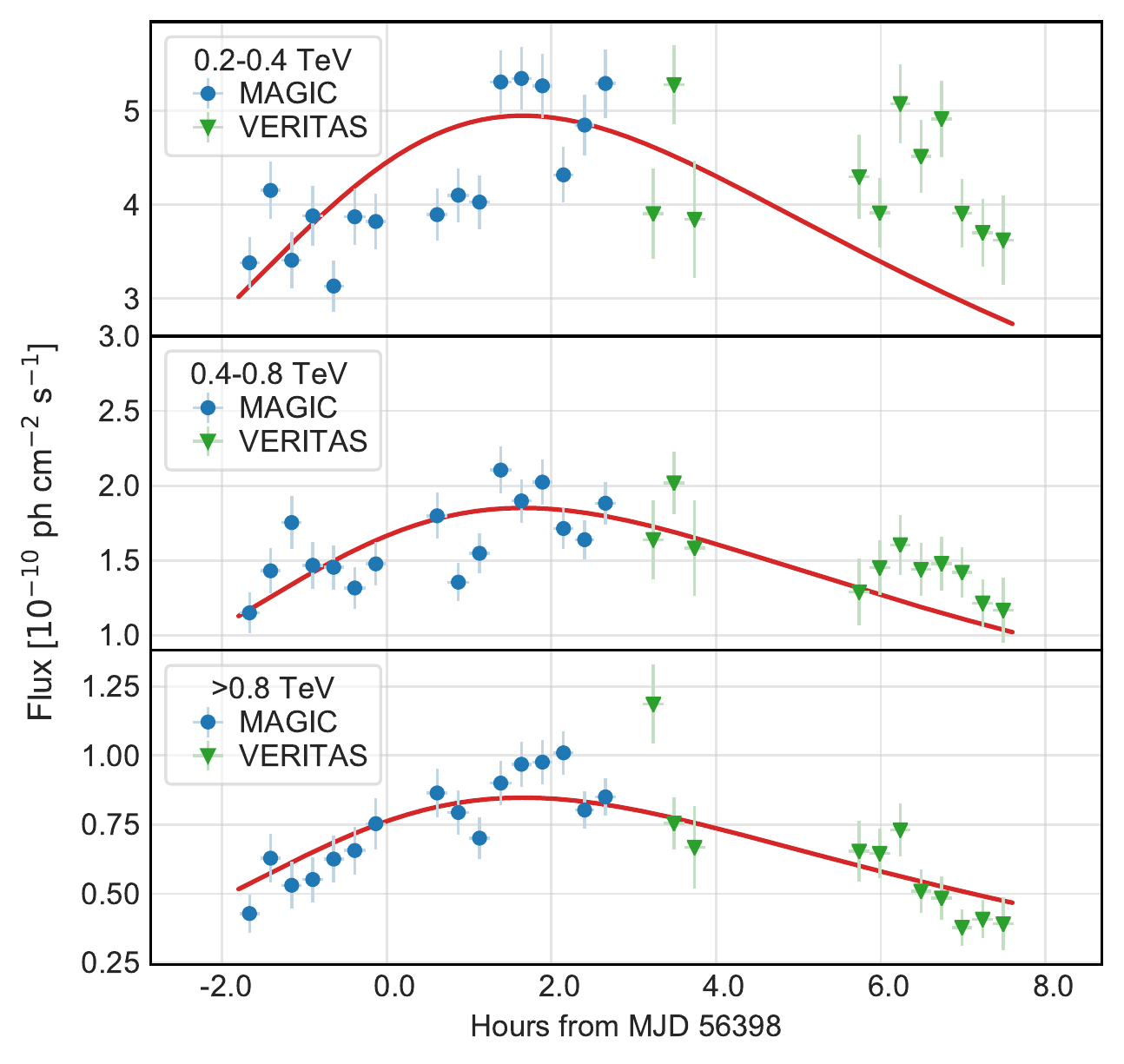}
\caption{Light curves from 2013 April 16 in three X-ray bands (left panel) and three VHE gamma-ray bands (right panel). The red curve is the resulting fit with the function defined by Eq.~\ref{eq:bump}, whose model parameters are reported in Table~\ref{tab:FitParameters16}.}
\label{fig:LC16}
\end{figure*}

\begin{deluxetable*}{c|c|c|c|c|c|r}
\tablecaption{Parameters resulting from the fit with Eq.~\ref{eq:flux_t} to the X-ray and VHE multi-band light curves for 2013 April 11-14.  \label{tab:FitParametersDailyOthers}}
\tablewidth{0pt}
\tablehead{
Band & {\it Offset}\tablenotemark{a}  & $Slope$ &   Flare  &  Flare  & Flare  & $\chi^2$/d.o.f \\
   &     & [h$^{-1}$] & Amplitude $A$& flux-doubling time\tablenotemark{b} [h] & $t_0$ [h] &   
}
\startdata
 \multicolumn7l{ 11 April 2013 }  \\ \cline{1-7}
3-7~keV   &  $0.326 \pm 0.004$ & $-0.057 \pm 0.002 $& - & - & - & 384 / 28 \\
7-30~keV  &  $0.205 \pm 0.004$ & $-0.058 \pm 0.003 $& - & - & - & 320 / 28 \\
30-80~keV &  $0.028 \pm 0.001$ & $ -0.051 \pm 0.007$& - & - & - & 37.7 / 28 \\
0.2-0.4 TeV & $4.37 \pm 0.10$ & $-0.051 \pm 0.006$ &  - & - & - & 74.7 / 35 \\
0.4-0.8 TeV & $1.50 \pm 0.04$ & $-0.055 \pm 0.006$ &  - & - & - & 58.1 / 37 \\
$>$0.8 TeV  & $0.50 \pm 0.02$ & $-0.058 \pm 0.007$ &  - & - & - & 44.0 / 37 \\
 \cline{1-7}\multicolumn7l{ 12 April 2013 }  \\ \cline{1-7}
3-7~keV   & $0.70 \pm 0.01  $ & $0.041 \pm 0.003$ & $0.26 \pm 0.07$ & $0.34 \pm 0.09$ &  $2.2  \pm 0.1 $ & 764 / 28 \\
7-30~keV  & $0.604 \pm 0.009$ & $0.029 \pm 0.004$ & $0.44 \pm 0.08$ & $0.32 \pm 0.05$ &  $2.21 \pm 0.08$ & 449 / 28 \\
30-80~keV & $0.122 \pm 0.003$ & $0.013 \pm 0.006$ & $0.66 \pm 0.12$ & $0.18 \pm 0.07$ &  $2.5  \pm 0.1 $ & 53.9 / 28 \\
0.2-0.4 TeV & $7.9  \pm 0.2 $ & $ 0.019 \pm 0.006$ &  - &  - & - & 131 / 37 \\
0.4-0.8 TeV & $3.51 \pm 0.09$ & $ 0.006 \pm 0.007$ &  - &  - & - & 146 / 42 \\
$>$0.8 TeV  & $1.81 \pm 0.04$ & $-0.001 \pm 0.006$ &  - &  - & - & 122 / 51 \\
 \cline{1-7}\multicolumn7l{ 13 April 2013 }  \\ \cline{1-7}
3-7~keV   &  $1.12 \pm 0.01$ & $0.15 \pm 0.04$ & $0.12  \pm 0.04 $ & $0.073 \pm 0.007$ & - & 237 / 11 \\
7-30~keV  &  $0.94 \pm 0.02$ & $0.24 \pm 0.05$ & $0.11  \pm 0.03 $ & $0.089 \pm 0.008$ & - & 165 / 11 \\
30-80~keV &  $0.17 \pm 0.01$ & $0.46 \pm 0.11$ & $0.053 \pm 0.039$ & $0.11  \pm 0.02 $ & - & 25.3 / 11 \\
0.2-0.4 TeV & $10.6 \pm 0.2 $ & $0.10  \pm 0.01 $ &    -             &      -            &      -    &  40.2 / 21 \\
0.4-0.8 TeV & $4.97 \pm 0.05$ & $0.079 \pm 0.006$ &  $0.23 \pm 0.05$ & $0.098 \pm 0.029$ & $2.52 \pm 0.05$ &  12.6 / 21 \\
$>$0.8 TeV  & $2.74 \pm 0.05$ & $0.10  \pm 0.01 $ &  $0.44 \pm 0.08$ & $0.17  \pm 0.04 $ & $2.36 \pm 0.07$ &  33.3 / 21 \\
 \cline{1-7}\multicolumn7l{ 14 April 2013 }  \\ \cline{1-7}
3-7~keV   & $0.49  \pm 0.03 $ & $-0.068 \pm 0.005$ & $0.48 \pm 0.06$ & $0.42 \pm 0.09$ & $2.95 \pm 0.09$ & 158 / 21 \\
7-30~keV  & $0.35  \pm 0.02 $ & $-0.071 \pm 0.004$ & $0.47 \pm 0.05$ & $0.35 \pm 0.08$ & $2.95 \pm 0.09$ & 74.6 / 21 \\
30-80~keV & $0.058 \pm 0.006$ & $-0.071 \pm 0.013$ & $0.50 \pm 0.23$ & $0.18 \pm 0.13$ & $3.5  \pm 0.3 $ & 50.8 / 21 \\
0.2-0.4 TeV & $8.0  \pm 0.3 $ & $-0.081 \pm 0.005$ &  - &  - & - & 75.8 / 41 \\
0.4-0.8 TeV & $3.3  \pm 0.1 $ & $-0.086 \pm 0.004$ &  - &  - & - & 60.4 / 46 \\
$>$0.8 TeV  & $1.54 \pm 0.05$ & $-0.100 \pm 0.004$ &  - &  - & - & 61.5 / 47 \\ 
\enddata
\tablenotetext{a}{For VHE bands in $10^{-10}$ ph cm$^{-2}$ s$^{-1}$, for X-ray bands in $10^{-9}$ erg cm$^{-2}$ s$^{-1}$.}
\tablenotetext{b}{Parameters $t_\mathrm{rise}$ and $t_\mathrm{fall}$ in Eq.~\ref{eq:flux_t} are set to be equal, and correspond to the flare flux-doubling time in this Table.}
\end{deluxetable*}

\begin{table}
\centering
\caption{Parameters resulting from the fit with Eq.~(\ref{eq:bump}) to the X-ray and VHE multi-band light curves from 2013 April 16. The three X-ray bands are fit with a single function, yielding $t_\mathrm{rise} = 1.4 \pm 0.2$~hrs, $t_\mathrm{fall} = 5.0 \pm 0.5$~hrs, break time $t_0 = -0.42 \pm 0.34$~hrs from midnight, with \mbox{$\chi^2 / \mathrm{d.o.f.} = 582 / 57$. } \label{tab:FitParameters16}}
\medskip
\begin{tabular}{r|c|r}
\hline
Band &   Flare  Amplitude $A$\tablenotemark{a}  & $\chi^2 / \mathrm{d.o.f.}$ \\
     \hline
3-7~keV &  $ 0.44 \pm 0.02 $ & -  \\
7-30~keV &  $ 0.38 \pm 0.02 $ &  -  \\
30-80~keV &  $ 0.083 \pm  0.007 $ &  - \\
0.2-0.4~TeV   & $4.2 \pm  0.2$ & 140 / 26\\
0.4-0.8~TeV   & $ 1.57 \pm 0.05 $ & 55 / 26 \\
$>$0.8~TeV  & $ 0.72 \pm 0.02 $ & 39 / 26  \\
\hline
\end{tabular}
\tablenotetext{a}{The flare amplitude $A$ is given in in $10^{-9}$ erg cm$^{-2}$ s$^{-1}$ for the X-ray bands, and $10^{-10}$ ph cm$^{-2}$ s$^{-1}$ for the VHE gamma-ray bands.}
\label{tab:LCFitApril16}
\end{table}

\clearpage

\section{Single-day flux-flux correlations}
\label{app:CorrFitOtherDays}

In this section we report the quantification of the VHE vs X-ray correlations for single-night data. In particular, we report the Pearson correlation coefficients and related significances, the values of the DCF, and the slopes and the $\chi^2$ values from the linear fits to the $\log F_\mathrm{VHE} - \log F_\mathrm{X-ray}$. These numbers are complementary to those reported in Table~\ref{tab:CorrelationCoefficients}, which report the same information for the nine-day data set (instead of single-night data sets).

\startlongtable
\begin{deluxetable*}{c|r|c|c|c|c|r}
\tablecaption{Single-night correlation coefficients and fit slopes for 2019 April 11-16. The statistical and systematic contributions to the total uncertainty are reported separately, and in this order. \label{tab:CorrelationCoefficientsDaily}}
\tablewidth{0pt}
\tablehead{
\colhead{VHE band} & \colhead{X-ray band} & \colhead{Pearson coeff.}  & \colhead{$N\sigma$ (Pearson)} & \colhead{DCF}   & Linear fit slope & $\chi^2$/d.o.f 
}
\startdata
 \multicolumn7l{ 11 April 2013 }  \\ \cline{1-7}
0.2-0.4 TeV & 3-7 keV & $0.83^{ + 0.05}_{ - 0.07}\pm 0.02$ & 6.1 & $0.90 \pm 0.21 \pm 0.007$ & $0.86 \pm 0.09 \pm 0.05$ &43 / 28 \\ 
          & 7-30 keV & $0.84^{ + 0.05}_{ - 0.07 }\pm 0.01$& 6.4 & $0.92 \pm 0.22\pm 0.003$ & $0.83 \pm 0.08 \pm 0.05 $ & 37 / 28 \\ 
          & 30-80 keV & $0.74^{ + 0.08 }_{- 0.10 }\pm 0.002$ & 4.9 & $1.35 \pm 0.37 \pm 0.02$ & $1.02 \pm 0.15 \pm 0.07$ &28 / 28 \\ 
0.4-0.8 TeV & 3-7 keV & $0.89^{ + 0.03 }_{- 0.05}\pm 0.04$ & 7.3 & $1.06 \pm 0.26\pm 0.02$ & $1.01 \pm 0.09\pm 0.11$ &26 / 28 \\ 
          & 7-30 keV & $0.86^{ + 0.04}_{ - 0.06}\pm 0.03$ & 6.7 & $1.03 \pm 0.26 \pm 0.03$ & $0.95 \pm 0.09\pm 0.10$ &30 / 28 \\ 
          & 30-80 keV & $0.64^{ + 0.10}_{ - 0.13}\pm 0.01$ & 4.0 & $1.29 \pm 0.45\pm 0.06$ & $1.08 \pm 0.15\pm 0.13$ &25 / 28 \\ 
$>$0.8 TeV & 3-7 keV & $0.79^{ + 0.06}_{ - 0.08}\pm 0.04$ & 5.6 & $1.04 \pm 0.28 \pm 0.01$ & $1.01 \pm 0.13 \pm 0.10$ & 28 / 28 \\ 
          & 7-30 keV & $0.78^{ + 0.07}_{ - 0.09}\pm 0.04$ & 5.4 & $1.03 \pm 0.28\pm 0.02$ & $0.96 \pm 0.12\pm 0.09$ &27 / 28 \\ 
          & 30-80 keV & $0.61^{ + 0.11 }_{- 0.14}\pm 0.02$ & 3.7 & $1.35 \pm 0.47 \pm 0.05$& $1.2 \pm 0.2 \pm 0.1$ &25 / 28 \\ 
\cline{1-7} \multicolumn7l{ 12 April 2013 }  \\ \cline{1-7}
0.2-0.4 TeV & 3-7 keV & $0.26^{ + 0.17}_{ - 0.19}\pm 0.10$ & 1.4 & $0.31 \pm 0.17\pm 0.11$ & $0.21 \pm 0.14 \pm 0.09$ &78 / 27 \\ 
          & 7-30 keV & $0.16^{ + 0.18}_{ - 0.19}\pm 0.09$ & 0.8 & $0.19 \pm 0.19\pm 0.10$ & $0.15 \pm 0.09 \pm 0.09$ &82 / 27 \\ 
          & 30-80 keV & $-0.19^{ + 0.19}_{ - 0.18 }\pm 0.07$& 1.0 & $-0.31 \pm 0.24\pm 0.12$ & $-1.03 \pm 0.39 \pm 14$ &66 / 27 \\ 
0.4-0.8 TeV & 3-7 keV & $0.05^{ + 0.18 }_{- 0.19}\pm 0.19$ & 0.3 & $0.06 \pm 0.16\pm 0.23$ & $0.02 \pm 0.16\pm 0.18$ &90 / 30 \\ 
          & 7-30 keV & $-0.07^{ + 0.19}_{ - 0.18}\pm 0.17$ & 0.4 & $-0.09 \pm 0.18 \pm 0.21$ & $-0.12 \pm 0.18 \pm 0.18$ &89 / 30 \\ 
          & 30-80 keV & $-0.37^{ + 0.17}_{ - 0.15}\pm 0.12$ & 2.1 & $-0.65 \pm 0.27 \pm 0.23$ & $-0.76 \pm 0.23 \pm 20$ &59 / 30 \\ 
$>$0.8 TeV & 3-7 keV & $0.19^{ + 0.17}_{ - 0.18}\pm 0.19$ & 1.0 & $0.27 \pm 0.25\pm 0.24$ & $0.15 \pm 0.14\pm 0.17$ &61 / 30 \\ 
          & 7-30 keV & $0.18^{ + 0.17}_{ - 0.18 }\pm 0.16$& 1.0 & $0.26 \pm 0.27\pm 0.21$ & $0.18 \pm 0.16 \pm 0.17$ &61 / 30 \\ 
          & 30-80 keV & $-0.04^{ + 0.19}_{ - 0.18 }\pm 0.09$& 0.2 & $-0.07 \pm 0.36\pm 0.17$ & $0.48 \pm 0.20\pm 0.79$ &61 / 30 \\ 
\cline{1-7} \multicolumn7l{ 13 April 2013$^{*}$}  \\ \cline{1-7}
0.2-0.4 TeV & 3-7 keV & $0.74^{ + 0.11 }_{- 0.17 }$ & 3.2 & 0.73 $\pm$ 0.24 & 0.98 $\pm$ 0.24 &52 / 12 \\ 
          & 7-30 keV & $0.70^{ + 0.13}_{ - 0.19}$ & 2.9 & 0.69 $\pm$ 0.24 & 0.75 $\pm$ 0.21 &60 / 12 \\ 
          & 30-80 keV & $0.65^{ + 0.14}_{ - 0.21}$ & 2.6 & 0.70 $\pm$ 0.28 & 0.58 $\pm$ 0.17 &49 / 12 \\ 
0.4-0.8 TeV & 3-7 keV & $0.86^{ + 0.06 }_{- 0.11 }$& 4.2 & 0.89 $\pm$ 0.25 & 0.94 $\pm$ 0.16 &19 / 12 \\ 
          & 7-30 keV & $0.81^{ + 0.08}_{ - 0.13}$ & 3.7 & 0.84 $\pm$ 0.25 & 0.72 $\pm$ 0.15 &24 / 12 \\ 
          & 30-80 keV & $0.76^{ + 0.10}_{ - 0.16}$ & 3.3 & 0.85 $\pm$ 0.28 & 0.50 $\pm$ 0.12 &23 / 12 \\ 
$>$0.8 TeV & 3-7 keV & $0.873^{ + 0.056}_{ - 0.094}$ & 4.5 & 0.87 $\pm$ 0.22 & 1.40 $\pm$ 0.23 &28 / 12 \\ 
          & 7-30 keV & $0.81^{ + 0.08}_{ - 0.13 }$& 3.8 & 0.81 $\pm$ 0.22 & 1.07 $\pm$ 0.23 &41 / 12 \\ 
          & 30-80 keV & $0.70^{ + 0.12}_{ - 0.19}$ & 2.9 & 0.76 $\pm$ 0.24 & 0.85 $\pm$ 0.21 &36 / 12 \\ 
\cline{1-7} \multicolumn7l{ 14 April 2013 }  \\    \cline{1-7}
0.2-0.4 TeV & 3-7 keV & $0.79^{ + 0.07 }_{- 0.09} \pm 0.003$ & 5.2 & $0.79 \pm 0.17 \pm 0.002$ & $0.82 \pm 0.15 \pm 0.05$ &130 / 25 \\ 
          & 7-30 keV & $0.78^{ + 0.07}_{ - 0.09} \pm 0.003$ & 5.2 & $0.79 \pm 0.16 \pm 0.002$ & $0.87 \pm 0.15 \pm 0.06$ &126 / 25 \\ 
          & 30-80 keV & $0.45^{ + 0.15}_{ - 0.18} \pm 0.002$ & 2.4 & $0.55 \pm 0.20  \pm 0.006$ & $1.5 \pm 0.4  \pm 0.1$ &76 / 25 \\ 
0.4-0.8 TeV & 3-7 keV & $0.78^{ + 0.07}_{ - 0.09 } \pm 0.006$ & 5.3 & $0.79 \pm 0.17 \pm 0.006$ & $0.79 \pm 0.18 \pm 0.11$ &138 / 27 \\ 
          & 7-30 keV & $0.78^{ + 0.07}_{ - 0.09} \pm 0.009$ & 5.3 & $0.80 \pm 0.17 \pm 0.006$ & $0.85 \pm 0.19 \pm 0.12$ &138 / 27 \\ 
          & 30-80 keV & $0.41^{ + 0.15}_{ - 0.17} \pm 0.007$ & 2.2 & $0.51 \pm 0.19 \pm 0.002$ & $1.7 \pm 0.5  \pm 0.3$ &94 / 27 \\ 
$>$0.8 TeV & 3-7 keV & $0.83^{ + 0.05}_{ - 0.07 } \pm 0.003$& 6.1 & $0.85 \pm 0.21 \pm 0.007$ & $0.94 \pm 0.18 \pm 0.10$ &107 / 27 \\ 
          & 7-30 keV & $0.81^{ + 0.06 }_{- 0.08} \pm 0.006$ & 5.7 & $0.83 \pm 0.19 \pm 0.003$ & $0.93 \pm 0.19  \pm 0.11$ &112 / 27 \\ 
          & 30-80 keV & $0.42^{ + 0.15 }_{- 0.17} \pm 0.007$ & 2.3 & $0.51 \pm 0.20 \pm 0.003$ & $1.6 \pm 0.4 \pm 0.2$ &86 / 27 \\ 
\cline{1-7} \multicolumn7l{ 15 April 2013 }  \\    \cline{1-7}
0.2-0.4 TeV & 3-7 keV & $0.77^{ + 0.07}_{ - 0.09} \pm 0.04$ & 5.3 & $0.82 \pm 0.20 \pm 0.04$ & $0.40 \pm 0.05 \pm 0.03$ &68 / 28 \\ 
          & 7-30 keV & $0.75^{ + 0.07}_{ - 0.10 } \pm 0.05$& 5.1 & $0.81 \pm 0.19  \pm 0.04$& $0.33 \pm 0.05 \pm 0.02$ &72 / 28 \\ 
          & 30-80 keV & $0.71^{ + 0.08}_{ - 0.11 } \pm 0.05$& 4.6 & $0.77 \pm 0.18 \pm 0.05$ & $0.29 \pm 0.05 \pm 0.02$ &80 / 28 \\ 
0.4-0.8 TeV & 3-7 keV & $0.80^{ + 0.06}_{ - 0.08 } \pm 0.08$& 5.8 & $0.86 \pm 0.19 \pm 0.08$ & $0.52 \pm 0.07 \pm 0.04$ &80 / 29 \\ 
          & 7-30 keV & $0.78^{ + 0.06}_{ - 0.09} \pm 0.08$ & 5.5 & $0.84 \pm 0.19 \pm 0.08$ & $0.41 \pm 0.06 \pm 0.03$ &88 / 29 \\ 
          & 30-80 keV & $0.73^{ + 0.08}_{ - 0.10} \pm 0.09$ & 5.0 & $0.81 \pm 0.18 \pm 0.09$ & $0.36 \pm 0.06  \pm 0.03$ &93 / 29 \\ 
$>$0.8 TeV & 3-7 keV & $0.95 \pm 0.02 \pm 0.02$ & 9.5 & $0.94 \pm 0.19 \pm 0.02$ & $0.91 \pm 0.07 \pm 0.05$ &77 / 29 \\ 
          & 7-30 keV & $0.94^{ + 0.02}_{ - 0.03 } \pm 0.02$& 9.1 & $0.94 \pm 0.19 \pm 0.02$ & $0.74 \pm 0.06 \pm 0.04$ &87 / 29 \\ 
          & 30-80 keV & $0.88^{ + 0.04}_{ - 0.05} \pm 0.03$ & 7.3 & $0.90 \pm 0.15 \pm 0.02$ & $0.63 \pm 0.06 \pm 0.03$ &90 / 29 \\ 
\cline{1-7} \multicolumn7l{ 16 April 2013 }  \\    \cline{1-7}
0.2-0.4 TeV & 3-7 keV & $0.53^{ + 0.15}_{ - 0.20 } \pm 0.05$& 2.5 & $0.62 \pm 0.22  \pm 0.05$ & $0.56 \pm 0.16 \pm 0.04$ &51 / 18 \\ 
          & 7-30 keV & $0.42^{ + 0.18}_{ - 0.22} \pm 0.06$ & 1.8 & $0.49 \pm 0.24  \pm 0.07$ & $0.39 \pm 0.16 \pm 0.05$ &62 / 18 \\ 
          & 30-80 keV & $0.21^{ + 0.22}_{ - 0.24 } \pm 0.08$& 0.9 & $0.47 \pm 0.61 \pm 0.17$ & $0.77 \pm 0.26 \pm 0.02$ &51 / 18 \\ 
0.4-0.8 TeV & 3-7 keV & $0.78^{ + 0.08}_{ - 0.11} \pm 0.06$ & 4.3 & $1.1 \pm 0.3 \pm 0.07$ & $0.67 \pm 0.13 \pm 0.09$ &19 / 18 \\ 
          & 7-30 keV & $0.72^{ + 0.10}_{ - 0.14} \pm 0.09$ & 3.7 & $1.0 \pm 0.3 \pm 0.09$ & $0.55 \pm 0.13 \pm 0.10$ &23 / 18 \\ 
          & 30-80 keV & $0.61^{ + 0.13}_{ - 0.17 } \pm 0.12$& 2.9 & $1.6 \pm 0.5  \pm 0.3$ & $0.68 \pm 0.15 \pm 0.11$ &18 / 18 \\ 
$>$0.8 TeV & 3-7 keV & $0.92^{ + 0.03}_{ - 0.05} \pm 0.02$ & 6.5 & $0.97 \pm 0.21 \pm 0.01$ & $1.6 \pm 0.2 \pm 0.09$ &19 / 18 \\ 
          & 7-30 keV & $0.88^{ + 0.04}_{ - 0.07} \pm 0.03$ & 5.8 & $0.94 \pm 0.23 \pm 0.03$ & $1.4 \pm 0.2 \pm 0.1$ &27 / 18 \\ 
          & 30-80 keV & $0.64^{ + 0.12}_{ - 0.17} \pm 0.05$ & 3.1 & $1.3 \pm 0.4 \pm 0.1$ & $1.8 \pm 0.4 \pm 0.06$ &32 / 18 \\ 
\enddata
\tablenotetext{*}{There are no simultaneous NuSTAR and VERITAS data on April 13, and hence there is no systematic error associated to the usage of the flux-scale factors. }
\end{deluxetable*}



\end{document}